\documentclass[12pt,letterpaper]{article}
\usepackage{amsmath,amssymb,array,calc,rotating,epsfig,psfrag,amscd, cite}

\makeatletter
\renewcommand\section{\@startsection {section}{1}{\z@}%
                                   {-3.5ex \@plus -1ex \@minus -.2ex}
                                   {2.3ex \@plus.2ex}%
                                   {\normalfont\large\bfseries}}
\renewcommand\subsection{\@startsection{subsection}{2}{\z@}%
                                     {-3.25ex\@plus -1ex \@minus -.2ex}%
                                     {1.5ex \@plus .2ex}%
                                     {\normalfont\bfseries}}
\makeatother

\let\non\nonumber

\let\s=\sigma

\newcommand{\bea}{\begin{eqnarray}}
\newcommand{\eea}{\end{eqnarray}}
\newcommand{\be}{\begin{equation}}
\newcommand{\ee}{\end{equation}}

\newcommand{\m}{\mu}
\newcommand{\n}{\nu}
\newcommand{\p}{\partial}

\newcommand{\C}[1]{$(\ref{#1})$}

\def\IZ{\relax\ifmmode\mathchoice
{\hbox{\cmss Z\kern-.4em Z}}{\hbox{\cmss Z\kern-.4em Z}}
{\lower.9pt\hbox{\cmsss Z\kern-.4em Z}} {\lower1.2pt\hbox{\cmsss
Z\kern-.4em Z}}\else{\cmss Z\kern-.4em Z}\fi}
\def\IR{\relax{\rm I\kern-.18em R}}

\def\one{{\hbox{ 1\kern-.8mm l}}}

\newlength{\bredde}
\def\slash#1{\settowidth{\bredde}{$#1$}\ifmmode\,\raisebox{.15ex}{/}
\hspace*{-\bredde} #1\else$\,\raisebox{.15ex}{/}\hspace*{-\bredde}
#1$\fi}

\newsavebox{\zzzbar}
\sbox{\zzzbar}
  {\setlength{\unitlength}{0.9em}
  \begin{picture}(0.6,0.7)
  \thinlines
  \put(0,0){\line(1,0){0.6}}
  \put(0,0.75){\line(1,0){0.575}}
  \multiput(0,0)(0.0125,0.025){30}{\rule{0.3pt}{0.3pt}}
  \multiput(0.2,0)(0.0125,0.025){30}{\rule{0.3pt}{0.3pt}}
  \put(0,0.75){\line(0,-1){0.15}}
  \put(0.015,0.75){\line(0,-1){0.1}}
  \put(0.03,0.75){\line(0,-1){0.075}}
  \put(0.045,0.75){\line(0,-1){0.05}}
  \put(0.05,0.75){\line(0,-1){0.025}}
  \put(0.6,0){\line(0,1){0.15}}
  \put(0.585,0){\line(0,1){0.1}}
  \put(0.57,0){\line(0,1){0.075}}
  \put(0.555,0){\line(0,1){0.05}}
  \put(0.55,0){\line(0,1){0.025}}
  \end{picture}}

\def\Im{{\rm Im ~}}

\newcommand{\ena}{\end{eqnarray}}
\newcommand{\beqa}{\begin{eqnarray}}
\newcommand{\eeqa}{\end{eqnarray}}

\newcommand{\g}{\gamma}

\def\g{\gamma}

\def\m{\mu}
\def\n{\nu}

\def\s{\sigma}

\numberwithin{equation}{section}

\begin{document}
\begin{titlepage}

\begin{center}

\hfill         \phantom{xxx}

\vskip 2 cm
{\Large \bf Supersymmetry constraints on the $\mathcal{R}^4$ multiplet in type
IIB on $T^2$}
\vskip 1.25 cm { Anirban Basu\footnote{email address:
    anirbanbasu@hri.res.in}}
{\vskip 0.5cm Harish-Chandra Research Institute,
Chhatnag Road, Jhusi, Allahabad 211019, India}

{\vskip 0.5cm and}

{\vskip 0.5cm Institute of Physics,
Sachivalaya Marg, Bhubaneswar, Orissa 751005, India}

\end{center}
\vskip 2 cm

\begin{abstract}
\baselineskip=18pt

We consider a class of eight derivative interactions in the effective action
of type IIB string theory compactified on $T^2$. These $1/2$ BPS interactions
have moduli dependent couplings. We impose the constraints of supersymmetry 
to show that each of these couplings satisfy a first order differential equation
on moduli space which relate it to other couplings in the same supermuliplet.
These equations can be iterated to give second order differential equations for the various couplings.
The couplings which only depend on the $SO(2) \backslash
SL(2,\mathbb{R})$ moduli satisfy Laplace equation on moduli space, and are
given by modular forms of $SL(2,\mathbb{Z})$. On the other hand, the ones that
only depend on the $SO(3) \backslash SL(3,\mathbb{R})$ moduli satisfy Poisson
equation on moduli space, where the source terms are given by other couplings in the
same supermultiplet. The couplings of the interactions which are charged under
$SU(2)$ are not automorphic forms of $SL(3,\mathbb{Z})$. Among the interactions we consider, the
$\mathcal{R}^4$ coupling depends on all the moduli.

\end{abstract}

\end{titlepage}

\pagestyle{plain}
\baselineskip=19pt

\section{Introduction}

Constructing the low energy effective action of string theory in a certain
background yields detailed information about the various symmetries of
the theory. The degrees of freedom of the effective action are the various
massless modes of the theory. Though in general, it is difficult to calculate
the effective action, there are certain cases where a class of terms can be
calculated exactly. Of course, this turns out to be possible because of the
large amount of symmetry the theory possesses. In this paper, we shall be
concerned with a particular example of this class of theories. We shall
consider type II string theory compactified on $T^2$, which has maximal
supersymmetry and is conjectured to have an exact $SL(2,\mathbb{Z}) \times SL(3,\mathbb{Z})$
symmetry~\cite{Hull:1994ys,Witten:1995ex}. The $SL(3,\mathbb{Z})$ symmetry is a symmetry of the action, while
the $SL(2,\mathbb{Z})$ symmetry is a symmetry of the equations of
motion. Considering M theory on $T^3$, the $SL(3,\mathbb{Z})$ has a geometric
origin as the group of large diffeomorphisms of the $T^3$, and
$SL(2,\mathbb{Z})$ arises from the modular transformations of the complexified
volume of the $T^3$. This theory is a particular example of toroidal
compactifications of type II string theory, which preserves maximal
supersymmetry. The moduli space is a coset space $H \backslash G$, where $G$
is a non--compact group, and $H$ is the maximal compact subgroup of
$G$~\cite{Cremmer:1980gs,Julia:1980gr}\footnote{Only for $d=8$, the moduli
  space factorizes into $(H_1 \backslash G_1) \otimes (H_2 \backslash G_2)$,
  where each factor satisfies this property.}. Non--perturbative effects break the continuous symmetry to a discrete
subgroup of $H$, which is the U-duality symmetry of the theory.  

The motivation for studying the theory in 8 dimensions arises from the fact
that a certain class of terms in the effective action is known in 10
dimensions, and explicit forms of the couplings and their
non--renormalization properties have been
analyzed~\cite{Green:1997tv,Green:1997di,Green:1997as,Green:1998by,Green:1999pu,Sinha:2002zr,Berkovits:2004px,Green:2005ba,Green:2006gt,Basu:2008cf}.
The 8 dimensional case is the next one in order of complexity. Hence, this
is the starting point for going down to lower dimensions. Our aim is to begin
with the action of $N=2$, $d=8$ supergravity which is obtained by
dimensionally reducing $d=11$ supergravity on $T^3$.  We then want to construct
a certain set of terms among the various higher derivative corrections to the 
supergravity action. The set of terms we want to consider are $1/2$ BPS and
satisfy non--renormalization properties, in particular, they receive
perturbative contributions only upto one loop. These terms arise at the 8
derivative level in the effective action. We would like to use the constraints coming
from supersymmetry to obtain equations satisfied by the moduli dependent
couplings of these interactions in the effective action. Various aspects of higher derivative
corrections in 8 and lower dimensions with maximal supersymmetry have been 
analyzed
in~\cite{Kiritsis:1997em,Pioline:1997pu,Pioline:1998mn,Pioline:2001jn,Basu:2007ru,Basu:2007ck,Gubay:2010nd},
and various properties of the couplings have been deduced.
This has led to explicit expressions for the $\mathcal{R}^4$
coupling in lower dimensions in~\cite{Green:2010wi,Green:2010kv}.

The effective action we shall construct is one particle irreducible, and hence
has infra--red divergences. However, the equations of motion are duality
invariant and coupled with the constraints of supersymmetry, certain terms
are amenable to 
a detailed analysis. In order to construct a class of such terms in the effective
action, we shall implement the Noether procedure to the required order in the
derivative expansion, also taking into account the corrected supersymmetry
transformations, generalizing the work of~\cite{Green:1998by}. In particular, we shall use the invariance of the action
under supersymmetry. We write the action and supersymmetry transformations as
\be S = S^{(0)} + \sum_{n=3}^\infty S^{(n)}, \quad \delta = \delta^{(0)} +
\sum_{n=3}^\infty \delta^{(n)}, \ee 
where $S^{(0)}$ and $\delta^{(0)}$ are the supergravity action and the
supersymmetry transformations of the various fields at the two derivative
level, respectively. There are arguments to suggest
that $S^{(1)}$ and $S^{(2)}$ vanish, and consequently so does $\delta^{(1)}$
and $\delta^{(2)}$. Thus, the first correction to the supergravity action is
given by $S^{(3)}$, and this is the term we want to focus on. Our convention
is that $S^{(n)}$ carries $2n+ 2$ derivatives. Thus using the Noether
procedure, one has to implement the relations
\be \delta^{(0)} S^{(n)} + \delta^{(n)} S^{(0)} + \sum_{p+q =n}\delta^{(p)}
S^{(q)}=0,\ee
upto a total derivative, for $n=0$ and $n\geq 3$.

We begin with an analysis of the field content and the action of $N=2, d=8$
supergravity. This is followed by a discussion of the supersymmetry
transformations at the two derivative level. In the next section, we discuss
the issue of gauge fixing the local symmetries of the moduli space, as well as
supersymmetry. We then construct the transformations of the moduli under U--duality. 
After that, we focus on the issue of constructing the effective action beyond
the two derivative level. To begin with, we construct a set of higher
derivative terms in the
effective action which are $1/2$ BPS, starting from on--shell linearized superspace.  
We then consider the role of supersymmetry in constraining these higher
derivative couplings. We look at a set of couplings which involves only the
$U(1) \backslash SL(2,\mathbb{R})$ moduli, and another set which involves only
the $SO(3) \backslash SL(3,\mathbb{R})$ couplings. We also consider a coupling
which involves all the moduli. We briefly discuss the systematics of the analysis for
lower dimensions very schematically. 

For the couplings which depend only on the $U(1) \backslash SL(2,\mathbb{R})$
moduli, we show that each coupling satisfies a first order differential
equation on the moduli space which relates it to another
coupling. From the explicit structure of the equations, we conclude that each
coupling satisfies Laplace equation on moduli space. The couplings are given
by automorphic forms of $SL(2,\mathbb{Z})$ with non--trivial weights, which is
determined by the $U(1)$ charges of the corresponding interactions.  
The couplings which depend only on the $SO(3)\backslash SL(3,\mathbb{R})$ moduli
behave in a more complicated way. These couplings also satisfy first order
differential equations on moduli space which relate it to other
couplings. However, the structure of the equations is such that, it follows
that each coupling satisfies Poisson equation on moduli space, with source
terms given by couplings in the same supermultiplet. Furthermore, the
couplings for the interactions which carry non--trivial $SU(2)$ charges are not
automorphic forms of $SL(3,\mathbb{Z})$, but transform in a complicated way.

It follows that supersymmetry does impose very strong constraints on the
structure of the higher derivative corrections. It would be interesting to
generalize the analysis to lower dimensions, and also to look at interactions
in the effective action which preserve less supersymmetry.

\section{The field content and the action of $N=2, d=8$ supergravity}

Let us first consider the field content of $N=2, d=8$
supergravity~\cite{Salam:1984ft},  
which is obtained by dimensionally reducing $d=11$ supergravity~\cite{Cremmer:1978km} on $T^3$.

\subsection{The bosonic degrees of freedom}

The bosonic fields are given by
\be \label{bose} e_\mu^{~a}, \quad L_m^{~i}, \quad L_U^{~u}, \quad A_\mu^{mU},
\quad B_{\mu\nu}^m, \quad C_{\mu\nu\rho}.\ee
In \C{bose}, $e_\mu^{~a}$ is the vielbein, and $\mu,a=0, \ldots, 7$,
where $\mu$ is the world index and $a$ is the local frame
index. There are 7 scalars in the theory which 
are parametrized by $L_m^{~i}$ and $L_U^{~u}$
which satisfy
\be {\rm det} L_m^{~i} = 1, \quad {\rm det} L_U^{~u} =1.\ee
Here $L_m^{~i}$ parametrizes elements of the coset space
$SO(3) \backslash SL(3,\mathbb{R})$, and so $m=1,2,3$ transforms as the 3 of
$SL(3,\mathbb{R})$ while $i=1,2,3$ transforms as the 3 of
$SO(3)$. Also $L_U^{~u}$ parametrizes elements of the coset space
$SO(2) \backslash SL(2,\mathbb{R})$, and so $U=1,2$ transforms as the 2 of
$SL(2,\mathbb{R})$ while $u=1,2$ transforms as the 2 of $SO(2)$. Thus
under $SL(3,\mathbb{R})$ and $SO(3)$ transformations, $L_m^{~i}$ which carries 5
degrees of freedom, transforms as
\be \label{sl3t} L_m^{~i} (x) \rightarrow O^i_{~j} (x) L_n^{~j} (x) R^n_{~m}, \ee 
where $O \in SO(3)$, and $R \in SL(3,\mathbb{R})$. Similarly, under $SL(2,\mathbb{R})$ and $SO(2)$
transformations, $L_U^{~u}$ which carries 2
degrees of freedom, transforms as
\be L_U^{~u} (x) \rightarrow N^u_{~v} (x) L_V^{~v} (x) S^V_{~U}, \ee  
where $N \in SO(2)$, and $S \in SL(2,\mathbb{R})$. Thus the classical moduli
space is
\be \Big( U(1) \backslash SL(2,\mathbb{R})\Big) \otimes \Big(SO(3) \backslash SL(3,\mathbb{R})\Big).\ee
There are 6 abelian gauge fields $A_\mu^{mU}$ which
transform as the $(3,2)$ of $SL(3,\mathbb{R}) \times
SL(2,\mathbb{R})$, and 3 two forms $B_{\mu\nu}^m$ which transform as
the $(3,1)$ of $SL(3,\mathbb{R}) \times SL(2,\mathbb{R})$. Finally,
the $U(1)$ invariant (anti)selfdual field strength $F_4^{\pm}$, where $F_4 = d C + \ldots$,
and $G_4^{\pm}$ which is defined by
\be \label{defG}
\frac{ie}{4!} G_{4\mu\nu\rho\lambda}^{\pm} =  \pm   \frac{\p \mathcal{L}}{\p
  F_4^{\mu\nu\rho\lambda \pm}}, \ee
form a doublet 
\be F^{U \pm}_{4} = \begin{pmatrix} F_4^\pm \\  G_4^\pm  \end{pmatrix} \ee 
under $SL(2,\mathbb{R})$, and is uncharged under $SL(3,\mathbb{R})$. Thus
the action is invariant under $SL(3,\mathbb{R})$, while only the
equations of motion are invariant under $SL(2,\mathbb{R})$. 
The theory has 128 bosonic degrees of freedom.

\subsection{The fermionic degrees of freedom}

Now let us consider the 256 fermionic degrees of freedom in the theory.
The fermions of $N=2$, $d=8$ supergravity are charged under $H$, but are uncharged
under $G$ of the coset spaces $H \backslash G$. Since we are now
considering spinors, $H$ is now $U(1) \times SU(2)$. We use $U(1)$
rather than $SO(2)$ for simplicity of manipulations because we shall consider
Weyl fermions, and $SU(2)$ because ${\rm spin}(3) = SU(2)$. In order to understand the dimensional 
reduction in the fermionic sector under
\be {\rm spin} (10,1) \rightarrow {\rm spin} (7,1) \times SU(2) , \ee
consider a 32 component Majorana fermion $\eta$ in $d=11$, which we decompose as
\be  \label{Maj11}
\eta  =  \begin{pmatrix} \psi_L \\ \psi_R  \end{pmatrix} = \begin{pmatrix} \psi_L \\ \s_2 \psi_L^*   
\end{pmatrix} , \ee   
where $\psi_L$ and $\psi_R$ are each 8 component chiral fermions of ${\rm spin}(7,1)$ in the 2 of $SU(2)$, and
$(\s_i)_A^{~B}$ are the Pauli matrices. Thus explicitly,
\be \eta = \begin{pmatrix} \psi_{LA} \\ \psi_{RA}  \end{pmatrix} = \begin{pmatrix} \psi_{LA} \\ -i \epsilon_{AB}\psi_L^{*B}   
\end{pmatrix}, \ee
where we have defined
\be (\psi_{LA})^* \equiv \psi_L^{*A}, \ee
and
\be \epsilon_{12} = -\epsilon_{21}=1 , \quad \epsilon^{12} = -\epsilon^{21} = 1.\ee

At various places, the spinor indices of $SU(2)$ will be raised and lowered using the relations
\be V_A = \epsilon_{AB} V^B, \quad V^A = - \epsilon^{AB} V_B .\ee
Our summation convention is
\be XY \equiv X^A Y_A = - X_A Y^A . \ee

The matrices $\hat\Gamma^{\hat{A}}$ in $d=11$ (where $\hat{A}=0, 1, \ldots, 10$ are the frame indices) 
satisfying
\be \{ \hat\Gamma_{\hat{A}} , \hat\Gamma_{\hat{B}} \} = 2 \eta_{\hat{A}\hat{B}},\ee
where our metric has mostly plus signature, decompose as
\bea \hat\Gamma^a = \Gamma^a \otimes 1, \non \\ \hat\Gamma^i = \Gamma_9 \otimes \s_i ,\eea
where $\Gamma_9$ is the $d=8$ chirality matrix defined by
\be \Gamma_9 = i \Gamma^0 \Gamma^1 \ldots \Gamma^7 . \ee 
For the $\Gamma^a$ matrices,  
we consider a chiral basis given by
\be  \Gamma^a = \begin{pmatrix} 0 & \gamma^a  \\  
\bar\gamma^a  & 0 \end{pmatrix}, \ee
where
\be \gamma^a \bar\gamma^b + \gamma^b \bar\gamma^a = 2 \eta^{ab},\ee
and $\eta^{ab} = {\rm diag}(-, +,\ldots, +)$. We consider an explicit basis for the $\gamma^a$
and $\bar\gamma^a$ matrices given by 
\bea \gamma^0 &=& 1 \otimes 1 \otimes 1 , \non \\ \gamma^1 &=& \s^2
\otimes \s^2 \otimes \s^2 , \non \\ \gamma^2 &=& 1 \otimes \s^3
\otimes \s^2 , \non \\
\gamma^3 &=& \s^3 \otimes \s^2 \otimes 1 , \non \\ \gamma^4 &=& \s^2
\otimes 1 \otimes \s^3 , \non \\
\gamma^5 &=& 1 \otimes \s^1 \otimes \s^2 , \non \\ \gamma^6 &=& \s^1
\otimes \s^2 \otimes 1 , \non \\ \gamma^7 &=& \s^2 \otimes 1 \otimes \s^1 , \eea 
and
\be \bar\gamma^0 = -\gamma^0, \quad \bar\gamma^I = \gamma^I ,\ee
for $I=1, \ldots , 7$. Thus $\Gamma^0$ is anti--hermitian, while
$\Gamma^I$ is hermitian. Also
\be \label{tbar} \bar\gamma^a = - \gamma^{aT} , \quad \g^{a*} = -\bar\g^a .\ee

Note that $\eta$ in \C{Maj11} satisfies the $d=11$ Majorana condition 
\be \eta = C_{11} \bar\eta^T ,\ee
where $C_{11}$ is the $d=11$ charge conjugation matrix given by
\be C_{11} = 1\otimes 1\otimes 1\otimes 1\otimes \s_2 ,\ee
where the factor of $\s_2$ acts on the $SU(2)$ indices. $C_{11}$ satisfies
\be C_{11} \hat\Gamma^{\hat{A}} C^{-1}_{11} = - \hat\Gamma^{\hat{A} T} ,\ee 
as expected. 

Now, it is natural to ask, what kind of fermion is $\eta$ from the $d=8$ point of view? 
Thus, we now think of it as a 16 component constrained Dirac spinor, with the $SU(2)$ indices coming
from the extended supersymmetry. To analyze this, we first need to consider a constrained Dirac spinor of $N=1, d=8$
supersymmetry. This is given by{\footnote{It follows from the discussion below that $\psi_L^*$ does transform as
$\psi_R$.}}
\be \label{Maj8} \psi = \begin{pmatrix} \psi_L \\ \psi_R  \end{pmatrix} = \begin{pmatrix} \psi_L \\ \psi_L^*   
\end{pmatrix}. \ee
This satisfies the $d=8$ Majorana condition
\be \psi = C_8 \bar\psi^T ,\ee
where $C_8$ is the $d=8$ charge conjugation matrix given by
\be C_8 = \begin{pmatrix} -1 & 0\\ 0 & 1  \end{pmatrix} ,\ee
which satisfies
\be \label{cc8}C_8 \Gamma^\mu C_8^{-1} = \Gamma^{\mu T}.\ee
Thus $\psi$ in \C{Maj8} is a pseudo--Majorana fermion{\footnote{This is called ``pseudo" because $C_8$ 
satisfies \C{cc8} with a $+$ sign on the right hand side, and not $-$~\cite{Kugo:1982bn,Awada:1985ag}.}}. Now, for the $N=2$ theory
\be \label{n=2f}
\psi_A= \begin{pmatrix} \psi_{LA} \\ \psi_{RA}  \end{pmatrix} = \begin{pmatrix} \psi_{LA} \\ -i \epsilon_{AB} \psi_L^{*B}   
\end{pmatrix}, \ee
satisfies the condition
\be \psi_A = -i\epsilon_{AB} \Gamma_9 C_8 \bar\psi^{BT} .\ee
Thus \C{n=2f} is an $SU(2)$ pseudo--Majorana fermion in the Weyl basis{\footnote{In general, $\epsilon_{AB}$ can be 
replaced by the symplectic form $\omega_{AB}$, leading to $Sp(2n)$ pseudo--Majorana fermions. This structure 
for extended supersymmetry is exactly as in 
$N=2, d=4$ where one has Majorana, instead of pseudo--Majorana fermions.}} (see~\cite{Sohnius:1985qm}
for example).    

To show that we are indeed working in a chiral basis for the $N=2$ fermions, 
consider an infinitesimal Lorentz transformation in $d=11$, under which a Majorana fermion transforms as
\be \delta \eta = \frac{1}{4} \xi_{\hat{A}\hat{B}} \hat\Gamma^{\hat{A}\hat{B}} \eta .\ee  
This leads to
\be \label{repack}
\delta \psi_A = \frac{1}{4} (\xi_{ab} \Gamma^{ab} + i \xi_i \s^i )_A^{~B} \psi_B ,\ee
where
\be \xi_i = \epsilon_{ijk} \xi_{jk} .\ee
Now, as before, label
\be \label{pseudoM} \psi_A  =  \begin{pmatrix} \psi_{LA} \\ \psi_{RA}  \end{pmatrix} , \ee 
which satisfies
\be \label{totrel} \psi_R = N \psi_L^* . \ee     
The fact that $\psi_L$ and $\psi_R$ have opposite chiralities leads to
\bea N (\xi_{ab} \g^{ab } +i \xi_i \s^i )^* N^{-1} = \xi_{ab} \bar\g^{ab} + i \xi_i \s^i . \eea
Thus
\be N = M \s_2 ,\ee
where $M$ satisfies
\be \label{eqnM} M (\gamma^a \bar\gamma^b - \gamma^b \bar\gamma^a )^* M^{-1} =
\bar\gamma^a \gamma^b - \bar\gamma^b \gamma^a .\ee 
Thus $N$ factorizes into the $SU(2)$ part and the spacetime part, and
\C{eqnM} is solved by
\be M = \gamma^0 = -\bar\gamma^0 = 1 ,\ee
and so
\be \psi_R = \s_2 \psi_L^* ,\ee
as before{\footnote{This argument also goes through for $N=1$ supersymmetry by looking at $d=8$ Lorentz transformations,
as there are no internal indices coming from extended supersymmetry}}. Also note that
\be \Gamma_9 \psi_L = -\psi_L, \quad \Gamma_9 \psi_R = \psi_R .\ee

Let us now tabulate the fermions of the $N=2, d=8$ theory.   
The fermions of negative chirality are given by 
\be \label{fermi} \psi_{\mu LA}, \quad \chi_{LA}^{i}, \quad
\lambda_{LA}.\ee
In \C{fermi}, the spin $3/2$ gravitini $\psi_{\mu LA}$ transform as the 2 of
$SU(2)$, while the spin $1/2$ fermions
$\chi^{i}_{LA}$ and $\lambda_{LA}$ transform as the 4 and 2 of $SU(2)$
respectively, and so
\be \s_i \chi^i_L = 0. \ee
Under $U(1)$, $\psi_{\mu LA}$, $\chi^{i}_{LA}$ and $\lambda_{LA}$ carry
charges $1/2$, $-1/2$, and $3/2$ respectively. The positive chirality
fermions are denoted by 
$\psi_{\mu RA}$, $\chi_{iRA}$ and
$\lambda_{RA}$, and are the conjugates of the negative chirality ones,
according to the discussion above. They carry $U(1)$ charges $-1/2, 1/2$,
and $-3/2$ respectively. The supersymmetry transformation parameter
$\epsilon_{LA}$ is in the 2 of $SU(2)$, and carries $U(1)$
charge $1/2$, while $\epsilon_{RA}$ carries $U(1)$ charge $-1/2$. 
Thus $\psi_{\mu LA}$, $\chi^{i}_{LA}$, and $\lambda_{LA}$ carry 160, 64,
and 32 degrees of freedom respectively. 

For arbitrary fermions $\psi_1$ and $\psi_2$, we define conjugation by
\be (\psi_1 \psi_2)^\dagger =  -\psi_2^\dagger \psi_1^\dagger ~\footnote{Thus, for example, 
$(\bar\psi_L \bar\g^\mu \p_\mu \psi_L)^\dagger = \bar\psi_L \bar\g^\mu \p_\mu \psi_L$ upto a total derivative. }.\ee

We also make use of the relations
\bea \label{defbar} \frac{1}{2} \bar\psi \Gamma^\mu \p_\mu \psi = \bar\psi_L
\bar\gamma^\mu \p_\mu \psi_L = \bar\psi_R \gamma^\mu
\p_\mu \psi_R ,\non \\ \frac{1}{2} \bar\psi_\mu
\Gamma^{\mu\nu\rho} \p_\nu\psi_\rho = \bar\psi_{\mu L}
\bar\gamma^{\mu\nu\rho} \p_\nu \psi_{\rho L} = \bar\psi_{\mu R}
\gamma^{\mu\nu\rho} \p_\nu\psi_{\rho R} ,\eea
on using \C{tbar}, and ignoring total derivatives. In \C{defbar}, we
have also defined
\be \bar\psi = \psi^\dagger \Gamma^0 , \quad \bar\psi_L = \psi_L^\dagger, 
\quad \bar\psi_R = -\psi_R^\dagger. \ee

\subsection{Relevant terms in the action}

Now let us consider some of the terms in the action which are relevant
for our purposes. They are given by
\bea \label{action} e^{-1} \mathcal{L}^{(0)} &=&  R -2 P_\mu^* P^\mu
- P_{\mu ij} P^{\mu ij} -\frac{1}{4} M_{UV} M_{mn} F_{2\mu\nu}^{mU}
F_2^{\mu\nu nV} -\frac{1}{12} M_{mn} F_{3\mu\nu\rho}^{~~~~m} F_3^{\mu\nu\rho n}
\non \\ && + \frac{i}{48} \Big( \frac{L^{~2}_{+}}{L^{~1}_{+}} F^+_{4\mu\nu\rho\lambda} F^{+\mu\nu\rho\lambda}_4 
- \frac{L^{~2}_{-}}{L^{~1}_{-}} F^-_{4\mu\nu\rho\lambda}
F^{-\mu\nu\rho\lambda}_4  \Big)
\non
\\ &&+4
\bar\psi_{\mu L} \bar\gamma^{\mu\nu\rho} {\mathcal{D}}_\nu
\psi_{\rho L} 
+4\bar\chi_{iL} \bar\gamma^\mu {\mathcal{D}}_\mu \chi^{i}_L +2\bar\lambda_{L}
\bar\gamma^\mu {\mathcal{D}}_\mu \lambda_L \non \\ && -\frac{1}{48\sqrt{2}} \Big[ \frac{F^+_{\mu\nu\rho\s}}{L^{~1}_+} 
\Big( \bar\psi^\lambda_R \g_{[\lambda} \bar\g^{\mu\nu\rho\s} \bar\g_{\tau]} \psi^\tau_L + \bar\psi^\lambda_L 
\bar\g^{\mu\nu\rho\s} \bar\g_\lambda \lambda_L - \bar\chi^i_L \bar\g^{\mu\nu\rho\s} \chi_{iR} \Big)
\non \\ && + \frac{F^-_{\mu\nu\rho\s}}{L^{~1}_-} \Big( \bar\psi^\lambda_L \bar\g_{[\lambda} \g^{\mu\nu\rho\s} \g_{\tau]} \psi^\tau_R 
+ \bar\psi^\lambda_R 
\g^{\mu\nu\rho\s} \g_\lambda \lambda_R - \bar\chi^i_R \g^{\mu\nu\rho\s} \chi_{iL} \Big) \Big] ,\eea
where the covariant derivatives are defined later.
Thus, we get that
\bea G_{4\mu\nu\rho\s}^+ = \frac{L^{~2}_{+}}{L^{~1}_{+}} F_{4\mu\nu\rho\s}^+ +\frac{i}{2\sqrt{2} L^{~1}_+} 
\Big( \bar\psi_{\lambda R} \g^{[\lambda} \bar\g_{\mu\nu\rho\s} \bar\g^{\tau]}
\psi_{\tau L} + \bar\psi^\lambda_L 
\bar\g_{\mu\nu\rho\s} \bar\g_\lambda \lambda_L - \bar\chi^i_L \bar\g_{\mu\nu\rho\s} \chi_{iR} \Big), \non 
\\  G_{4\mu\nu\rho\s}^- = \frac{L^{~2}_{-}}{L^{~1}_{-}} F_{4\mu\nu\rho\s}^- -\frac{i}{2\sqrt{2} L^{~1}_-} 
\Big( \bar\psi_{\lambda L} \bar\g^{[\lambda} \g_{\mu\nu\rho\s} \g^{\tau]}
\psi_{\tau R} 
+ \bar\psi^\lambda_R 
\g_{\mu\nu\rho\s} \g_\lambda \lambda_R - \bar\chi^i_R \g_{\mu\nu\rho\s} \chi_{iL} \Big)  , \eea
where $L^{~U}_{\pm}$ is defined shortly.

In \C{action}, the various field strengths are defined
by\footnote{Square brackets are normalized with unit weight.}
\bea \label{deff} \frac{1}{2} F_{2\mu\nu}^{mU} &=&  \p_{[ \mu} A_{\nu ]}^{mU}, \non
\\ \frac{1}{3} F_{3
  \mu\nu\lambda m} &=&  \p_{[\mu} B_{\nu\lambda ] m } +\frac{1}{2}
\epsilon_{mnp} \epsilon_{UV} A_{[\mu}^{nU} F_{2\nu\lambda ] }^{pV} , \non
\\ \frac{1}{4}  F_{4\mu\nu\lambda\rho} &=& \p_{[\mu} C_{\nu\lambda\rho
]} + \frac{3}{2} \Big( B_{m[\mu\nu } -\epsilon_{mnp} A^{2n}_{[\mu} A^{1p}_\nu  \Big) F_{2\lambda\rho ]}^{m1} 
. \eea

We have also defined the (anti)selfdual parts of any 4 form $X_4$ by
\be X_4^{\pm} = \frac{1}{2} (X_4 \pm i * X_4),\ee
where{\footnote{We have that
\be \epsilon_{01\ldots 7} =1, \ee
for the frame indices.}}
\be * X_{4abcd} = \frac{1}{4!}
\epsilon_{abcd}^{~~~~~efgh} X_{4efgh}.\ee 
Note that
\bea \frac{1}{4!} 
\epsilon_{abcd}^{~~~~~efgh} \gamma^{abcd}  = i \gamma^{efgh} , 
\non \\ \frac{1}{4!} 
\epsilon_{abcd}^{~~~~~efgh} \bar\gamma^{abcd}  = -i 
\bar\gamma^{efgh},\eea
leading to
\bea \gamma^{\mu\nu\lambda\rho} X^-_{4\mu\nu\lambda\rho} = \gamma^{\mu\nu\lambda\rho} X_{4\mu\nu\lambda\rho},
 \quad \gamma^{\mu\nu\lambda\rho} X^+_{4\mu\nu\lambda\rho} = 
 0, \non \\ \bar\gamma^{\mu\nu\lambda\rho} X^+_{4\mu\nu\lambda\rho} = \bar\gamma^{\mu\nu\lambda\rho} X_{4\mu\nu\lambda\rho}, 
\quad \bar\gamma^{\mu\nu\lambda\rho} X^-_{4\mu\nu\lambda\rho} = 
 0.\eea

Also the $SO(3) \backslash SL(3,\mathbb{R})$ and $SO(2) \backslash SL(2,\mathbb{R})$ 
moduli are contained in 
\be M_{mn} = L_m^{~i} L_n^{~j} \delta_{ij},\ee
and
\be M_{UV} = L_U^{~u} L_V^{~v} \delta_{uv}, \ee
respectively. 

To understand the structure of the kinetic terms of the moduli 
parametrizing $SO(3) \backslash SL(3,\mathbb{R})$, consider $L$ which parametrizes the 
elements of the coset space $H \backslash G$. We use
the Cartan decomposition (see~\cite{Salam:1981xd} for example)
\be -L d L^{-1}  = P + Q ,\ee
where $Q$ is in $H$, while $P$ is in $H\backslash G$. 
Thus $P$ and $Q$ are invariant under the global transformation $L
\rightarrow LG$. Then the kinetic terms for the moduli can be expressed in
terms of $P$, while $Q$ gives rise to the composite $H$ gauge field.    
Thus, for the coset space $SO(3) \backslash SL(3,\mathbb{R})$, 
we have that
\be - \label{cartan}L_{mi} \p_\mu L^{~m}_j = P_{\mu ij} + Q_{\mu ij},\ee
where $P_{\mu ij}$ is the symmetric, and $Q_{\mu ij}$ is the
anti--symmetric part of the left hand side of \C{cartan}. $P_{\mu ij}$
is automatically traceless because ${\rm det} L_m^{~i} =1$.

It is also useful for our purposes to write the kinetic terms for the
moduli only in terms of the matrix $M$. This is done by
noting that
\be P_{\mu ij} = - \frac{1}{2} L_{mi} L_{nj} \p_\mu M^{mn},\ee
which leads to
\be - P_{\mu ij} P^{\mu ij} = \frac{1}{4} \p_\mu M_{mn}
\p^\mu M^{mn} = \frac{1}{4} {\rm Tr} (\p_\mu M \p^\mu M^{-1}).\ee 

To understand the structure of the moduli fields parametrizing $SO(2)
\backslash SL(2,\mathbb{R})$, it is convenient to consider the complex basis
\be L^{~U}_{\pm} = \frac{1}{\sqrt{2}} ( L^{~U}_{1} \pm i L^{~U}_{2}), \ee 
where the subscripts in $L^{~U}_{\pm}$ label the $U(1)$ charges. Thus, we
also have that
\be (L^{~U }_{\pm})^* = L^{~U}_{\mp}, \quad \epsilon_{UV} L^{~U}_{-}
L^{~V}_{+} =i .\ee
The kinetic term for the moduli can be expressed in terms of the
$SL(2,\mathbb{R})$ invariant combination $P_\mu$ defined by (it carries $U(1)$
charge 2)
\be P_\mu = -\epsilon_{UV} L^{~U}_{+} \p_\mu L^{~V}_{+},\ee
while the composite $U(1)$ gauge field is given by the $SL(2,\mathbb{R})$ invariant
combination
\be Q_\mu = - \epsilon_{UV} L^{~U}_{-} \p_\mu L^{~V}_{+}.\ee

Let us consider the transformations of the various fields under
infinitesimal $U(1) \times SU(2)$ gauge transformations. Under a
$U(1)$ gauge transformation, a field
$\Phi_q$ carrying $U(1)$ charge $q$ transforms as
\be \delta \Phi_q (x) = i q \Sigma (x) \Phi_q (x),\ee
and thus the gauge field $Q_\mu$ transforms as
\be \delta Q_\mu =  \p_\mu \Sigma. \ee
Thus the covariant derivative
\be \mathcal{D}_\mu \Phi_q = \p_\mu \Phi_q - i q Q_\mu \Phi_q 
\ee
transforms as
\be \delta \mathcal{D}_\mu \Phi_q =iq \Sigma \mathcal{D}_\mu \Phi_q .\ee

To consider $SU(2)$ gauge transformations, let us define
\be A_{\mu i} = \frac{1}{2} \epsilon_{ijk} Q_{\mu jk}. \ee
Thus under a gauge transformation\footnote{This also leads to
\be \delta L_{mi} = -\epsilon_{ijk} \theta_j L_{mk} ,\ee 
on using $\delta (L^{~m}_{i} L_{mj})=0$.}
\be \delta L^{~m}_{i} = -\epsilon_{ijk} \theta_j L^{~m}_{k},\ee
we have that
\be \delta A_{\mu i} = \p_\mu \theta_i + \epsilon_{ijk} A_{\mu j}
\theta_k . \ee
So, for $\Psi = (\psi_{\mu }, \lambda)$, the gauge transformation
\be \delta \Psi = \frac{i}{2} \theta^i \s^i \Psi , \ee  
leads to the covariant derivative
\be \mathcal{D}_\mu \Psi = \p_\mu \Psi - \frac{i}{2} A_\mu^{~i} \s^i
\Psi ,\ee
which transforms as
\be \delta \mathcal{D}_\mu \Psi = \frac{i}{2} \theta^i \s^i
\mathcal{D}_\mu \Psi .\ee
For $\chi^i$ satisfying $\s_i \chi^i =0$, we also have that
\be \chi^i = i \epsilon^{ijk} \s^j \chi^k ,\ee
leading to
\bea \delta \chi^i &=& -\epsilon_{ijk} \theta_j \chi^k + \frac{i}{2} \theta_j \s_j \chi^i 
\non \\ &=& -i \theta^j \s^i \chi^j + \frac{3i}{2} 
\theta^j \s^j \chi^i ,\eea
on using
\be \label{useeqn}\s_i \chi_{j} - \s_j \chi_{i} = -i \epsilon_{ijk} \chi_{k} . \ee
So the covariant derivative
\be \mathcal{D}_\mu \chi^i = \p_\mu \chi^i + i A_\mu^{~j} \s^i \chi^j
- \frac{3i}{2} A_\mu^{~j} \s^j \chi^i  , \ee
transforms as
\be \delta \mathcal{D}_\mu \chi^i = -i \theta^j \s^i \mathcal{D}_\mu \chi^j + \frac{3i}{2} 
\theta^j \s^j \mathcal{D}_\mu \chi^i ,\ee
on using the Schouten identity
\be \epsilon^{ijk} \chi^l  + \epsilon^{ilj} \chi^k + \epsilon^{jlk} \chi^i
+ \epsilon^{kli} \chi^j = 0. \ee

Thus, for the various fermionic interactions in \C{action}, we have that
\bea \label{covder} {\mathcal{D}}_\mu \psi_{\nu L} &=& D_\mu \psi_{\nu L} - \frac{i}{2}
A_\mu^{~i} \s^i \psi_{\nu L}  - \frac{i}{2} Q_\mu \psi_{\nu L} , \non \\  {\mathcal{D}}_\mu
\lambda_L &=& D_\mu \lambda_L - \frac{i}{2}
A_\mu^{~i} \s^i \lambda_L - \frac{3i}{2} Q_\mu \lambda_L ,
\non \\ \mathcal{D}_\mu
\chi^i_L &=& D_\mu \chi^i_L + i A_\mu^{~j} \s^i \chi^j_L - \frac{3i}{2}
A_\mu^{~j} \s^j \chi^i_L +\frac{i}{2} Q_\mu \chi^i_L , \non \\ 
{\mathcal{D}}_\mu \psi_{\nu R} &=& D_\mu \psi_{\nu R} - \frac{i}{2}
A_\mu^{~i} \s^i \psi_{\nu R}  + \frac{i}{2} Q_\mu \psi_{\nu R} , \non \\  {\mathcal{D}}_\mu
\lambda_R &=& D_\mu \lambda_R - \frac{i}{2}
A_\mu^{~i} \s^i \lambda_R +\frac{3i}{2} Q_\mu \lambda_R ,
\non \\ \mathcal{D}_\mu
\chi^i_R &=& D_\mu \chi^i_R + i A_\mu^{~j} \s^i \chi^j_R - \frac{3i}{2}
A_\mu^{~j} \s^j \chi^i_R -\frac{i}{2} Q_\mu \chi^i_R ,\eea
where $D_\mu$ is the ordinary covariant derivative. 

It is convenient for our purposes to redefine field strengths that are invariant under
$G$, and carry specific charges under $H$. For the 2 form field strengths, we define
\bea F_{2\mu\nu}^i = \epsilon_{UV} F_{2\mu\nu}^{mU} L^{~V}_+ L_m^{~i}, \non \\ 
F_{2\mu\nu}^{*i} = \epsilon_{UV} F_{2\mu\nu}^{mU} L^{~V}_- L_m^{~i},\eea
which carry charges $1$ and $-1$ respectively under $U(1)$, and are in the $3$ of $SU(2)$. For the 3 form
field strengths, we define 
\be F_{3\mu\nu\rho i} = L^{~m}_{i} F_{3\mu\nu\rho m}, \ee
which is uncharged under $U(1)$, and is in the $3$ of $SU(2)$.
Finally, for the 4 form field strengths, we define the selfdual field strength
\be T^+_{\mu\nu\rho\s} = \epsilon_{UV} L^{~U}_{-} F^{+V}_{4\mu\nu\rho\s} , \ee
which carries $U(1)$ charge $-1$, and the anti--selfdual field strength
\be T^-_{\mu\nu\rho\s} = \epsilon_{UV} L^{~U}_{+} F^{-V}_{4\mu\nu\rho\s} ,\ee
which carries $U(1)$ charge 1. Both $T^\pm$ are uncharged under $SU(2)$.

Thus, we have that
\bea T_{4\mu\nu\rho\s}^+ =  \frac{i}{L^{~1}_{+}}\Big[ F_{4\mu\nu\rho\s}^+ +\frac{L^{~1}_-}{2\sqrt{2}} 
\Big( \bar\psi_{\lambda R} \g^{[\lambda} \bar\g_{\mu\nu\rho\s} \bar\g^{\tau]}
\psi_{\tau L} + \bar\psi^\lambda_L 
\bar\g_{\mu\nu\rho\s} \bar\g_\lambda \lambda_L - \bar\chi^i_L \bar\g_{\mu\nu\rho\s} \chi_{iR} \Big)\Big], \non 
\\  T_{4\mu\nu\rho\s}^- = -\frac{i}{L^{~1}_{-}} \Big[ F_{4\mu\nu\rho\s}^- +\frac{L^{~1}_+}{2\sqrt{2} } 
\Big( \bar\psi_{\lambda L} \bar\g^{[\lambda} \g_{\mu\nu\rho\s} \g^{\tau]}
\psi_{\tau R} + \bar\psi^\lambda_R 
\g_{\mu\nu\rho\s} \g_\lambda \lambda_R - \bar\chi^i_R \g_{\mu\nu\rho\s} \chi_{iL} \Big) \Big].
\eea

\section{Deriving the supergravity action and the supersymmetry transformations}

In order to construct the relevant terms in the $d=8$ action as well as the supersymmetry transformations of the various fields,
we start from the action and the supersymmetry transformations of the $d=11$ supergravity theory. 
The action is given by~\cite{Cremmer:1978km}
\bea \label{act11d}
V^{-1} \mathcal{L}_{11} &=& \frac{R (\omega)}{4} - \frac{1}{48} F_{MNPQ} F^{MNPQ}
+ \frac{1}{2} \bar\eta_M \hat\Gamma^{MNP} D_{N} \Big( 
\frac{\omega + \hat\omega}{2}\Big)\eta_{P} \non \\  &&- \frac{1}{192} (\bar\eta_{R} 
\hat\Gamma^{RSMNPQ} \eta_S + 12 \bar\eta^M 
\hat\Gamma^{NP} \eta^Q)(F_{MNPQ} +\hat{F}_{MNPQ}) \non \\ && 
+ \frac{2}{(144)^2} V^{-1} \epsilon^{M_1 \ldots M_{11}} F_{M_1 \ldots M_4} F_{M_5 \ldots M_8} C_{M_9 M_{10} M_{11}} .\eea
We denote the the local frame and world indices by $\hat{A},\hat{B},\ldots$ and $M,N,\ldots$ respectively.
In \C{act11d}, $V_M^{~\hat{A}}$, $C_{MNP}$ and $\eta_M$ are the vielbein, the 3 form potential, and 
the gravitino respectively. We also have that
\be D_M \Big(\frac{\omega + \hat\omega}{2}\Big) \eta_N =  \p_M \eta_N + \frac{1}{8} (\omega +
\hat\omega)_M^{~~\hat{A}\hat{B}} \hat\Gamma_{\hat{A}\hat{B}} \eta_N  .\ee
We work in the second order formalism where $\omega_M^{~~\hat{A}\hat{B}}$ is an independent field which satisfies its equation of 
motion (see~\cite{VanNieuwenhuizen:1981ae} for example), leading to 
\be \label{defomega}
\omega_M^{~~\hat{A}\hat{B}} = \omega_M^{~~\hat{A}\hat{B}} (V) + K_M^{~~\hat{A}\hat{B}} .\ee
In \C{defomega}, $\omega_M^{~~\hat{A}\hat{B}} (V)$ is the standard spin connection of pure gravity given by
\bea \omega_M^{~~\hat{A}\hat{B}} (V) &=& \frac{1}{2} V^{\hat{A}N} (\p_M V_N^{~\hat{B}} - \p_N V_M^{~\hat{B}}) - \frac{1}{2} V^{\hat{B}N} 
(\p_M V_N^{~\hat{A}} -
\p_N V_M^{~\hat{A}} ) \non \\ && -\frac{1}{2} V^{\hat{A}P} V^{\hat{B}Q} (\p_P  V_{Q\hat{C}} - \p_Q V_{P\hat{C}}) V_M^{~\hat{C}},\eea 
while the contorsion tensor $K_M^{~~\hat{A}\hat{B}}$ is given by
\be \label{contorsion}
K_{M\hat{A}\hat{B}} = \frac{1}{4} \Big[ -\bar\eta^{\hat{C}} \hat\Gamma_{M\hat{A}\hat{B}\hat{C}\hat{D}} \eta^{\hat{D}} 
+ 2 \Big(\bar\eta_M \hat\Gamma_{\hat{B}} \eta_{\hat{A}} - \bar\eta_M \hat\Gamma_{\hat{A}}\eta_{\hat{B}}
+ \bar\eta_{\hat{B}} \bar\Gamma_M \eta_{\hat{A}} \Big)\Big] .\ee

The supercovariant spin connection in \C{act11d} is given by 
\bea \label{supcov11w}
\hat{\omega}_M^{~~\hat{A}\hat{B}} &=& \omega_M^{~~\hat{A}\hat{B}}+ \frac{1}{4} \bar\eta_{\hat{C}} 
\hat\Gamma_M^{~~\hat{A}\hat{B}\hat{C}\hat{D}} \eta_{\hat{D}} \non \\
&=& \omega_M^{~~\hat{A}\hat{B}} (V) + \frac{1}{2} \Big(\bar\eta_M \hat\Gamma^{\hat{B}} \eta^{\hat{A}} 
- \bar\eta_M \hat\Gamma^{\hat{A}} \eta^{\hat{B}}
+ \bar\eta^{\hat{B}} \hat\Gamma_M \eta^{\hat{A}} \Big) .\eea

For the 3 form potential, it is easier to work with the frame indices to perform the dimensional reduction, and so the
supercovariant 4 form field strength $\hat{F}_{\hat{A}\hat{B}\hat{C}\hat{D}}$ is given by
\be \label{supcov114} \hat{F}_{\hat{A}\hat{B}\hat{C}\hat{D
}} = F_{\hat{A}\hat{B}\hat{C}\hat{D}} + 3 \bar\eta_{[\hat{A}} \hat\Gamma_{\hat{B}\hat{C}} \eta_{\hat{D}]} ,\ee
where
\be F_{\hat{A}\hat{B}\hat{C}\hat{D}} = 4 \p_{[\hat{A}} C_{\hat{B}\hat{C}\hat{D}]} + 12 \omega_{[\hat{A}\hat{B}}^{~~~~\hat{E}} (V)
C_{\hat{C}\hat{D}]\hat{E}} .\ee

Let us now consider the supersymmetry transformations of the various fields. 
Apart from the fermionic trilinear terms in the supervariation of the fermions, the other
transformations can be directly obtained from~\cite{Salam:1984ft}, as mentioned in detail below. In order to obtain the
fermionic trilinear terms in the supervariation of the fermions, we consider maximal supergravity in $d=11$ and obtain 
them by dimensional reduction, given the complete supervariations of the theory. 
We mention only those steps which are relevant for our manipulations.      
The supersymmetry transformations and the local Lorentz transformations of the $d=11$ theory are
\bea \label{11dtrans}  
V^{~M}_{\hat{B}} \delta^{(0)} V_{M\hat{A}} &=& - \bar\xi \hat\Gamma_{\hat{A}} \eta_{\hat{B}}  + \lambda_{\hat{A}\hat{B}},\non \\ 
\delta^{(0)} C_{\hat{A}\hat{B}\hat{C}} 
&=& - \frac{3}{2} \bar\xi \hat\Gamma_{[\hat{A}\hat{B}}
\eta_{\hat{C}]} -3 C_{\hat{D}[\hat{A}\hat{B}} V^{~M}_{\hat{C}]} \delta^{(0)} V_M^{~\hat{D}} , \non \\ 
\delta^{(0)} \eta_{\hat{A}} &=& \hat{D}_{\hat{A}} \xi - \frac{1}{144} \Big( \hat\Gamma^{\hat{B}\hat{C}\hat{D}\hat{E}}_{~~~~~~\hat{A}} 
- 8 \hat\Gamma^{\hat{C}\hat{D}\hat{E}} \delta^{\hat{B}}_{~\hat{A}}
\Big) \xi \hat{F}_{\hat{B}\hat{C}\hat{D}\hat{E}} \non \\ && 
- V^{~M}_{\hat{A}} (\delta^{(0)} V_{\hat{M}}^{~\hat{B}}) \eta_{\hat{B}} + \frac{1}{4} \lambda_{\hat{B}\hat{C}} 
\hat\Gamma^{\hat{B}\hat{C}} \eta_{\hat{A}} ,\eea
where $\xi$ is the supersymmetry parameter. $\lambda_{\hat{A}\hat{B}}$ is the
parameter for local Lorentz transformations.
The supercovariant derivative $\hat{D}_M \xi$ is given by
\be \hat{D}_M \xi = \Big( \p_M + \frac{1}{4} \hat\omega_M^{~~\hat{A}\hat{B}} \hat\Gamma_{\hat{A}\hat{B}} \Big) \xi.\ee

Let us briefly mention the relation between the various fields we have 
and those used by~\cite{Salam:1984ft}\footnote{correcting several typos}. 
While they work directly in terms of the two moduli of 
$U(1) \backslash SL(2,\mathbb{R})$ and thus the $SL(2,\mathbb{R})$ covariance is not manifest, we maintain the explicit 
covariance by working in terms of $L^{~U}_{\pm}$. In fact, we shall later gauge fix the $U(1)$ transformation, which will
force the $SL(2,\mathbb{R})$ transformations to be realized non--linearly on the various fields. Our formulae then
exactly reduce to the ones obtained by~\cite{Salam:1984ft}. We now mention the relations needed to go from their 
formulae (SS) to ours. We set $\kappa =1$, as well as  
\be e^{2\phi_{SS}} = U_2, \quad B_{SS} = -\frac{U_1}{2} , \quad e_\mu^{~a SS} = e_\mu^{~a} , \quad L_m^{~i SS} = L_m^{~i} ,\ee
and so
\be P_{\mu ij}^{SS} = P_{\mu ij} , \quad Q_{\mu ij}^{SS} = -Q_{\mu ij} .\ee
For the two 1 form potentials, we set
\be \quad A_\mu^{mSS} = \frac{A_\mu^{m1}}{2} , \quad  
B_\mu^{mSS} = \frac{A_\mu^{m2}}{2}  ,\ee 
leading to
\bea F_{\mu\nu}^{mSS} = 
\frac{1}{2} F_{2\mu\nu}^{m1} , \quad G_{\mu\nu}^{mSS} =  \frac{1}{2} (F_{2\mu\nu}^{m2} -  U_1 F^{m1}_{2\mu\nu} ). \eea
For the 2 form potential, we set
\be  2 B_{\mu\nu m}^{SS} = B_{\mu\nu m} - \frac{1}{2} \epsilon_{mnp} (A_{\mu}^{2n} A_{\nu }^{1p} 
- A_{\nu }^{2n} A_{\mu}^{1p}),  \ee
leading to
\be G_{\mu\nu\rho m}^{SS} = \frac{1}{2} F_{3\mu\nu\rho m}. \ee
For the 3 form potential, we set 
\be B_{\mu\nu\lambda}^{SS} = \frac{1}{2} C_{\mu\nu\lambda} ,\ee
leading to
\be G_{\mu\nu\lambda\rho}^{SS} = \frac{1}{2} F_{4\mu\nu\lambda\rho} .\ee

We shall later need to construct certain quartic fermion couplings in the $d=8$ theory starting from \C{act11d}.  
For that, we shall need to know the relations between the fields $F_{\hat{A}\hat{B}\hat{C}\hat{D}}$ and the fields
$G_4^{SS}, G_3^{SS}, G_2^{SS}$ and the scalar $B^{SS}$. They are
\bea G_{\mu\nu\lambda\rho}^{SS} &=& e^{-4 \phi^{SS}/3} e_\mu^{~a} e_\nu^{~b} e_\lambda^{~c} e_\rho^{~d} F_{abcd} , \non \\  
G_{\mu\nu\rho m}^{SS} &=& e^{-\phi^{SS}/3} e_\mu^{~a} e_\nu^{~b} e_\rho^{~c} L_m^{~i} F_{abci}, \non \\  
G_{\mu\nu m}^{SS} &=& \frac{1}{2} e^{2\phi^{SS}/3} \epsilon_m^{~np} e_\mu^{~a} e_\nu^{~b} L_n^{~i} L_p^{~j} F_{abij}, \non \\ 
\p_\mu B_{SS} &=& \frac{1}{6} e^{5\phi^{SS}/3} e_\mu^{~a} \epsilon^{ijk} F_{aijk}.\eea

To obtain the fermions, we set
\be \label{fer8} \psi_\mu^{SS} = \psi_\mu , \quad \chi_i^{SS} = \chi_i + \s_i \Gamma_9 \frac{\lambda}{3} ,\ee
where $\psi_\mu$, $\chi_i$ and $\lambda$ are $SU(2)$ pseudo--Majorana fermions in the Weyl basis as discussed before.
Thus for the action in \C{action}, this gives us
\be \label{scaleact} e^{-1} \mathcal{L}^{(0)} = 4 (e^{SS})^{-1} \mathcal{L}_{SS}.\ee

Note that the complete set of Chern--Simons terms in the action is given by
\bea \label{CS}
 4 \mathcal{L}_{SS} &=& \frac{1}{12^3} \epsilon^{\mu_1 \ldots \mu_8} \Big[ 3 B_{SS} G^{SS}_{\mu_1 \ldots \mu_4} G^{SS}_{\mu_5 \ldots
\mu_8} - 8\epsilon^{ijk} G^{SS}_{\mu_1 \mu_2 \mu_3 i} G^{SS}_{\mu_4 \mu_5 \mu_6 j}B^{SS}_{\mu_7 \mu_8 k} \non \\ &&
+ 288 F^{SS}_{\mu_1 \mu_2 i} B^{SS}_{\mu_3 \mu_4 i} \Big( (\p_{\mu_5} B^{SS}_{\mu_6 j}) B^{SS}_{\mu_7 \mu_8 j} - \frac{1}{3}
G^{SS}_{\mu_5\mu_6 \mu_7 j} B^{SS}_{\mu_8 j}\Big)  \non \\ && - 96 B^{SS}_{\mu_2 \mu_3 \mu_4}
(\p_{\mu_5} B^{SS}_{\mu_6 i}) G^{SS}_{\mu_1 \mu_7 \mu_8 i} \Big], \eea
where we have repeatedly integrated by parts, and used the Bianchi identities
\bea \p_{[\lambda} G^{SS}_{\mu\nu\rho\s]} = 4 F^{mSS}_{[\lambda\mu} G^{SS}_{\nu\rho\s]m}, \non \\ \p_{[\lambda} G^{SS}_{\mu\nu\rho]m} =
3\epsilon_{mnp} F^{nSS}_{[\lambda\mu} G^{pSS}_{\nu\rho]} . \eea

In \C{action}, we have only kept the first term in \C{CS}. The other terms in \C{CS}
are independent of $F_{4\mu\nu\rho\s}$, and so do not 
contribute to $G_{4\mu\nu\rho\s}$.

Of course, the values of $L^{~U}_{\pm}$ have to be substituted using the ones obtained later in \C{gfu}. Some of these 
calculations have
an overlap with~\cite{Andrianopoli:1996ve}, who work in a covariant formalism.

Note that the relation between the $d=11$ fermionic fields and the $d=8$ fermionic fields which are relevant for our purposes 
are given in equations (29) and (34) of~\cite{Salam:1984ft}\footnote{Equation (34) should read
\be \xi = e^{-\kappa\phi/6} \epsilon .\ee}. In particular, the $d=8$ fermions are given by~\cite{Cremmer:1979up,Salam:1984ft}
\bea \label{deffermions} \eta_a = e^{\phi_{SS}/6} \Big( \psi_a - \frac{1}{6} \Gamma_a \lambda \Big) , \non \\  \eta_i 
= e^{\phi_{SS}/6} \Big( \chi_i + \s_i \Gamma_9 \frac{\lambda}{3} \Big) ,\eea
where we have also used \C{fer8}.

This is useful in constructing the only other term in the action \C{action} which contributes to the 
definition of $G_{4\mu\nu\rho\s}$  
apart from those already mentioned before. This term is given by
\bea e^{-1}_{SS} \mathcal{L}_{SS}  =
-\frac{1}{96} \Big( \bar\eta_{\hat{A}} \Gamma^{\hat{A} \hat{B} abcd} \eta_{\hat{B}} 
+ 12 \bar\eta^a \Gamma^{bc} \eta^d \Big) F_{abcd} , \eea
where $F_{abcd}$ is a particular component of $F_{\hat{A}\hat{B}\hat{C}\hat{D}}$.

Let us now mention the local Lorentz transformation parameters in $d=8$ which are obtained directly from $d=11$, as 
described by~\cite{Salam:1984ft}. Dimensional reduction on $T^3$ breaks the $SO(10,1)$ symmetry of the frame indices to $SO(3) \times SO(7,1)$,
which is implemented by a gauge choice $V_m^{~a} =0$. Preserving this gauge choice, as well as requiring that $L^{-1} \delta^{(0)}
L$ is in $SO(3) \backslash SL(3,\mathbb{R})$ fixes $\lambda_{ia}$, and the local
Lorentz transformations parameters $\lambda_{ab}'$ and $\lambda_{ij}'$ of $SO(7,1)$ and $SO(3)$ respectively
in terms of $\lambda_{ab}$ and $\lambda_{ij}$. These relations are given by
\bea \lambda_{ia} &=& -\bar\epsilon \Gamma_a \Big( \chi_i +  \s_i \Gamma_9 \frac{\lambda}{3} \Big) ,\non \\
\lambda_{ab}' &=& \lambda_{ab} + \frac{1}{6} \bar\epsilon \Gamma_{ab} \lambda , \non \\ \lambda_{ij}' &=& 
\lambda_{ij} - \frac{1}{2} \bar\epsilon \Gamma_9 \Big( \s_i \chi_j - \s_j \chi_i \Big)- \frac{i}{3} \epsilon_{ijk}
\bar\epsilon \s^k \lambda .\eea   
We shall construct the complete supervariations of the various fermions from \C{11dtrans}, remembering to parametrize
the residual local Lorentz transformations by $\lambda_{ab}'$ and $\lambda_{ij}'$.   
Thus, the $d=8$ supersymmetry and local Lorentz transformations are given by
\bea \delta^{(0)} \eta_i = \ldots + \frac{1}{4} \Big( \lambda_{ab}' \Gamma^{ab} + \lambda_{jk}' \s^{jk} \Big) 
\eta_i + \lambda_{ij}' \eta_j , \non \\ \delta^{(0)} \eta_a = 
\ldots + \frac{1}{4} \Big( \lambda_{bc}' \Gamma^{bc} + \lambda_{ij}' \s^{ij} \Big) 
\eta_a + \lambda_{ab}' \eta_b , \eea 
where the $\ldots$ are the supersymmetry transformations given in \C{susytran}.

It should be noted that the supersymmetry transformations for all the fields
given in \C{susytran} are not simply 
obtained from those in~\cite{Salam:1984ft} by substituting the various expressions above. This is because they have already gauge fixed 
the $U(1)$ transformations, while our transformations are manifestly gauge covariant. Thus their transformations are the same as what we 
have only for the fields that are $U(1)$ invariant. For the other fields, their transformations have extra terms, which we shall describe 
later when we fix the $U(1)$ gauge symmetry. These extra terms, which are not gauge invariant, take a very simple form at the end, though
they look very complicated to start with. Various cancellations which are a consequence of supersymmetry are responsible for this 
simplification. In appendix \C{canc}, we outline the nature of these cancellations, which is quite intricate. 
Thus, the complete supersymmetry transformations of the $d=8$ theory are given
by \C{susytran}.

\section{Gauge fixing the local symmetry transformations}

We have two sets of moduli, each of which parametrizes a coset space $H
\backslash G$. We first gauge fix $H$ to obtain the physical degrees of
freedom in $L$. We then show how the $G$ symmetry is realized non--linearly on
the moduli. Finally, we consider the gauge fixed supersymmetry
transformations.

\subsection{Gauge fixing $L$}

So far we have parametrized the elements of the
coset space $H \backslash G$ in terms of $L$. We shall now gauge fix $L$ to
obtain the physical degrees of freedom. To do so, we use the Iwasawa
decomposition to represent $L$ as
\be \label{iwd} L = HKN,\ee
where $H$ is a matrix of gauge transformations, $K$ is a diagonal
matrix of unit determinant, and $N$ is an upper triangular matrix with
diagonal entries unity. Thus to work in a fixed gauge, we simply
remove the factor of $H$ in \C{iwd}. 

Let us now mention the moduli of type IIB string theory on $T^2$. The
complex structure $U$ of the $T^2$ parametrizes the moduli space
$SO(2) \backslash SL(2,\mathbb{R})$. The 5 degrees of freedom
parametrizing the moduli space $SO(3) \backslash SL(3,\mathbb{R})$ include
the complexified coupling $\tau$ obtained from 10 dimensions, and the
Kahler structure $T$ of the $T^2$ defined by 
\be T = B_N + i V, \ee
where $V$ is the volume of the $T^2$ in the string frame. The remaining
modulus is $B_R$, where $B_N$ ($B_R$) is obtained from the $NS$--$NS$
($R$--$R$) 2 form in 10 dimensions. Thus, the $SO(2) \backslash
SL(2,\mathbb{R})_\tau$ of S--duality, and the $SO(2) \backslash SL(2,\mathbb{R})_T$ 
of T--duality are
intertwined in the $SO(3) \backslash SL(3,\mathbb{R})$ moduli space. We now express
the components of $L$ in terms of these degrees of freedom.

For the $U(1) \backslash SL(2,\mathbb{R})$ moduli space, we take
\be L_V^{~v} =  \begin{pmatrix} \sqrt{U_2} & 0 \\ 0 & 1/\sqrt{U_2}  
\end{pmatrix}  \begin{pmatrix} 1 & -U_1/U_2 \\ 0 & 1  
\end{pmatrix}  = \frac{1}{\sqrt{U_2}} \begin{pmatrix} U_2 & -U_1 \\ 0 & 1  
\end{pmatrix} ,\ee
which leads to
\be L_v^{~V} =  \frac{1}{\sqrt{U_2}} \begin{pmatrix} 1 & U_1 \\ 0 & U_2  
\end{pmatrix} ,\ee
and
\be \label{SL2M}
M_{UV} = \frac{1}{U_2} \begin{pmatrix} \vert U \vert^2 & -U_1 \\ -U_1 & 1  
\end{pmatrix} . \ee 
Thus we get that
\be  \label{gfu} \begin{pmatrix} L^{~1}_{+} & L^{~2}_{+} \\ L^{~1}_{-} & L^{~2}_{-}  
\end{pmatrix}  = \frac{1}{\sqrt{2 U_2}} \begin{pmatrix} 1 & U \\ 1 & \bar{U}  
\end{pmatrix} ,\ee
yielding
\bea P_\mu = -\frac{\p_\mu U}{2 U_2} , \non \\ Q_\mu =  - \frac{\p_\mu
U_1}{2 U_2}. \eea
So in the action, we get that
\be - 2 P_\mu^* P^\mu = -\frac{\p_\mu U \p^\mu \bar{U}}{2 U_2^2}. \ee

For the $SO(3) \backslash SL(3,\mathbb{R})$ moduli space, we take
\bea \label{SL3v} L_m^{~i} &=& \begin{pmatrix} \nu^{-1/3} & 0 & 0 \\ 
0 & \sqrt{\tau}_2 \nu^{1/6} & 0  \\ 0 & 0 &  \nu^{1/6} / \sqrt\tau_2 \end{pmatrix} 
\begin{pmatrix} 1 &  -\sqrt{\frac{\nu}{\tau_2}} {\rm Im} B &
  \sqrt{\frac{\nu}{\tau_2}}{\rm Re} B 
\\ 0 & 1 &  -\tau_1 /\tau_2  \\ 0 & 0 & 1 \end{pmatrix} \non \\ 
& =& \frac{\nu^{1/6}}{\sqrt{\tau_2}} \begin{pmatrix} \sqrt{\frac{\tau_2 }{\nu}} &   -{\rm Im} B &  
{\rm Re} B  \\ 0 & \tau_2  & - \tau_1  \\ 0 & 0 &  1 
\end{pmatrix} ,\eea
where $\nu = (\tau_2 V^2)^{-1}$, and $B = B_R + \tau B_N$.

This leads to\footnote{This is the same as~\cite{Kiritsis:1997em} on sending $\tau_1 \rightarrow - \tau_1$
and $B_N \rightarrow - B_N$.}
\be  \label{SL3M}
M_{mn} = \frac{\nu^{1/3}}{\tau_2} \begin{pmatrix}  \tau_2/\nu + \vert B
\vert^2 & -{\rm Re}(\bar{\tau} B) & {\rm Re} B \\ -{\rm Re}(\bar{\tau} B) & \vert \tau
\vert^2  &- \tau_1 \\ {\rm Re} B & -\tau_1  & 1  \end{pmatrix}. \ee

It is useful to see how the $U(1) \backslash SL(2,\mathbb{R})_\tau$
and $U(1) \backslash SL(2,\mathbb{R})_T$ subspaces of \C{SL3v} are intertwined. To see the 
$U(1) \backslash SL(2,\mathbb{R})_\tau$ subspace, we drop the $B$
dependence for simplicity, and focus on $\tau$ and $V$, leading to
\be \label{1tau} L_m^{~i} = \nu^{1/6} \begin{pmatrix} \nu^{-1/2} & 0 & 0 \\ 
0 & \sqrt{\tau}_2  & -\tau_1 / \sqrt{\tau}_2  \\ 0 & 0 &  1/ \sqrt\tau_2 \end{pmatrix} ,\ee
where $\nu$ is S--duality invariant.  To see the 
$U(1) \backslash SL(2,\mathbb{R})_T$ subspace, we drop the $B_R$ and $\tau_1$
dependence for simplicity, and focus on $T$, $\tau_2$ and $V$, leading to
\be \label{1t} L_m^{~i} = e^{-\hat\phi/3} \begin{pmatrix} \sqrt{T}_2 & -T_1 / \sqrt{T}_2 & 0 \\ 
0 & 1/ \sqrt{T}_2  & 0  \\ 0 & 0 &  e^{\hat\phi} \end{pmatrix} ,\ee
where
\be \label{changecoord} e^{-2 \hat\phi} = \tau_2^2 V \ee
is the T--duality invariant $d=8$ dilaton.

\subsection{Non--linearly realized $G$ symmetry}

Having gauge fixed $H$, let us now see how the $G$ symmetry is
realized non--linearly on the moduli. First consider the $U(1)
\backslash SL(2,\mathbb{R})$ moduli space, where
\be L^{~U}_{\pm} \rightarrow e^{\pm i\Sigma} L^{~V}_{\pm} S_V^{~U}. \ee 
It is sufficient for our purposes to look at infinitesimal
transformations. Thus taking
\be  S_V^{~U} = \begin{pmatrix} 1 + \alpha & \beta  \\ \gamma & 
1 - \alpha  \end{pmatrix},\ee
where $\alpha, \beta, \gamma$, and $\delta$ are infinitesimal real
parameters, requiring reality of $L^1_{\pm}$ in \C{gfu}, we get that
\be \Sigma = - \gamma U_2 . \ee 
Also including the constraints coming from $L^2_{\pm}$, we get that
\be \delta U = \beta -  2 \alpha U - \gamma U^2 .\ee
It is also easy to write down the finite transformations. Taking
\be  S_V^{~U} = \begin{pmatrix} d & b  \\ c & a 
\end{pmatrix},\ee
where $a,b,c,d \in \mathbb{R}$ and $ad-bc =1$, the above analysis leads to
\be {\rm tan} \Sigma  = - \frac{c U_2}{c U_1 +d} ,\ee
thus leading to the $SL(2,\mathbb{R})$
transformation
\be U \rightarrow \frac{a U + b}{c U + d}.\ee

For the $SO(3) \backslash SL(3,\mathbb{R})$ moduli, in \C{sl3t}, we take
\be \label{msl3} R^n_{~m} = \begin{pmatrix} 1 + \alpha & b & c  \\ d & 1 +  \beta &
  f \\ g &  h &
1 - \alpha  -\beta \end{pmatrix}, \ee
where $\alpha, \beta, b,c, d,f,g$, and $h$ are infinitesimal real
parameters, and
\be \label{small} O^i_{~j} = \delta_{ij} + \epsilon_{ijk} \theta_k .\ee 
Thus preserving the gauge choice $L_2^{~1} = L_3^{~1} = L_3^{~2} =0$
in \C{SL3v}, we get that 
\bea \label{smth} \theta_1 &=& c {\rm Im} B - f \tau_2 , \non \\ \theta_2 &=& c
\sqrt{\frac{\tau_2}{\nu}}, \non \\ \theta_3 &=& - \frac{1}{\sqrt{\nu
    \tau_2}} (b + c \tau_1).\eea
Thus, the transformations of the remaining non--vanishing elements of $L_m^{~i}$ lead to
\bea \label{smallt} \delta \nu &=& - 3 \nu \Big(\alpha -  c {\rm Re}B + \frac{(b + c
  \tau_1)}{\tau_2}  {\Im B}\Big) ,\non \\ \delta \tau &=& (\alpha + 2
\beta  - c B)\tau - b B -h + f \tau^2 , \non \\ \delta B &=& (2\alpha
+ \beta)B - \frac{i(b+c\tau)}{\nu} + g - d \tau + f \tau B - c B^2 .\eea

One can consider finite transformations as well. In order to do so, we 
use the definition of the finite form of the matrix $O$ in \C{small} given by
\be \label{usedef} O^i_{~j} =  \frac{1}{2} {\rm Tr} (g^{-1} \s_i g \s_j) , \ee
where $g$ is an element of the $SU(2)$ group, and the trace is in the
fundamental representation of $SU(2)$. This follows from the
defining equation for the transformation matrices $D$ of the adjoint
representation of any group, given by
\be g^{-1} T_a g = D_a^{~b} (g) T_b ,\ee
where $g$ is an element of the group, and $T_a$ are the generators in
the fundamental representation. Thus, \C{usedef} follows for $SU(2)$,
where
\be g = {\rm cos} \frac{\vert \vec\theta \vert}{2} + i \frac{\vec\s
  \cdot \vec\theta}{\vert \vec\theta \vert} {\rm sin} \frac{\vert \vec\theta \vert}{2},\ee
and we have chosen the normalization in \C{usedef} to obtain \C{small} in the
infinitesimal limit. This leads to
\be O^i_{~j} (\theta)= \delta_{ij} {\rm cos} \vert \vec\theta \vert + 
\epsilon_{ijk} \theta_k \frac{{\rm sin} \vert \vec\theta \vert }{\vert
  \vec\theta \vert} + \frac{2 \theta_i \theta_j}{\vert \vec\theta
  \vert^2} {\rm sin}^2 \frac{\vert \vec\theta \vert}{2} . \ee 
This representation is called the axis--angle representation in literature. 
It consists of an $SO(3)$ rotation about an axis given by
\be \vec{n} = \frac{\vec\theta}{\vert \vec\theta \vert} ,\ee
by an angle $\vert \vec\theta \vert$. 
One can then proceed exactly along the lines of the above discussion
to obtain the finite transformations. Thus the finite transformations are
given by the solution of a set of involved equations, which are difficult to manipulate. 
Thus to obtain the finite transformations, we use a different representation of $SO(3)$ rotation 
matrices.  

We parametrize an arbitrary $SO(3)$ rotation by successive rotations around the $2,3$ and $1$ axes 
by angles $-\phi_2,\phi_3$ and $\phi_1$ respectively, and so
\be O = R_{23} (\phi_1) R_{12} (\phi_3) R_{13} (-\phi_2),\ee
where
\be R_{23} (\phi_1)= \begin{pmatrix} 1 & 0 & 0\\ 
0 & {\rm cos} \phi_1 & {\rm sin} \phi_1 
\\ 0 & -{\rm sin} \phi_1  &  {\rm cos} \phi_1  \end{pmatrix} ,\ee
for example. These angles are called the Tait--Bryan angles in literature\footnote{Also 
called yaw, pitch and roll. These angles are not to be confused with the Euler angles, which involve 
three rotations as well, but the first and third are about the same axis.}. Thus we have that
\be O^i_{~j} = \begin{pmatrix} {\rm cos} \phi_2 {\rm cos} \phi_3 & {\rm sin} \phi_3
  & -  {\rm sin} \phi_2  {\rm cos} \phi_3 \\ -{\rm cos} \phi_1  {\rm cos} \phi_2 {\rm sin} \phi_3 + {\rm sin} \phi_1 
{\rm sin} \phi_2 & {\rm cos} \phi_1 {\rm cos} \phi_3 & {\rm cos} \phi_1 {\rm sin} \phi_2 {\rm sin} \phi_3 + {\rm sin}
\phi_1 {\rm cos} \phi_2 
\\ {\rm sin} \phi_1 {\rm cos} \phi_2 {\rm sin} \phi_3  +  {\rm cos} \phi_1 {\rm sin} \phi_2
& -{\rm sin} \phi_1 {\rm cos} \phi_3 & - {\rm sin} \phi_1 {\rm sin} \phi_2 {\rm sin} \phi_3 + {\rm cos} \phi_1 
{\rm cos} \phi_2 \end{pmatrix} . \ee
Thus, for small angles,
\be \phi_i = \theta_i .\ee 
We further define
\be \mu^j_{~m} = L_n^{~j} R^n_{~m} . \ee
Thus, preserving the gauge choices $L_2^{~1} = L_3^{~1} =0$, we get that
\bea \label{tb1} {\rm sin} \phi_2 &=& \frac{\mu^1_{~3} \mu^2_{~2} - 
\mu^1_{~2} \mu^2_{~3}}{\sqrt{(\mu^1_{~3} \mu^2_{~2} - \mu^1_{~2} \mu^2_{~3} )^2 + (\mu^2_{~2} \mu^3_{~3} 
- \mu^3_{~2} \mu^2_{~3})^2}} , \non \\
{\rm cos} \phi_2 &=& \frac{\mu^2_{~2} \mu^3_{~3} 
- \mu^3_{~2} \mu^2_{~3}}{\sqrt{(\mu^1_{~3} \mu^2_{~2} - \mu^1_{~2} \mu^2_{~3} )^2 + (\mu^2_{~2} \mu^3_{~3} 
- \mu^3_{~2} \mu^2_{~3})^2}}, \non \\ {\rm sin} \phi_3 &=& \frac{\mu^3_{~2} \mu^1_{~3} 
- \mu^1_{~2} \mu^3_{~3}}{\sqrt{(\mu^1_{~3} \mu^2_{~2} - \mu^1_{~2} \mu^2_{~3} )^2 + (\mu^2_{~2} \mu^3_{~3} 
- \mu^3_{~2} \mu^2_{~3})^2 + (\mu^1_{~2} \mu^3_{~3} - \mu^1_{~3} \mu^3_{~2})^2}}, \non \\ {\rm cos} \phi_3 &=& 
\frac{\sqrt{(\mu^1_{~3} \mu^2_{~2} - \mu^1_{~2} \mu^2_{~3} )^2 + (\mu^2_{~2} \mu^3_{~3} 
- \mu^3_{~2} \mu^2_{~3})^2 }}{\sqrt{(\mu^1_{~3} \mu^2_{~2} - \mu^1_{~2} \mu^2_{~3} )^2 + (\mu^2_{~2} \mu^3_{~3} 
- \mu^3_{~2} \mu^2_{~3})^2 + (\mu^1_{~2} \mu^3_{~3} - \mu^1_{~3} \mu^3_{~2})^2}}.\eea
Finally, preserving the gauge choice $L_3^{~2} =0$, we get that
\bea \label{tb2} {\rm sin} \phi_1 &=& -\frac{\mu^2_{~3} {\rm sec} \phi_3}{\sqrt{(\mu^2_{~3} {\rm sec} \phi_3)^2 
+ (\mu^1_{~3} {\rm sin} \phi_2 + \mu^3_{~3} {\rm cos} \phi_2)^2}}, \non \\ {\rm cos} \phi_1 &=& 
\frac{\mu^1_{~3} {\rm sin} \phi_2 + \mu^3_{~3} {\rm cos} \phi_2}{\sqrt{(\mu^2_{~3} {\rm sec} \phi_3)^2 
+ (\mu^1_{~3} {\rm sin} \phi_2 + \mu^3_{~3} {\rm cos} \phi_2)^2}}, \eea
where we have used the relation
\be \mu^1_{~3} {\rm cos} \phi_2 - \mu^3_{~3} {\rm sin} \phi_2  = - \mu^2_{~3} {\rm tan} \phi_3. \ee
There is an overall sign ambiguity in obtaining \C{tb1} and \C{tb2}, 
which is fixed by taking the small $\phi_i$ limit, and matching 
with \C{smth}. Thus the angles $\phi_1, \phi_2$ and $\phi_3$ are fixed, which when 
inserted into the expressions given below determine the complete set of transformations.  
The remaining expressions are obtained by varying the non--vanishing elements of the vielbein leading to
\bea \nu' &=& (O^1_{~j} \mu^j_{~1})^{-3} , \non \\ \tau' &=& -\frac{(O^3_{~j} - i O^2_{~j} ) 
\mu^j_{~2}}{O^3_{~j} \mu^j_{~3}}, \non \\ B' &=& \frac{(O^3_{~j}  - i O^2_{~j} ) \mu^j_{~1}}{O^3_{~j} 
\mu^j_{~3}} ,\eea
where we have used
\be (O^1_{~i} \mu^i_{~1}) (O^2_{~j} \mu^j_{~2} ) (O^3_{~k} \mu^k_{~3} )=1 .\ee
It is straightforward to write down these explicit expressions. We keep them implicit as they are 
quite complicated and shall be calculated explicitly later.

The coordinates $\nu,\tau$ and $B$ in \C{smallt} we have
chosen to parametrize the $SO(3) \backslash SL(3,\mathbb{R})$ moduli
space are a natural choice from the $U(1) \backslash
SL(2,\mathbb{R})_\tau$ point of view. To see this, we take the
corresponding $SL(2,\mathbb{R})_\tau$ subspace in \C{msl3} given by setting
$\alpha = b=c=d=g=0$. Then \C{smallt} yields that $\delta \nu =0$ as
expected. Also we get that
\bea \tau \rightarrow \frac{\hat{a} \tau + \hat{b}}{\hat{c} \tau + \hat{d}}, \quad B
\rightarrow \frac{B}{\hat{c} \tau + \hat{d}} , \eea 
where $\hat{a} = 1 + \beta, \hat{b} = -h, \hat{c} = -f$, and $\hat{d}
= 1 -\beta$ in the infinitesimal limit, as required.  

Another useful parametrization of the $SO(3) \backslash
SL(3,\mathbb{R})$ moduli space is by the coordinates $T, \xi$ and $e^{-2\hat\phi}$, where
\be \xi = - B_R + i \tau_1 V ,\ee
which is natural from the $U(1) \backslash
SL(2,\mathbb{R})_T$ point of view which follows from the discussion below.
Thus the coordinates $\hat\phi, T$ and $\xi$ are related to $\nu, B$ and $\tau$ by \C{changecoord} and  
\bea \label{longrel} T &=& \frac{1}{\tau_2} \Big( {\rm Im} B +i \tau_2 V \Big) , \non \\ 
\xi &=& \frac{1}{\tau_2} \Big( {\rm Im} (B \bar\tau)+ i \tau_1 \tau_2 V \Big). \eea

This leads to
\be \label{SL3v2} L_m^{~i} = \begin{pmatrix} e^{-\hat\phi/3} \sqrt{T_2} & -
  e^{-\hat\phi/3} T_1/\sqrt{T_2}   & e^{2\hat\phi/3} {\rm Im} (\xi \bar{T})/T_2 \\ 
0 & e^{-\hat\phi/3}/ \sqrt{T_2}  & - e^{2\hat\phi/3} {\rm Im} \xi/T_2 
\\ 0 & 0 &  e^{2\hat\phi/3} \end{pmatrix} ,\ee
and thus
\be M_{mn} = \frac{e^{4 \hat\phi/3}}{T_2} \begin{pmatrix} e^{-2\hat\phi} 
\vert T \vert^2 + ({\rm Im} (\xi \bar{T}))^2 
/ T_2
& -(e^{-2\hat\phi} T_1 + \xi_2 {\rm Im} (\xi \bar{T}) / T_2 )  & {\rm Im} (\xi \bar{T}) \\ 
-(e^{-2\hat\phi} T_1 + \xi_2 {\rm Im} (\xi \bar{T}) / T_2 )
& (\xi_2^2 + T_2 e^{-2\hat\phi})/ T_2  & - \xi_2  
\\ {\rm Im}(\xi \bar{T})  & -\xi_2  &  T_2 \end{pmatrix}. \ee
Proceeding as before, for infinitesimal transformations, we get that
\bea \theta_1 &=& e^{-\hat\phi} \frac{(c T_1 -f)}{\sqrt{T_2}}, \non \\
\theta_2 &=& c e^{-\hat\phi} \sqrt{T_2}, \non \\ \theta_3 &=& - (b T_2
+ c \xi_2 ), \eea
as well as
\bea \label{smallT} \delta \hat\phi &=& \frac{3}{2} \Big( -(\alpha + \beta) + c
\frac{{\rm Im} \xi \bar{T}}{T_2} - f \frac{{\rm Im} \xi}{T_2} \Big),
\non \\ \delta  T &=& (\alpha -\beta) T - d - f\xi + b T^2 + c T \xi ,
\non \\ \delta \xi &=& -g -hT + (\beta + 2 \alpha) \xi + c \xi^2 + bT
\xi + icT e^{-2\hat\phi} -if e^{-2\hat\phi}.\eea

The finite transformations are given by
\bea e^{2\hat\phi'} &=& (O^3_{~j} \mu^j_{~3})^3 , \non \\ T' &=& -\frac{(O^2_{~j} -i O^1_{~j}) 
\mu^j_{~1}}{O^2_{~j} \mu^j_{~2}} , \non \\ \xi' &=& (O^1_{~i} \mu^i_{~1}) \Big[ (O^3_{~j} \mu^j_{~2} )(O^2_{~k} 
\mu^k_{~1}) - (O^3_{~j} \mu^j_{~1} )(O^2_{~k} \mu^k_{~2} )-i (O^1_{~j} \mu^j_{~1} )(O^3_{~k} \mu^k_{~2} 
)\Big] . \eea

Consider the $SL(2,\mathbb{R})_T$ subspace in \C{msl3} given by setting
$\alpha = -\beta, c=f=g=h=0$. Then \C{smallT} yields that $\delta \hat\phi =0$ as
expected. Also we get that
\bea T \rightarrow \frac{\hat{a} T + \hat{b}}{\hat{c} T + \hat{d}}, \quad \xi
\rightarrow \frac{\xi}{\hat{c} T + \hat{d}} , \eea 
where $\hat{a} = 1 + \alpha, \hat{b} = -d, \hat{c} = -b$, and $\hat{d}
= 1 -\alpha$ in the infinitesimal limit. The fermions also transform
accordingly by gauge transformations given by the $\theta_i$. 

A similar analysis for $N=8, d=4$ supergravity has been carried out in~\cite{Kallosh:2008ic}.

\subsection{Gauge fixed supersymmetry transformations}

Now let us consider the gauge fixed supersymmetry transformations of
the moduli. In order to maintain the choices of gauge \C{gfu} and
\C{SL3v}, we have to make additional gauge transformations with field
dependent parameters. First consider the moduli for the $SO(2) \backslash
SL(2,\mathbb{R})$ coset~\cite{Schwarz:1983qr}. From \C{susytran}, preserving the reality of
$L^{~1}_{\pm}$ gives us that
\be \label{fixS} \Sigma_\epsilon = - \frac{i}{2} (\bar\epsilon_L \lambda_R -
\bar\epsilon_R \lambda_L). \ee 
Also including the effect of this compensating gauge
transformation on $L^{~2}_\pm$, we get that
\be \delta^{(0)} U = -2i U_2 \bar\epsilon_R \lambda_L , \quad \delta^{(0)} \bar{U}
= 2i U_2 \bar\epsilon_L \lambda_R .\ee

We next consider the moduli for the $SO(3) \backslash
SL(3,\mathbb{R})$ coset, where including compensating gauge
transformations, we get that
\be \label{var3} \delta^{(0)} L_m^{~i} = \Lambda_{ij} L_m^{~j} - \epsilon^{ijk}
\theta_j^{\epsilon} L_m^{~k} ,\ee
where
\be \Lambda_{ij} =  -\frac{1}{2} \Big( \bar\epsilon_L (\s_i \chi_{jR} +
\s_j \chi_{iR} ) -
\bar\epsilon_R (\s_i \chi_{jL} +\s_j \chi_{iL} )\Big).\ee
Thus $\Lambda_{ij}$ is real and satisfies $\Lambda_{ij} =
\Lambda_{ji}$, as well as
\be \Lambda_{11} + \Lambda_{22} + \Lambda_{33} =0 .\ee
So $\Lambda_{ij}$ has 5 independent components which we can take to be
$\Lambda_{11}, \Lambda_{12}, \Lambda_{13}, \Lambda_{22}$, and
$\Lambda_{23}$. For our purposes, it is very convenient to explicitly use
\be \Lambda_{ij} - \frac{1}{3} \delta_{ij} \Lambda_{kk} \ee
in place of $\Lambda_{ij}$ in \C{var3}. While it gives no extra information,
the tracelessness is automatic, and we do not have to choose an explicit  basis for the 5
independent components.

Proceeding as before, preserving the gauge choice $L_2^{~1} = L_3^{~1} = L_3^{~2} =0$
in \C{SL3v} or \C{SL3v2}, we get that
\bea \label{fixt} \theta_1^\epsilon &=& - \Lambda_{23}, \non \\ \theta_2^\epsilon
&=& \Lambda_{13}, \non \\ \theta_3^\epsilon &=& - \Lambda_{12},\eea
which together with the remaining supervariations in \C{SL3v} gives us
\bea \delta^{(0)} \nu &=& - 2 \Big( \Lambda_{11} -\frac{1}{2} (\Lambda_{22} + \Lambda_{33})\Big) \nu 
, \non \\ \delta^{(0)} \tau &=& i
\tau_2 (\Lambda_{22} - \Lambda_{33} + 2 i \Lambda_{23}), \non \\
\delta^{(0)} B &=& 2 \Big( \Lambda_{13} \sqrt{\frac{\tau_2}{\nu}} - \Lambda_{23}
       {\rm Im} B \Big) + i \Big( (\Lambda_{22} - \Lambda_{33}) {\rm Im}
       B - 2 \Lambda_{12} \sqrt{\frac{\tau_2}{\nu}}\Big).\eea
In the coordinate system given by \C{SL3v2}, we get that
\bea \label{varmoduli} \delta^{(0)} \hat\phi &=& \Big(\Lambda_{33}
-\frac{1}{2} (\Lambda_{11} + \Lambda_{22}) \Big)
,
\non \\ \delta^{(0)} T &=& T_2 \Big( -2 \Lambda_{12} + i (\Lambda_{11} -
\Lambda_{22})\Big) , \non \\ \delta^{(0)} \xi
&=& -2 (\sqrt{T_2} \Lambda_{13} e^{-\hat\phi} + \Lambda_{12} {\rm Im}
\xi) + i \Big( (\Lambda_{11} - \Lambda_{22}){\rm Im} \xi -2
\Lambda_{23} \sqrt{T_2} e^{-\hat\phi}\Big) .\eea 
This also leads to extra terms in the supersymmetry transformations
for the other fields, of which, the fermions are relevant for our
purposes. For the fermions, the extra terms in the supervariation
which have to be added to \C{susytran} are 
\bea \label{addsusytran} \hat\delta \lambda_L &=& \frac{3i}{2} \Sigma_\epsilon \lambda_L +
\frac{i}{2} \theta^i_\epsilon \s^i \lambda_L ,
\non \\ \hat\delta \lambda_R &=& - \frac{3i}{2} \Sigma_\epsilon 
\lambda_R +\frac{i}{2} \theta^i_\epsilon \s^i \lambda_R, \non \\
\hat\delta \psi_{\mu L} &=& \frac{i}{2} \Sigma_\epsilon \psi_{\mu L}
+\frac{i}{2} \theta^i_\epsilon \s^i \psi_{\mu L}, \non \\ \hat\delta
\psi_{\mu R}  &=& -\frac{i}{2} \Sigma_\epsilon \psi_{\mu R}
+\frac{i}{2} \theta^i_\epsilon 
\s^i \psi_{\mu R}, \non
\\ \hat\delta \chi^i_L &=& -\frac{i}{2} \Sigma_\epsilon \chi^i_L -i
\theta^j_\epsilon \s^i 
\chi^j_L + \frac{3i}{2} 
\theta^j_\epsilon \s^j \chi^i_L, \non \\ \hat\delta \chi^i_R &=&
\frac{i}{2} \Sigma_\epsilon \chi^i_R -i \theta^j_\epsilon \s^i \chi^j_R + \frac{3i}{2} 
\theta^j_\epsilon \s^j \chi^i_R ,\eea
where $\Sigma_\epsilon$ and $\theta^i_\epsilon$ are given by \C{fixS} and
\C{fixt} respectively. Thus from now onwards, the complete supersymmetry transformations of the supergravity theory
will be denoted as $\delta^{(0)}$. In particular, $\delta^{(0)}$ for the various fermions is given by the sum of 
\C{susytran} and \C{addsusytran}.

\section{Transformations of the moduli under U--duality}

In the above discussion, we have gauge fixed the $H$ symmetry transformations to obtain the physical 
degrees of freedom. This led to transformations
of the various fields where the $G$ symmetry is realized non--linearly. 

We now focus on these transformations of the moduli in detail. While the transformations
of the moduli parametrizing the coset space $SO(2) / SL(2,\mathbb{R})$ were easy to write down explicitly as given above,
this was not the case for the moduli parametrizing the coset space $SO(3) / SL(3,\mathbb{R})$ due to the gauge redundancy.
So for this purpose it is
easier to use the matrix $M$, which is gauge invariant. In fact, under $L \rightarrow {\mathcal{O}} L S$, we have that      
\be \label{changeg} M_{mn} = L_m^{~i} L_m^{~j} \delta_{ij} \rightarrow (S^T M S)_{mn} .\ee
Thus, we obtain the transformations of the moduli using \C{changeg}, and we also change the notation slightly 
from before, for later convenience.

\subsection{Transformation under $SL(2,\mathbb{Z})$}

For the $SL(2,\mathbb{R})$ transformations, taking
\be \label{sl2tt}
S = \begin{pmatrix} \mathcal{A} & -\mathcal{C}  \\ -\mathcal{B} & \mathcal{D} \end{pmatrix},\ee
where $\mathcal{A}, \mathcal{B}, \mathcal{C}$ and $\mathcal{D}$ are real numbers satisfying $\mathcal{A}
\mathcal{D} - \mathcal{B} \mathcal{C}=1$, \C{changeg} gives us that
\be U_1' = \frac{1}{2} \Big( \frac{\mathcal{A}U +\mathcal{B}}{\mathcal{C}U+ \mathcal{D}} + 
\frac{\mathcal{A}\bar{U} +\mathcal{B}}{ \mathcal{C}\bar{U} +\mathcal{D}} \Big) , \quad
U_2' = \frac{U_2}{\vert \mathcal{C}U +\mathcal{D} \vert^2}, \ee
leading to
\be \label{autSL2}
U' = \frac{\mathcal{A}U+ \mathcal{B}}{\mathcal{C}U+ \mathcal{D}} ,\ee
as before.

\subsection{Transformation under $SL(3,\mathbb{Z})$}

For the $SL(3,\mathbb{R})$ transformations, we take
\be  \label{sl3tt} S = \begin{pmatrix} \mathcal{A} & -\mathcal{C} & \mathcal{J}  \\ -\mathcal{B} & \mathcal{D} &
  -\mathcal{F} \\ \mathcal{H} &  -\mathcal{E} & \mathcal{G} \end{pmatrix} , \ee
where $\mathcal{A},\mathcal{B},\mathcal{C},\mathcal{D},\mathcal{E},\mathcal{F},\mathcal{G},\mathcal{H}$ and $\mathcal{J}$
are real numbers satisfying ${\rm det} S =1$.

Again, using \C{changeg}, we get that
\bea \label{sl3trans}
\nu' &=& \frac{1}{\nu^2 \tau_2^3} \Big[\nu \tau_2 \vert \mathcal{C} \Xi_3 - \mathcal{J} \Xi_2  \vert^2 +  
[\nu{\rm Im} (\Xi_2 \bar\Xi_3) ]^2 
\Big]^{3/2} \non \\ &=&\frac{1}{\nu^2 \tau_2^3} \Big[\nu \tau_2 \vert \Upsilon_1 \tau + \Upsilon_2 \vert^2 + \nu^2 
\Big(\Upsilon_1 {\rm Im} (\bar{B} \tau) - \Upsilon_2 {\rm Im} B - \Upsilon_3 \tau_2 \Big)^2 
\Big]^{3/2}, \non \\ \tau_1' &=& \frac{\mathcal{C} \mathcal{J} \tau_2 + \nu 
{\rm Re}(\Xi_2 \bar\Xi_3 )}{\mathcal{J}^2 
\tau_2 + \nu \vert \Xi_3 \vert^2} , \non \\ 
\tau_2' &=& \frac{\Big[\nu \tau_2 \vert \mathcal{C} \Xi_3 - \mathcal{J} \Xi_2  \vert^2 +  
[\nu{\rm Im} (\Xi_2 \bar\Xi_3) ]^2 
\Big]^{1/2}}{\mathcal{J}^2 
\tau_2 + \nu \vert \Xi_3 \vert^2} \non \\ &=&\frac{\Big[\nu \tau_2 \vert \Upsilon_1 \tau + \Upsilon_2 \vert^2 + \nu^2 
\Big(\Upsilon_1 {\rm Im} (\bar{B} \tau) - \Upsilon_2 {\rm Im} B - \Upsilon_3 \tau_2 \Big)^2 
\Big]^{1/2}}{\mathcal{J}^2 
\tau_2 + \nu \vert \Xi_3 \vert^2}, \non \eea
\bea
{\rm Re} B' &=& 
\frac{\mathcal{A J} \tau_2 + \nu {\rm Re}(\Xi_1 \bar\Xi_3 )}{\mathcal{J}^2 
\tau_2 + \nu \vert \Xi_3  \vert^2} ,  \\ {\rm Im} B' &=& \frac{\tau_2 
\nu {\rm Re}(\mathcal{J} \Xi_1 - \mathcal{A} \Xi_3 )(\mathcal{J} \bar\Xi_2
- \mathcal{C} \bar\Xi_3) + \nu^2 {\rm Im}(\Xi_1 \bar\Xi_3) {\rm Im} (\Xi_2 \bar\Xi_3)}{(\mathcal{J}^2 
\tau_2 + \nu \vert \Xi_3 \vert^2 )\Big[ \nu \tau_2 \vert \mathcal{C} \Xi_3 - \mathcal{J} \Xi_2  \vert^2 +  
[\nu{\rm Im} (\Xi_2 \bar\Xi_3)]^2 \Big]^{1/2}} \non \\ &=& 
\frac{\nu\tau_2 \Omega_1 /2 - \nu^2 \Omega_2 /4}{(\mathcal{J}^2 
\tau_2 + \nu \vert \Xi_3 \vert^2 )\Big[\nu \tau_2 \vert \Upsilon_1
\tau + \Upsilon_2 \vert^2 + \nu^2 
\Big(\Upsilon_1 {\rm Im} (\bar{B} \tau) - \Upsilon_2 {\rm Im} B - \Upsilon_3 \tau_2 \Big)^2 
\Big]^{1/2}}  \non ,
\eea
where
\bea  
\Upsilon_1 &=& \mathcal {CF} - \mathcal{DJ}, \non \\ \Upsilon_2 &=& \mathcal{CG} - \mathcal{EJ}, 
\non \\ \Upsilon_3 &=& \mathcal{DG} - \mathcal{EF}  , \non \\ \Upsilon_4 &=& \mathcal{AF} - \mathcal{BJ}, 
\non \\ \Upsilon_5 &=& \mathcal{AG} - \mathcal{JH}, \non \\  \Upsilon_6 &=& \mathcal{BG} - \mathcal{HF}\eea
involves various subdeterminants of \C{sl3tt}, and
\bea \Xi_1 &=& \mathcal{A} B + \mathcal{B} \tau + \mathcal{H}, \non \\ \Xi_2 &=& 
\mathcal{C} B + \mathcal{D} \tau + \mathcal{E}, \non \\ \Xi_3 &=& \mathcal{J} B + \mathcal{F}\tau + \mathcal{G}
.\eea

Also in the expression for ${\rm Im} B'$, we have used
\bea \Omega_1 &=& (\mathcal{J} \Xi_1 -\mathcal{A} \Xi_3)(\mathcal{J} \bar\Xi_2 - \mathcal{C} \bar\Xi_3) 
+ (\mathcal{J} \bar\Xi_1 - \mathcal{A} \bar\Xi_3)(\mathcal{J} \Xi_2 - \mathcal{C} \Xi_3)  
\non \\ &=& (\Upsilon_1 \bar\tau + \Upsilon_2)
(\Upsilon_4 \tau + \Upsilon_5) + (\Upsilon_1 \tau + \Upsilon_2)
(\Upsilon_4 \bar\tau + \Upsilon_5) , \non \\ 
\Omega_2 &=&(\Xi_1 \bar\Xi_3 - \Xi_3 \bar\Xi_1)(\Xi_2 \bar\Xi_3 - \Xi_3 \bar\Xi_2) \non \\ &=& 
-4 \Big( \Upsilon_1 {\rm Im} (B\bar\tau) + \Upsilon_2 {\rm Im} B + \Upsilon_3 {\rm Im} \tau \Big) 
\Big( \Upsilon_4 {\rm Im} (B\bar\tau) + \Upsilon_5 {\rm Im} B + \Upsilon_6 {\rm Im} \tau \Big),\eea
for brevity. 

Thus we have that
\bea \tau' &=& \Big[ \frac{\mathcal{C}^2 
\tau_2 + \nu \vert \Xi_2 \vert^2}{\mathcal{J}^2 
\tau_2 + \nu \vert \Xi_3 \vert^2} \Big]^{1/2} e^{i\theta_\tau}, \non \eea
\bea
B' &=& \frac{ e^{i\theta_B}}{\Big[ \mathcal{J}^2 \tau_2 + \nu \vert \Xi_3 \vert^2 \Big] \Big[ \nu \tau_2 \vert \mathcal{C} \Xi_3 - \mathcal{J} \Xi_2  \vert^2 +[ \nu 
{\rm Im}(\Xi_2 \bar\Xi_3)]^2 \Big]^{1/2}} \times \non \\ &&   \Big[ \Big(\nu \tau_2 \vert \mathcal{C} \Xi_3 - \mathcal{J} \Xi_2  \vert^2 +[ \nu 
{\rm Im}(\Xi_2 \bar\Xi_3)]^2 \Big) \Big(\mathcal{A} \mathcal{J} \tau_2 +  \nu {\rm Re} (\Xi_1 \bar\Xi_3)\Big)^2 \non \\ && + \Big( \tau_2 \nu 
{\rm Re} (\mathcal{J} \Xi_1 -\mathcal{A} \Xi_3 )(\mathcal{J} \bar\Xi_2
- \mathcal{C} \bar\Xi_3)+ \nu^2 {\rm Im} (\Xi_1 \bar\Xi_3) {\rm Im} (\Xi_2 \bar\Xi_3) \Big)^2 \Big]^{1/2}  
 ,\eea
where
\bea {\rm tan} \theta_\tau &=& \frac{\Big[ \nu \tau_2 \vert \mathcal{C} \Xi_3 - \mathcal{J} \Xi_2  \vert^2 + [\nu 
{\rm Im}(\Xi_2 \bar\Xi_3)]^2  \Big]^{1/2}}{\mathcal{C} \mathcal{J} \tau_2 + \nu {\rm Re}
( \Xi_2 \bar\Xi_3  )}, \non \\ {\rm tan} \theta_B &=& \frac{\tau_2 \nu {\rm Re}(\mathcal{J} \Xi_1 - \mathcal{A} \Xi_3 )(\mathcal{J} \bar\Xi_2
- \mathcal{C} \bar\Xi_3) + \nu^2 {\rm Im}(\Xi_1 \bar\Xi_3) {\rm Im} (\Xi_2 \bar\Xi_3) }{[\mathcal{A J} \tau_2 + \nu {\rm Re}(\Xi_1 \bar\Xi_3 )][\nu \tau_2 \vert \mathcal{C} \Xi_3 - \mathcal{J} \Xi_2  \vert^2 + [\nu 
{\rm Im}(\Xi_2 \bar\Xi_3)]^2 ]^{1/2}}. \eea

All this can be written compactly as
\bea \label{tsl3}
\nu' &=& \frac{1}{\nu^2\tau_2^3} (\xi_{22} \xi_{33} - \xi^2_{23})^{3/2}, \non  \\\tau' &=& \frac{\xi_{23} + i \sqrt{\xi_{22} \xi_{33} - \xi^2_{23}}}{\xi_{33}}, \non \\ B' &=& 
\frac{\xi_{13} + i (\xi_{12} \xi_{33} - \xi_{13} \xi_{23})/\sqrt{\xi_{22} \xi_{33} - \xi^2_{23}} }{\xi_{33}},\eea
where
\bea \xi_{12} &=& \mathcal{AC} \tau_2 + \nu {\rm Re} (\Xi_1 \bar\Xi_2), \non \\ \xi_{13} &=& \mathcal{AJ} \tau_2 
+ \nu {\rm Re} (\Xi_1 \bar\Xi_3), \non \\ \xi_{23} &=& \mathcal{CJ} \tau_2 + \nu {\rm Re} (\Xi_2 \bar\Xi_3),
\non \\ \xi_{22} &=& {\mathcal{C}}^2 \tau_2 + \nu \vert \Xi_2 \vert^2  , \non \\ \xi_{33} &=& 
{\mathcal{J}}^2 \tau_2 + \nu \vert \Xi_3 \vert^2 . \eea

One can also write down the expressions for the transformed moduli in terms of the coordinates $e^{2\hat\phi}, T$
and $\xi$ on moduli space. This gives 
\bea \label{trans2}
e^{2\hat\phi'} &=& e^{2\hat\phi} \eta_{33}^{3/2}, \non \\ T' &=& \frac{(\eta_{12} \eta_{33} - \eta_{13} \eta_{23}) 
+i \sqrt{\eta_{33}} e^{-2\hat\phi}}{\eta_{22} \eta_{33} - \eta_{23}^2} , \non \\ \xi' &=& \frac{(\eta_{12} \eta_{23} 
- \eta_{13}\eta_{22}) +i\eta_{23} e^{-2\hat\phi}/\sqrt{\eta_{33}}}{\eta_{22} \eta_{33} - \eta^2_{23}},\eea
where
\bea \eta_{12} &=& \frac{e^{-2\hat\phi} }{T_2} {\rm Re} (\mathcal{A} T+ \mathcal{B})(\mathcal{C} \bar{T} + \mathcal{D}) 
+  \Big( \mathcal{A} \frac{{\rm Im} (\xi \bar{T})}{T_2} + \mathcal{B} \frac{\xi_2}{T_2} 
+ \mathcal{H} \Big) \Big( \mathcal{C} \frac{{\rm Im} (\xi \bar{T})}{T_2} + \mathcal{D} \frac{\xi_2}{T_2} 
+ \mathcal{E} \Big), \non \\ \eta_{13} &=&\frac{e^{-2\hat\phi} }{T_2} {\rm Re} (\mathcal{A} T+ \mathcal{B})
(\mathcal{J} \bar{T} + \mathcal{F}) 
+  \Big( \mathcal{A} \frac{{\rm Im} (\xi \bar{T})}{T_2} + \mathcal{B} \frac{\xi_2}{T_2} 
+ \mathcal{H} \Big) \Big( \mathcal{J} \frac{{\rm Im} (\xi \bar{T})}{T_2} + \mathcal{F} \frac{\xi_2}{T_2} 
+ \mathcal{G} \Big), \non \\  \eta_{23} &=&\frac{e^{-2\hat\phi} }{T_2} {\rm Re} (\mathcal{C} T+ \mathcal{D})
(\mathcal{J} \bar{T} + \mathcal{F}) 
+  \Big( \mathcal{C} \frac{{\rm Im} (\xi \bar{T})}{T_2} + \mathcal{D} \frac{\xi_2}{T_2} 
+ \mathcal{E} \Big) \Big( \mathcal{J} \frac{{\rm Im} (\xi \bar{T})}{T_2} + \mathcal{F} \frac{\xi_2}{T_2} 
+ \mathcal{G} \Big), \non \\  
\eta_{22} &=& \frac{e^{-2\hat\phi} }{T_2} \vert \mathcal{C} T 
+ \mathcal{D} \vert^2 + \Big( \mathcal{C} \frac{{\rm Im} (\xi \bar{T})}{T_2} + \mathcal{D} \frac{\xi_2}{T_2} 
+ \mathcal{E} \Big)^2  \non \\ \eta_{33} &=& \frac{e^{-2\hat\phi} }{T_2} \vert \mathcal{J} T 
+ \mathcal{F} \vert^2 + \Big( \mathcal{J} \frac{{\rm Im} (\xi \bar{T})}{T_2} + \mathcal{F} \frac{\xi_2}{T_2} 
+ \mathcal{G} \Big)^2 . \eea

It is easy to check that the transformations \C{tsl3} and \C{trans2} 
reproduce \C{smallt} and \C{smallT}
in the infinitesimal limit.

This continuous symmetry of supergravity is broken in string theory by non--perturbative effects. The full theory
has a discrete U--duality symmetry $SL(2,\mathbb{Z}) \times SL(3,\mathbb{Z})$. Under this symmetry, the various fields
continue to transform as above, the only difference being that \C{sl2tt} and \C{sl3tt} have integer entries. 

\section{A class of interactions in the higher derivative action from on--shell linearized superspace}

In order to construct higher derivative corrections to supergravity in the low energy effective
action, consider $N=2, d=8$ superspace with superderivatives given by
\be D_L^A = \frac{\p}{\p \theta_{LA}} + i \bar\theta_L^A \bar\gamma^\m
\p_\mu , \quad D_R^A = - \frac{\p}{\p \theta_{RA}} +i \bar\theta_R^A
\gamma^\mu \p_\mu . \ee
The superspace fermionic coordinates $\theta_L$ and $\theta_R$ carry $U(1)$
charges $1/2$ and $-1/2$ respectively, and are in the $2$ of $SU(2)$.
The degrees of freedom of the supergravity multiplet are contained in linearized superfields
of this superspace. In particular, they are
contained in a chiral superfield $W$ and a linear superfield
$L_{ABCD}$~\cite{Siegel:1995px,Berkovits:1997pj}.

\subsection{Chiral and linear superfields}
 
The chiral superfield $W$ satisfies
\be \label{defc} D_{RA} W =0, \quad D_{LA} \bar{W} =0.\ee
It carries charge 2 under $U(1)$ and is uncharged under $SU(2)$.
The linear superfield $L_{ABCD}$ which satisfies the reality condition 
$L^{ABCD} \equiv (L_{ABCD})^*$ is totally symmetric in its
$SU(2)$ indices, and satisfies
\be \label{defl}  D_{L(A} L_{BCDE )}=0 , \quad D_{R (A}
L_{BCDE)} =0. \ee
Thus, $L_{ABCD}$ is in the $5$ of $SU(2)$, and is uncharged under $U(1)$. 

Since we are dealing with a theory with maximal supersymmetry, it is expected that all 
the fields will be part of a single supermultiplet. This is indeed the case, 
because the chiral and linear superfields are not independent, and satisfy the on--shell relations
\be D_{LA} \gamma^{\mu\nu} D_{LB} W = D_R^C \bar\gamma^{\mu\nu} D_R^D
L_{ABCD}, \quad D_{RA} \bar\gamma^{\mu\nu} D_{RB} \bar{W} = D_L^C \gamma^{\mu\nu} D_L^D
L_{ABCD}. \ee

We now write down some of the components of the chiral and linear superfields at low
orders in $\theta_L, \theta_R$ (ignoring various
numerical factors), which will be relevant for our
purposes. The lowest component of the chiral superfield $W$ is given by
\be  \epsilon_{UV} L^{~U}_{+} \delta L^{~V}_{+} \ee
which in the gauge \C{gfu}, yields
\be 
 \epsilon_{UV} L^{~U}_{+} \delta L^{~V}_{+} = \frac{\delta U}{2 U_2}. \ee
Thus we have that 
\bea \label{contR}
W &=& \frac{\delta U}{U_2} + \bar\theta_R \lambda_L + \Big[ (\bar\theta_R \s^i \gamma^{\mu\nu}
\theta_L)\hat{F}^i_{2\mu\nu} + (\bar\theta_R \g^{\mu\nu\lambda\rho} \theta_L)
\hat{T}^-_{\mu\nu\lambda\rho} \Big] \non \\ && + (\bar\theta_R \s^i \gamma^{\mu\nu}
\theta_L) (\bar\theta_R \s^i \p_\mu \psi_{\nu L}) + (\bar\theta_R \s^i
\gamma^{\mu\nu} \theta_L )(\bar\theta_R \s^i \gamma^{\lambda\rho} \theta_L)
R_{\mu\nu\lambda\rho} + \ldots .\eea

In order to avoid proliferation of indices while writing down the degrees of freedom in the
linear superfield $L_{ABCD}$, it is convenient for our purposes to define 
the real superfield $L_{ij}$ by
\be L_{ij} = \Big[ (\s_{(i})^{AB} (\s_{j)})^{CD} - \frac{1}{3} \delta_{ij} (\s_k)^{AB} (\s_k)^{CD}
\Big] L_{ABCD} + {\rm h.c.}.\ee 
Thus $L_{ij}$ is in the $5$ of $SU(2)$, and is uncharged under $U(1)$. 
The lowest component of $L_{ij}$ is given by
\be l_{ij} = \frac{1}{2} \Big[ L^{~m}_{i} \delta L_{mj} + L^{~m}_{j} \delta L_{mi} - \frac{2}{3} \delta_{ij} L^{~m}_{k}
\delta L_{mk} \Big], \ee
leading to
\bea \label{compexp}
L_{ij} &=& l_{ij} + [\bar\theta_L \s_{(i} \chi_{j)R} - \bar\theta_R \s_{(i} \chi_{j)L}] 
+ i \epsilon_{kl(i} \hat{P}^\mu_{j)k} (\bar\theta_L \bar\g_\mu \s_l \theta_L) + \Big[ (\bar\theta_L \s_{(i} \bar\g_{\mu\nu\lambda}
\theta_L) \hat{F}_{j)3}^{\mu\nu\lambda} \non \\ && +
i(\bar\theta_R \s_{(i} 
\gamma^{\mu\nu} \theta_L) \hat{F}^*_{j)2\mu\nu}  + i(\bar\theta_L \s_{(i} \bar\gamma^{\mu\nu} \theta_R) 
\hat{F}_{j)2\mu\nu} \non \\ &&+ i(\bar\theta_R \s_{(i} \gamma^{\mu\nu} \theta_L)  (\bar\theta_L \s_{j)} \p_\mu \psi_{\nu R})
-i (\bar\theta_L \s_{(i} \bar\gamma^{\mu\nu} \theta_R )(\bar\theta_R \s_{j)} \p_\mu \psi_{\nu L})
\non \\ && + (\bar\theta_R \gamma^{\mu\nu} \s_{(i} \theta_L) (\bar\theta_L \s_{j)} \bar\gamma^{\lambda\rho} \theta_R ) 
R_{\mu\nu\lambda\rho} -{\rm trace} \Big] +\ldots .\eea

To write down explicit expressions for $l_{ij}$ in a fixed gauge, we choose coordinates $T, \xi$ and $e^{-2\hat\phi}$ on the 
$SO(3)/SL(3,\mathbb{R})$ moduli space. This is because the couplings when expanded at weak string
coupling, have perturbative
contributions which are functions of only $T$ and $\bar{T}$. 
Thus $\xi$ appears only in the non--perturbative part of the various couplings. Working
in the gauge \C{SL3v2}, and using
\be  L^{~m}_{i} = \begin{pmatrix} e^{\hat\phi/3} / \sqrt{T_2} & 
e^{\hat\phi/3} T_1/ \sqrt{T_2}    & e^{\hat\phi/3} \xi_1 / \sqrt{T_2}  \\ 
 0 & e^{\hat\phi/3} \sqrt{T_2}  
& e^{\hat\phi/3} \xi_2 /  \sqrt{T_2} \\ 0 &  0 & e^{-2\hat\phi/3} \end{pmatrix} ,\ee
we see that the 5 independent components of $l_{ij}$ are given by
\bea l_{11} &=& \frac{\delta T_2}{2 T_2} - \frac{\delta \hat\phi}{3} , \non \\ l_{22} &=& 
-\frac{\delta T_2}{2 T_2} - \frac{\delta \hat\phi}{3}, \non \\ l_{12} &=&
-\frac{\delta T_1}{2 T_2} , \non \\
l_{13} &=& \frac{e^{\hat\phi}}{2 T_2^{3/2}} (\xi_2 \delta T_1 - T_2 \delta \xi_1) , \non \\ 
l_{23} &=& - e^{\hat\phi} \frac{\sqrt{T_2}}{2} \delta \Big( \frac{\xi_2}{T_2} \Big) . \eea

In order to construct a class of terms in the higher derivative 
effective action, we need to construct superspace actions using the chiral and linear superfields.  

\subsection{Superactions}

First let us construct terms in the effective action involving the chiral superfield $W$. It is given by
\be \label{actionW}
\int d^8 x e \int d^{16} \bar\theta_R f (W) +{\rm h.c.},\ee
where
\bea  \label{measureW}
d^{16} \bar\theta_R &=& \epsilon^{\alpha_1 \ldots \alpha_8} \epsilon^{\beta_1 \ldots \beta_8}  
(d\bar\theta_{R \alpha_1}^{A_1} \ldots d\bar\theta_{R\alpha_8}^{A_8}) 
(d\bar\theta_{R\beta_1 A_1} \ldots d\bar\theta_{R\beta_8  A_8} ) .\eea
Using the Schouten identity
\be \epsilon^{[\alpha_1 \ldots \alpha_8} \epsilon^{\beta_1 ] \ldots \beta_8} =0, \ee
it follows that \C{measureW} is proportional to $(D_{+L})^8 (D_{-L})^8$, where $(D_{+L})^8$ and
$(D_{-L})^8$ are both spacetime scalars. Thus \C{actionW} is invariant under supersymmetry 
transformations\footnote{All such discussions are true upto a total derivative. The $\pm$ $SU(2)$ indices
are defined by
\be V_\pm = V_1 \pm i V_2 . \ee} using
$\delta W \sim D_{\pm L} W$. 

In order to construct terms in the effective action using the linear superfield $L_{ABCD}$, 
let us consider the combination 
\be \widetilde{L} = L_{++++} + 4 \zeta L_{+++-} + 6 \zeta^2 L_{++--} + 4 \zeta^3 L_{+---}
+ \zeta^4 L_{----} ,\ee
where $\zeta$ is a complex parameter.
Now from \C{defl}, we get that 
\bea && D_{+L} L_{++++} =0, \non \\ 
&& D_{-L} L_{++++} + 4 D_{+L} L_{+++-} = 0, \non \\ && 2 D_{-L} L_{+++-} + 3 D_{+L} L_{++--} = 0
, \non \\ && 2 D_{+L} L_{+---} + 3 D_{-L} L_{++--} = 0 , \non \\
&& D_{+L} L_{----} + 4 D_{-L} L_{+---} = 0, \non \\ && D_{-L} L_{----} = 0 , \eea
and similarly for $D_R$, leading to
\bea \label{specrel} D_{+L} g (\widetilde{L}) = -\zeta D_{-L} g(\widetilde{L}) , \non \\ 
D_{+R} g (\widetilde{L}) = -\zeta D_{-R} g(\widetilde{L}) . \eea
This ability to interchange $D_{\pm}$ when acting on $g(\widetilde{L})$ is useful to write down a superspace action
involving the linear superfield. 

Such a superspace action is given by~\cite{Berkovits:1997pj}
\be \label{actionL} \int d^8 x  e \int d^8 \bar\theta_R d^8 \bar\theta_L  \Big[ \oint_0 d \zeta 
g (\widetilde{L},\zeta) + {\rm h.c.} \Big] , \ee 
where 
\bea  \label{measureL} d^8 \bar\theta_R d^8 \bar\theta_L &=& 
\epsilon^{\alpha_1 \ldots \alpha_8} \epsilon^{\beta_1 \ldots \beta_8} (d\bar\theta_{R\alpha_1}^{A_1} \ldots 
d\bar\theta_{R\alpha_8}^{A_8}) (d\bar\theta_{L\beta_1 A_1} \ldots d\bar\theta_{L\beta_8 A_8} ), \eea
and the contour integral in the complex $\zeta$ plane is around the origin. In order to to show that \C{actionL} is
invariant under supersymmetry, 
we note that \C{measureL} gives 8 powers of $D_L$ and 8 powers of $D_R$, while a supervariation of 
$\widetilde{L}$ yields one more factor of $D_L$ and $D_R$. However, using \C{specrel}, it follows that $D_{+L}$
and $D_{-L}$ (and also $D_{+R}$ and $D_{-R}$) 
can be interchanged, and so finally we end up with 9 powers of the same superderivative which vanishes,
and thus the action is invariant. Thus using the linear superfield, we get the action
\be \int d^8 x  e \int d^8 \bar\theta_R d^8 \bar\theta_L  \Big[ g (L_{ABCD} ) + {\rm h.c.}\Big] ,\ee
where
\be g (L_{ABCD}) = \oint_0 d \zeta 
g (\widetilde{L},\zeta) . \ee

We finally redefine
\be g (L_{ij}) \equiv g (L_{ABCD}) + {\rm h.c.} ,\ee
giving us the superspace action 
\be \label{intlin}
\int d^8 x e \int d^8 \bar\theta_R d^8 \bar\theta_L  g (L_{ij} )  \ee
involving the linear superfield.

We now consider higher derivative corrections to supergravity coming from the superspace Lagrangian
\be \label{totact}
e^{-1} \mathcal{L} = \Big[ \int d^{16} \bar\theta_R f (W) +{\rm h.c.} \Big] 
+ \int d^8 \bar\theta_R d^8 \bar\theta_L  g (L_{ij} ) .\ee

\section{Higher derivative corrections and supersymmetry constraints} 

We next consider the role of supersymmetry in constraining the various
couplings which arise as the coefficients involving the moduli of the various
interactions in the effective action. We first consider a set of couplings which
involve only the $SO(2) \backslash  SL(2,\mathbb{R})$ moduli, and then the
ones which involve only the $SO(3) \backslash SL(3,\mathbb{R})$ moduli. Finally
we consider a coupling which involves all the moduli.

\subsection{Couplings involving only the  $SO(2) \backslash  SL(2,\mathbb{R})$ moduli}

The set of couplings we shall consider are the ones obtained from linearized superspace. 
Thus let us consider interactions in \C{totact}, involving the chiral
superfield. We shall see that these couplings
are automorphic forms of $SL(2,\mathbb{Z})$. In order to construct these interactions,   
we shall make use of the definitions
\bea \label{manylambda} \lambda_L^{16} &\equiv& \frac{1}{8! 9!}
\epsilon^{\alpha_1 \ldots \alpha_8} \epsilon^{\beta_1 \ldots \beta_8} 
\epsilon_{A_1 B_1} \ldots \epsilon_{A_8 B_8} ( 
\lambda_{L \alpha_1}^{A_1} \ldots \lambda_{L \alpha_8}^{A_8} ) ( \lambda_{L \beta_1}^{B_1} 
\ldots \lambda_{L \beta_8}^{B_8} ) \non \\ &=& (\lambda_{L1}^1 \lambda_{L1}^2) \ldots (\lambda_{L8}^1 
\lambda_{L8}^2) , 
\non \\ ( \lambda_L^{15} )^{\beta_8}_{B_8} &\equiv& \frac{2!}{7!9!} \epsilon^{\alpha_1 \ldots \alpha_8} 
\epsilon^{\beta_1 \ldots \beta_8} \epsilon_{A_1 B_1} \ldots \epsilon_{A_8 B_8} (\lambda_{L \alpha_1}^{A_1} \ldots 
\lambda_{L \alpha_8}^{A_8} ) (\lambda_{L \beta_1}^{B_1} \ldots \lambda_{L \beta_7}^{B_7} ) , \non \\
(\lambda_L^{14})^{\beta_7 \beta_8}_{B_7 B_8} &\equiv& \frac{3!}{6!9!} \epsilon^{\alpha_1 \ldots \alpha_8} \epsilon^{\beta_1 \ldots 
\beta_8} \epsilon_{A_1 B_1} \ldots \epsilon_{A_8 B_8} (\lambda_{L \alpha_1}^{A_1} \ldots
\lambda_{L \alpha_8}^{A_8}) (\lambda_{L\beta_1}^{B_1} \ldots \lambda_{L \beta_6}^{B_6} ),\eea

such that
\bea \label{moredef}
(\lambda_L^{15} )^\alpha_A \lambda^B_{L\beta} &=& \delta_A^B \delta^\alpha_\beta \lambda_L^{16} , \non \\
(\lambda_L^{14})^{\alpha\beta}_{AB} \lambda^C_{L\gamma} &=& (\lambda_L^{15})^\beta_B \delta^\alpha_\gamma \delta^C_A 
- (\lambda_L^{15})^\alpha_B \delta^\beta_\gamma \delta^C_A + (\lambda_L^{15})^\beta_A \delta^\alpha_\gamma \delta^C_B 
-(\lambda_L^{15})^\alpha_A \delta^\beta_\gamma \delta^C_B ,
\eea

leading to
\bea (\lambda_L^{15} )^\alpha_A \lambda^A_{L\alpha} &=& 16 \lambda_L^{16} , \non \\ 
(\lambda_L^{14})^{\alpha\beta}_{AB} \lambda^A_{L\alpha} &=& 21 (\lambda_L^{15})^\beta_B , \non \\
(\lambda_L^{14})^{\alpha\beta}_{AB} \lambda^C_{L\gamma} \lambda^D_{L\delta} &=& \lambda_L^{16} (\delta^\alpha_\gamma
\delta^\beta_\delta - \delta^\beta_\gamma \delta^\alpha_\delta ) (\delta^C_A \delta^D_B + \delta^C_B \delta^D_A ). \eea

Note that the interactions in \C{manylambda} are particular examples of the general form
\be (\lambda_L^{8+r})^{\beta_{r+1} \ldots \beta_8}_{B_{r+1} \ldots B_8} \equiv \frac{\Gamma (10-r)}{\Gamma(r+1) \Gamma(10)}
\epsilon^{\alpha_1 \ldots \alpha_8} \epsilon^{\beta_1 \ldots \beta_8} \epsilon_{A_1 B_1} \ldots \epsilon_{A_8 B_8} ( 
\lambda_{L \alpha_1}^{A_1} \ldots \lambda_{L \alpha_8}^{A_8} ) ( \lambda_{L \beta_1}^{B_1} 
\ldots \lambda_{L \beta_r}^{B_r} ) ,\ee
for $0 \leq r \leq 8$, and are the only ones we need.

We now consider interactions involving sixteen fermions in $S^{(3)}$. In particular, we consider interactions of the form
$\lambda_L^{16}$ and $\bar\psi_{L\mu} \bar\g^\mu \lambda_L^{15} $. 
These interactions mix with no other interactions in $S^{(3)}$ under
the supersymmetry transformations $\delta^{(0)}$ of the type discussed below. To consider these terms in the effective action, 
we take  a subset of terms in \C{actionW} given by   
\be S^{(3)} = \int d^8 x e \Big[ f^{(12,-12)} (U,\bar{U}) \lambda_L^{16} + f^{(11,-11)}(U,\bar{U}) \hat{F} \lambda_L^{14}
  \Big] + \ldots,\ee
where\footnote{Note that the contribution of the type
\be \hat{T}^- \lambda_L^{14} \equiv  \hat{T}^-_{\mu\nu\lambda\rho} 
(\g^{\mu\nu\lambda\rho})_{\alpha\beta} \epsilon^{AB} (\lambda_L^{14})^{\alpha\beta}_{AB} \ee
vanishes because 
\be \epsilon^{AB} (\lambda_L^{14})^{\alpha\beta}_{AB} =0.\ee}
\be \hat{F} \lambda_L^{14} \equiv \hat{F}^i_{2\mu\nu} (\s^i \g^{\mu\nu}  )^{AB}_{\alpha \beta} 
(\lambda_L^{14})^{\alpha\beta}_{AB} .\ee 

This leads to
\be \label{impact}
\mathcal{L}^{(3)} = e  \Big[ f^{(12,-12)} (U,\bar{U}) \lambda_L^{16} + f^{(11,-11)} (U,\bar{U})
\bar\psi_{\mu L} \bar\g^\mu \lambda_L^{15}  + \ldots \Big], \ee
where we have used $\hat{F}_2 \sim \bar\psi_L \lambda_L$ from \C{manysupcov} in the expression 
for $\hat{F} \lambda_L^{14}$. We have also rescaled
$f^{(11,-11)}$, and used the identity
\be \bar\g_\nu \g^{\mu\nu} = -7 \bar\g^\mu 
. \ee 

Now consider the supervariation under $\delta^{(0)}$ of \C{impact} into terms of the form 
$\lambda_L^{16} \epsilon_L \psi_R$. Thus
\bea \label{varact1}
\delta^{(0)} \mathcal{L}^{(3)} &=& (\delta^{(0)} e) f^{(12,-12)}\lambda_L^{16} +e \Big[
f^{(12,-12)} \delta^{(0)} \lambda_L^{16} \non \\ &&
+ (\delta^{(0)} f^{(11,-11)}) \bar\psi_{\mu L} \bar\g^\mu \lambda_L^{15}
+ f^{(11,-11)} \delta^{(0)} (\bar\psi_{\mu L} \bar\g^\mu \lambda_L^{15})\Big] + \ldots .\eea
These supervariations can be evaluated using the supersymmetry transformations \C{susytran}\footnote{The
$U(1)$ violating terms due to gauge fixing also have to be added, as discussed before.} leading to
\bea \label{suvar1}
(\delta^{(0)} e) f^{(12,-12)}\lambda_L^{16} &=& - e f^{(12,-12)} \lambda_L^{16}
(\bar\epsilon_R \g^\mu \psi_{\mu R}), \non \\ (\delta^{(0)} f^{(11,-11)}) \bar\psi_{\mu L} 
\bar\g^\mu \lambda_L^{15} &=&  -2  U_2 \frac{\p f^{(11,-11)}}{\p U} \lambda_L^{16}
(\bar\epsilon_R \g^\mu \psi_{\mu R}), \non \\ f^{(11,-11)} \delta^{(0)} (\bar\psi_{\mu L} 
\bar\g^\mu \lambda_L^{15}) &=& 11i f^{(11,-11)} \lambda_L^{16} (\bar\epsilon_R \g^\mu \psi_{\mu R}) ,\non \\
f^{(12,-12)} \delta^{(0)} \lambda_L^{16} &=& \Big( \frac{21}{4} + \frac{35}{8}\Big) f^{(12,-12)} 
\lambda_L^{16} (\bar\epsilon_R \g^\mu \psi_{\mu R}). \eea
The two contributions to the last equation in \C{suvar1} come from $\delta^{(0)} \lambda_L \sim \hat{F}_2
\epsilon_L$ and $\delta^{(0)} \lambda_L \sim \hat{T}^- \epsilon_L$ respectively.  
Thus, \C{varact1} gives us
\be \label{noadd}
\delta^{(0)} \mathcal{L}^{(3)} = e \Big[ 2i D_{11} f^{(11,-11)} + \frac{69}{8} f^{(12,-12)}
\Big] \lambda_L^{16} (\bar\epsilon_R \g^\mu \psi_{\mu R}),\ee
where $D_{11}$ is given by \C{covderSL2}.

Let us consider possible supervariations $\delta^{(3)}$ which acting on terms in $\mathcal{L}^{(0)}$, the supergravity action,
might also contribute to $\lambda_L^{16} \epsilon_L \psi_R$. 
Based on the $U(1)$ invariance of $\mathcal{L}^{(0)}$, it is easy to
see that there can be no
such terms in $\mathcal{L}^{(0)}$. Thus \C{noadd} does not receive any
more contributions and we get that
\be \label{1eqn}D_{11} f^{(11,-11)} = \frac{69i}{16} f^{(12,-12)}.\ee 

Next consider the supervariation under $\delta^{(0)}$ of \C{impact} into terms of the form 
$\lambda_L^{16} \epsilon_R \lambda_R$. This gives us 
\bea \label{varact2}
\delta^{(0)} \mathcal{L}^{(3)} = e \Big[ (\delta^{(0)} f^{(12,-12)} ) \lambda_L^{16} 
+ f^{(12,-12)} \delta^{(0)} \lambda_L^{16}
+ f^{(11,-11)} (\delta^{(0)} \bar\psi_{\mu L}) \bar\g^\mu \lambda_L^{15}\Big] + \ldots .\eea
Again using \C{susytran}, we get
\bea \label{suvar2} (\delta^{(0)} f^{(12,-12)} ) \lambda_L^{16} &=& 2i U_2 \frac{\p f^{(11,-11)}}{\p \bar{U}} 
\lambda_L^{16} (\bar\epsilon_L \lambda_R) , \non \\ f^{(12,-12)} \delta^{(0)} \lambda_L^{16} &=& 12 f^{(12,-12)} 
\lambda_L^{16} (\bar\epsilon_L \lambda_R) , \non \\  f^{(11,-11)}
(\delta^{(0)} \bar\psi_{\mu L}) \bar\g^\mu \lambda_L^{15} &=& 14i f^{(11,-11)} 
\lambda_L^{16} (\bar\epsilon_L \lambda_R). \eea

The last equation in \C{suvar2} involves many contributions coming from \C{susytran}. It can be deduced
using the identities
\bea \frac{i}{27} (\bar\g_\mu \s^i \lambda_L)^A_\alpha (\bar\g^\mu)^{\alpha\beta} (\lambda_L^{15})_{A\beta}
(\bar\epsilon_L \s^i \lambda_R ) &=& 0 , \non \\ -\frac{11i}{54} ( \s^i \lambda_R)^A_\alpha 
(\bar\g^\mu)^{\alpha\beta} (\lambda_L^{15})_{A\beta} (\bar\epsilon_L \s^i \bar\g_\mu \lambda_L ) &=& 
\frac{44i}{9} \lambda_L^{16} (\bar\epsilon_L \lambda_R) , \non \\ \frac{i}{54} ( \bar\g_{\mu\nu} \s^i 
\lambda_R)^A_\alpha (\bar\g^\mu)^{\alpha\beta} (\lambda_L^{15})_{A\beta} (\bar\epsilon_L \s^i \bar\g^\nu 
\lambda_L ) &=& -\frac{28i}{9} \lambda_L^{16} (\bar\epsilon_L \lambda_R) ,\non \\ 
\frac{5i}{54} ( \s^i \epsilon_R)^A_\alpha (\bar\g^\mu)^{\alpha\beta} (\lambda_L^{15})_{A\beta} (\bar\lambda_L 
\s^i \bar\g_\mu \lambda_L ) &=& -\frac{20i}{9} \lambda_L^{16} (\bar\epsilon_L \lambda_R) , \non \\
-\frac{i}{54} ( \bar\g_{\mu\nu} \s^i \epsilon_R)^A_\alpha (\bar\g^\mu)^{\alpha\beta} (\lambda_L^{15})_{A\beta} 
(\bar\lambda_L \s^i \bar\g^\nu \lambda_L ) &=& \frac{28i}{9} \lambda_L^{16} (\bar\epsilon_L \lambda_R) ,
\non \\ -\frac{i}{12} \epsilon_{R\alpha}^A (\bar\g^\mu)^{\alpha\beta} (\lambda_L^{15})_{A\beta} 
(\bar\lambda_L \bar\g_\mu \lambda_L ) &=& \frac{2i}{3} \lambda_L^{16} (\bar\epsilon_L \lambda_R) ,
\non \\ \frac{i}{144} (\bar\g^{\nu\rho}\epsilon_R)_\alpha^A (\bar\g^\mu)^{\alpha\beta} (\lambda_L^{15})_{A\beta} 
(\bar\lambda_L \bar\g_{\mu\nu\rho} \lambda_L ) &=& \frac{7i}{3} \lambda_L^{16} (\bar\epsilon_L \lambda_R) , 
\non \\
\frac{i}{12} (\bar\g^\mu \lambda_L )_\alpha^A (\bar\g^\mu)^{\alpha\beta} (\lambda_L^{15})_{A\beta} 
(\bar\epsilon_L \lambda_R ) &=& \frac{32i}{3} \lambda_L^{16} (\bar\epsilon_L \lambda_R) , \non \\ 
-\frac{i}{36} (\bar\g^\nu \lambda_L)_\alpha^A (\bar\g^\mu)^{\alpha\beta} (\lambda_L^{15})_{A\beta} 
(\bar\epsilon_L \bar\g_{\mu\nu} \lambda_R ) &=& 0, \non \eea
\bea -\frac{i}{432} (\s^i \bar\g^{\mu\nu\lambda\rho}
\epsilon_R)_\alpha^A (\bar\g_\mu)^{\alpha\beta} (\lambda_L^{15})_{A\beta} (\bar\lambda_R
\g_{\nu\lambda\rho} \s^i \lambda_R ) &=& \frac{35i}{3} \lambda_L^{16} (\bar\epsilon_L \lambda_R),\non \\ 
-\frac{i}{72} (\s^i \bar\g_{\lambda\rho} \epsilon_R)_\alpha^A (\bar\g_\mu)^{\alpha\beta} 
(\lambda_L^{15})_{A\beta} (\bar\lambda_L \bar\g^{\mu\lambda\rho} \s^i \lambda_L ) &=& -14i \lambda_L^{16} 
(\bar\epsilon_L \lambda_R) .\eea

Thus, \C{varact2} gives
\be \label{2eqn} \delta^{(0)} \mathcal{L}^{(3)} = e \Big[ -2 \bar{D}_{-12} f^{(12,-12)}
+14i f^{(11,-11)}\Big] \lambda_L^{16} (\bar\epsilon_L \lambda_R) .\ee

One might think there can be a term of the form $\lambda_L^{15} \chi_L$ which
might contribute for $\delta^{(0)} \chi_L \sim \epsilon_R \lambda_L
\lambda_R$. Such a term, which does not follow from linearized superspace,
would have to be of the form
\be (\lambda^{14}_L )^{\alpha\beta}_{AB} (\s_i \lambda_L)^A_\alpha \chi^{iB}_{L\beta}\ee
which vanishes using \C{moredef}.

Once again, we consider modified supersymmetry transformations $\delta^{(3)}$ which acting on terms in $\mathcal{L}^{(0)}$
might contribute to $\lambda_L^{16} \epsilon_R \lambda_R$. The only
possibility is a term of the form $\lambda_L^2 \lambda_R^2$ in $\mathcal{L}^{(0)}$.
This term in the supergravity action is given by
\be \label{needalso} \mathcal{L}^{(0)} = \frac{1}{96}  e
\Big[ 30 (\bar\lambda_R \g^\mu \lambda_R) (\bar\lambda_R \g_\mu \lambda_R)
+ (\bar\lambda_R \g^{\mu\nu\rho} \lambda_R)(\bar\lambda_R \g_{\mu\nu\rho} \lambda_R)\Big]\ee
as deduced in \C{lambda4}. Now \C{needalso} can vary into $\lambda_L^{16}
\epsilon_R \lambda_R$ for $\delta^{(3)} \lambda_R \sim \lambda_L^{14} \epsilon_R$.
In general, it is difficult to write down complete expressions for the
corrected supersymmetry transformations $\delta^{(3)}$ for any field. For the
case we need, let us consider the supervariation given
by\footnote{A contribution of the type
\be g_4 (U,\bar{U})(\lambda_L^{14})^{\beta\gamma}_{AB} (\g_{\mu\nu\lambda\rho})_{\beta\gamma}
  (\bar\g^{\mu\nu\lambda\rho} \epsilon_R)^B_\alpha\ee
vanishes because $\g_{\mu\nu\lambda\rho}^T = \g_{\mu\nu\lambda\rho}$.}
\bea \label{modsusyvar}
\delta^{(3)} \lambda_{R\alpha A} &=& (\lambda_L^{14})^{\beta\gamma}_{AB}
\Big[ g_1 (U,\bar{U})(\g_\mu)_{\g\alpha} (\g^\mu \epsilon_R)^B_\beta + g_2 
(U,\bar{U})(\g_{\mu\nu\rho})_{\g\alpha} (\g^{\mu\nu\rho}
  \epsilon_R)^B_\beta \non \\ && + g_3 (U,\bar{U}) (\g_{\mu\nu})_{\beta\gamma}
  (\bar\g^{\mu\nu} \epsilon_R)^B_\alpha\Big].\eea
Though \C{modsusyvar} looks complicated, it is the simplest set of terms that
one can try based on the symmetries. Of course, we shall not consider the set
of all possible supervariations due to its complexity, but we shall restrict
ourselves to showing how this set of terms is good enough to lead to
strong constraints on the structure of the equations based on general
considerations. So from now onwards, our aim will be to obtain the structure of
the equations, and not bother about the coefficients.  
Acting on \C{needalso}, we note that it gives
\be \label{modt} \delta^{(3)} \mathcal{L}^{(0)} = 252 i e g \lambda_L^{16} (\bar\epsilon_L \lambda_R) ,\ee
where
\be \label{lincom} g = g_1 - 6 g_2 + 4 g_3.\ee
It seems difficult to make stronger statements given that there are 3
undetermined moduli dependent coefficients in \C{modsusyvar}. However, we
shall now see that the constraints imposed by the closure of the superalgebra 
prove strong enough to determine what is needed for our purpose.

Because we are dealing with a theory of maximal supersymmetry, there exists no
off--shell formulation of the theory. In fact, the closure of the superalgebra
is only upto the equations of  motion of the various fields, and various local
symmetry transformations. This is also true for $\lambda_R$. Thus, 
\be \label{allterms} [\delta_1, \delta_2] \lambda_R = \Big( [\delta^{(0)}_1, \delta^{(0)}_2]
+ [\delta^{(0)}_1, \delta^{(3)}_2] + [\delta^{(3)}_1, \delta^{(0)}_2] + \ldots
\Big) \lambda_R \ee
closes only upto the equation of motion of $\lambda_R$, and other local
symmetries. We shall use this to our advantage and use the equation of motion
of $\lambda_R$ to constrain $g_1, g_2$ and $g_3$. 

Let us first consider closure at the level of supergravity. From the various
expressions in \C{susytran}, we get that
\bea \label{needshort} 
[\delta^{(0)}_1 , \delta^{(0)}_2] \lambda_R &=& -\bar\g^\mu \epsilon_{L2}
(\bar\epsilon_{L1} \p_\mu \lambda_R) + \frac{1}{4} \bar\g^{\mu\nu}\s^i
\epsilon_{R2}
(\bar\epsilon_{R1} \g_\nu \s^i \p_\mu \lambda_R) - \frac{1}{48}
\bar\g^{\mu\nu\lambda\rho}
\epsilon_{R2} (\bar\epsilon_{R1} \g_{\nu\lambda\rho} \p_\mu \lambda_R) \non \\&&
- (1 \leftrightarrow 2) .\eea
Using the Fierz and Schouten identities repeatedly, we rewrite \C{needshort} as 
\bea  \label{close1} [\delta^{(0)}_1 , \delta^{(0)}_2] \lambda_R &=& (\bar\epsilon_{L1}
\bar\g^\mu \epsilon_{L2}) \p_\mu \lambda_R + \Big[ -\frac{7}{16} (\bar\epsilon_{L1}
\bar\g^\mu \epsilon_{L2}) \bar\g_\mu   +\frac{1}{96}
(\bar\epsilon_{L1}
\bar\g^{\mu\nu\lambda} \epsilon_{L2}) \bar\g_{\mu\nu\lambda} \Big]\slash\p
\lambda_R \non \\ && -\frac{1}{8} \Big[ \epsilon_{R1} (\bar\epsilon_{R2} \slash\p
\lambda_R) + \frac{1}{2} \bar\g^{\mu\nu} \epsilon_{R1} (\bar\epsilon_{R2}
\g_{\mu\nu} \slash\p \lambda_R)\Big] -(1 \leftrightarrow 2).\eea
While the first term on the right hand side of \C{close1} is the standard
derivative term, the remaining terms must vanish, leading to the free equation
of motion for $\lambda_R$. 

Now let us focus on the first corrections to \C{close1} obtained from
\C{allterms} on using only the terms given in \C{modsusyvar}, along
with the ones which yield the $SL(2,\mathbb{Z})$ covariant
derivative. Considering only the $O(\epsilon_{R1} \epsilon_{L2})$ term, we get that
\bea \label{caldet}\Big( [\delta^{(0)}_1, \delta^{(3)}_2] + [\delta^{(3)}_1, \delta^{(0)}_2]\Big)
\lambda_R &=& \Big( 2i U_2 \frac{\p}{\p U} +\frac{45}{4}\Big)g \Big[  -\frac{7}{2} (\bar\epsilon_{L1}
\bar\g^\mu \epsilon_{L2}) \bar\g_\mu  \lambda_L^{15} \non \\ && +\frac{1}{12}
(\bar\epsilon_{L1}
\bar\g^{\mu\nu\lambda} \epsilon_{L2}) \bar\g_{\mu\nu\lambda}
\lambda_L^{15}
- \frac{1}{2} \bar\g^{\mu\nu} \epsilon_{R1} (\bar\epsilon_{R2}
\g_{\mu\nu}  \lambda_L^{15})\Big] \non \\ && = (2 D_{11}g) \Big[  -\frac{7}{2} (\bar\epsilon_{L1}
\bar\g^\mu \epsilon_{L2}) \bar\g_\mu  \lambda_L^{15}  +\frac{1}{12}
(\bar\epsilon_{L1}
\bar\g^{\mu\nu\lambda} \epsilon_{L2}) \bar\g_{\mu\nu\lambda}
\lambda_L^{15}\non \\ && 
- \frac{1}{2} \bar\g^{\mu\nu} \epsilon_{R1} (\bar\epsilon_{R2}
\g_{\mu\nu}  \lambda_L^{15})\Big]+ \frac{1}{4} \delta^{(3)}_{\hat\epsilon}
\lambda_R ,\eea
where $\delta^{(3)}_{\hat\epsilon}$ is the supersymmetry transformation \C{modsusyvar} with
parameter
\be \label{locsusy} \hat\epsilon = (\bar\epsilon_{R2} \lambda_L) \epsilon_{R1} .\ee
The choice \C{locsusy} is uniquely determined once the appropriate
$SL(2,\mathbb{Z})$ weight has been assigned to $g$. 

In \C{caldet}, exactly the same linear combination of $g_1, g_2$ and $g_3$
given by \C{lincom} appears as the one in \C{modt}. Thus the closure of the
superalgebra on $\lambda_R$ is good enough to provide us precisely the information we
need. Thus, upto a local supersymmetry transformation,  considering the terms
of the form $(\bar\epsilon_{L1}
\bar\g^\mu \epsilon_{L2}), (\bar\epsilon_{L1}
\bar\g^{\mu\nu\lambda} \epsilon_{L2})$ and $\epsilon_{R1} (\bar\epsilon_{R2}
\g_{\mu\nu}  \ldots )$ in \C{close1} and \C{caldet}, we get the equation of motion
\bea \label{match1} \slash\p \lambda_R + 16 (D_{11} g) \lambda_L^{15} + \ldots =0,\eea
which we match with the equation of motion obtained from the action \C{action}
and \C{impact}
\be \label{match2} \slash\p \lambda_R -\frac{i}{2} f^{(12,-12)} \lambda_L^{15} =0 .\ee
Note that this cannot be the complete analysis. This is because \C{caldet}
does not contribute to the free equation of motion obtained from the term
$\epsilon_{R1} (\bar\epsilon_{R2}\ldots )$ in \C{close1}. Thus there must be
other supervariations which will also contribute, and which will yield the
final equation we need. Even without worrying about the other possible
contributions, from \C{match1} and \C{match2} we get that
\bea 16 D_{11} g + \ldots = -\frac{i}{2} f^{(12,-12)} ,\eea
where the $\ldots$ denote the other contributions. Based on $SL(2,\mathbb{Z})$
covariance, we get that
\be \label{3eqn}g \sim f^{(11,-11)},\ee
which must be also true of the other contributions. 

Note that there are more constraints that can be obtained from imposing the
closure of the superalgebra acting on $\lambda_R$. We looked at those terms
that involve the $\lambda_L$ equation of motion. There are several other such
terms, for example, the gravitino equation of motion also arises from the same
closure. This leads to very strong constraints on the couplings.   

Now, from \C{1eqn}, \C{2eqn}, \C{modt} and \C{3eqn}, we get that
\be D_{11} f^{(11,-11)} \sim f^{(12,-12)}, \quad \bar{D}_{-12} f^{(12,-12)}
\sim f^{(11,-11)},\ee
leading to
\bea \label{susyfix}4 \bar{D}_{-12} D_{11} f^{(11,-11)} = a f^{(11,-11)}, \non
\\4 D_{11} \bar{D}_{-12} f^{(12,-12)} = a f^{(12,-12)} .\eea
Though we have not determined the coefficient $a$, clearly it can be determined based on
the arguments we have made, and taking into account all the terms. Thus, \C{susyfix} 
is completely fixed by supersymmetry. 

The equations \C{susyfix} have a unique solution on the fundamental domain of
$SL(2,\mathbb{Z})$
given the boundary condition
that the couplings have a power law behavior in $U_2$ for large $U_2$ based on physical
considerations. In fact, the solutions must be given by \C{nzw} in appendix \C{SL2aut}
for $m=11$ and $12$, for some choice of $s$. Thus the value of $a$ is also
determined by the value of $s$.

In order to determine $s$, we simply use data from string perturbation
theory. The $U$ dependent one loop amplitude for the $\mathcal{R}^4$
interaction is known to be given by \C{finexp}. The $\mathcal{R}^4$
interaction is obtained in the effective action from linearized superspace
using \C{contR}. Thus based on the $SL(2,\mathbb{Z})$ covariance of the
various couplings, they must be related to one another by the action of the
$SL(2,\mathbb{Z})$ covariant derivatives \C{covderSL2}. Thus $s=1$, which
fixes
\be a = -121.\ee

Thus, supersymmetry completely determines the moduli dependent couplings of
some of the interactions in the effective action, which we have obtained using linearized
superspace. These interactions were obtained using the chiral superfield, hence the
couplings are independent of the $SO(3)\backslash SL(3,\mathbb{R})$ moduli. In
fact, the couplings which have non--zero weights under $SL(2,\mathbb{Z})$
transformations, are the coefficient functions of interactions charged under
$U(1)$ and so cannot receive contributions from interactions constructed
from the linear superfield, which is neutral under $U(1)$. This will also be
true the other way round when we shall consider couplings which transform
non--trivially under $SL(3,\mathbb{Z})$ transformations, which are coefficient
functions of interactions that carry $SU(2)$
charge. They will depend only on the $SO(3) \backslash SL(3,\mathbb{R})$
moduli. 
 
Thus the only interactions which can have couplings that depend on both the
$U(1)\backslash SL(2,\mathbb{R})$ and $SO(3)\backslash SL(3,\mathbb{R})$
moduli are those that are uncharged under $U(1)$ as well as $SU(2)$. Among the
interactions that follow from linearized superspace, one such interaction is
the $\mathcal{R}^4$ interaction. Thus our above discussion fixes only 
the $U,\bar{U}$ moduli dependence of this coupling.

Note that all the other couplings we have determined have some striking
differences from the $\mathcal{R}^4$ coupling. These couplings receive
contributions only from one loop in string theory, and there are not other
perturbative or non--perturbative contributions. Also, unlike the
$\mathcal{R}^4$ coupling, they do not have an infra--red logarithmic divergence at one
loop, because of the absence of the ${\rm ln} U_2$ term in its expression.

There is a direct relationship between the $U(1)$ charge of a specific
interaction, and the weight of its coupling. The coupling of an interaction
which carries $U(1)$ charge $q$, is an $SL(2,\mathbb{Z})$ automorphic form of weight $(q/2,-q/2)$.

\subsection{Couplings involving only the  $SO(3) \backslash  SL(3,\mathbb{R})$ moduli}

We next consider a set of couplings that involve only the $SO(3)\backslash
SL(3,\mathbb{R})$ moduli.
To begin with, we shall consider a set of 16 fermion interactions arising from the part of the action involving the linear superfield 
in \C{totact}. We shall see, that compared to the discussion above, the analysis is considerably more complicated.

We look at a small subset of interactions in $S^{(3)}$ which
mix with no other interactions under 
supersymmetry transformations $\delta^{(0)}$ of the type we shall
consider. This will lead
to a coupled set of linear differential equations for the various couplings we
consider, which will lead to Poisson equations on moduli
space for the various individual couplings.

As before, the equations obtained from $\delta^{(0)} S^{(3)}$ shall also receive contributions from $\delta^{(3)} S^{(0)}$. However, unlike the above analysis involving 
only the chiral superfield, there are several possible terms in the supergravity action $S^{(0)}$ which can contribute. 
This is because the superaction involving the linear superfield involves the integral $\int d\theta_L^8 d\theta_R^8$ 
which 
is real, and yields those 16 fermion interactions in $S^{(3)}$ such that there are several contributions from 
$\delta^{(3)} S^{(0)}$. Of course, the procedure to calculate them is exactly
the same as above. We shall only constrain the structure of the equations
using supersymmetry, but we shall not fix the various coefficients. In
particular, we shall be very schematic in our discussion of the $\delta^{(3)}
S^{(0)}$ contributions.   
In principle, they can all be fixed using only supersymmetry, but the
calculations get very tedious.        

In order to avoid the large number of indices associated with these interactions, we shall adopt a simple notation.
The $SU(2)$ spin $3/2$ fermions $\chi_L$ and $\chi_R$ in the various interactions shall always arise in the 
combination $\s_{(i} \chi_{j)L}$ and $\s_{(i} \chi_{j)R}$. We shall simply denote them
\be \s_{(i} \chi_{j)L} \equiv \chi_L , \quad \s_{(i} \chi_{j)R} \equiv \chi_R ,\ee
and drop the various $i,j$ indices when there is no scope for confusion.

For these fermions, we define
\bea \chi_L^8 \chi_R^8 &\equiv& \epsilon^{\alpha_1 \ldots \alpha_8} \epsilon^{\beta_1 \ldots \beta_8} \epsilon_{A_1 B_1}
\ldots \epsilon_{A_8 B_8} \Big( (\s_{(i_1} \chi_{j_1 )})_{L\alpha_1}^{A_1} \ldots (\s_{(i_8} \chi_{j_8 )})_{L\alpha_8}^{A_8} \Big)
\non \\ && \times 
\Big( (\s_{(k_1} \chi_{l_1 )})_{R\beta_1}^{B_1} \ldots (\s_{(k_8} \chi_{l_8 )})_{R\beta_8}^{B_8}\Big), \non \\ 
\chi_L^8 (\chi_R^7 )^{\beta_8}_{B_8} &\equiv& \epsilon^{\alpha_1 \ldots \alpha_8} \epsilon^{\beta_1 \ldots \beta_8} \epsilon_{A_1 B_1}
\ldots \epsilon_{A_8 B_8} \Big( (\s_{(i_1} \chi_{j_1 )})_{L\alpha_1}^{A_1} \ldots (\s_{(i_8} \chi_{j_8 )})_{L\alpha_8}^{A_8} \Big)
\non \\ && \times 
\Big( (\s_{(k_1} \chi_{l_1 )})_{R\beta_1}^{B_1} \ldots (\s_{(k_7} \chi_{l_7 )})_{R\beta_7}^{B_7}\Big), \non \\
(\chi_L^7 )^{\alpha_8}_{A_8} \chi_R^8 &\equiv& \epsilon^{\alpha_1 \ldots \alpha_8} \epsilon^{\beta_1 \ldots \beta_8} \epsilon_{A_1 B_1}
\ldots \epsilon_{A_8 B_8} \Big( (\s_{(i_1} \chi_{j_1 )})_{L\alpha_1}^{A_1} \ldots (\s_{(i_7} \chi_{j_7)})_{L\alpha_7}^{A_7} \Big)
\non \\ && \times 
\Big( (\s_{(k_1} \chi_{l_1 )})_{R\beta_1}^{B_1} \ldots (\s_{(k_8} \chi_{l_8)})_{R\beta_8}^{B_8}\Big) , \non \\
\chi_L^8 (\chi_R^6 )^{\beta_7 \beta_8}_{B_7 B_8} &\equiv& \epsilon^{\alpha_1 \ldots \alpha_8} \epsilon^{\beta_1 \ldots \beta_8} 
\epsilon_{A_1 B_1} \ldots \epsilon_{A_8 B_8} \Big( (\s_{(i_1} \chi_{j_1 )})_{L\alpha_1}^{A_1} \ldots (\s_{(i_8} \chi_{j_8 )})_{L\alpha_8}^{A_8} 
\Big) \non \\ && \times 
\Big( (\s_{(k_1} \chi_{l_1 )})_{R\beta_1}^{B_1} \ldots (\s_{(k_6} \chi_{l_6 )})_{R\beta_6}^{B_6}\Big), \non \\
(\chi_L^6 )^{\alpha_7 \alpha_8}_{A_7 A_8} \chi_R^8 &\equiv& \epsilon^{\alpha_1 \ldots \alpha_8} \epsilon^{\beta_1 \ldots \beta_8} 
\epsilon_{A_1 B_1}
\ldots \epsilon_{A_8 B_8} \Big( (\s_{(i_1} \chi_{j_1 )})_{L\alpha_1}^{A_1} \ldots (\s_{(i_6} \chi_{j_6)})_{L\alpha_6}^{A_6} \Big)
\non \\ && \times 
\Big( (\s_{(k_1} \chi_{l_1 )})_{R\beta_1}^{B_1} \ldots (\s_{(k_8} \chi_{l_8)})_{R\beta_8}^{B_8}\Big). \eea

First let us consider interactions of the form 
\bea \chi_L^8 \chi_R^8, \quad \bar\psi_{L\mu} \bar\g^\mu \chi_L^7 \chi_R^8,
\quad 
\bar\psi_{R\mu} \g^\mu \chi_L^8 \chi_R^7, \quad (\bar\psi_{L\mu} \bar\g^\mu
\chi_L^7)(\bar\psi_{R\nu} \g^\nu \chi_R^7), \quad (\bar\psi_{L\mu}
\bar\g^\mu)^2 \chi_L^6 \chi_R^8  \eea 
in $S^{(3)}$. These interactions
can all be obtained from \C{totact}, on using \C{compexp}. 
In order to write down these interactions, we consider
a subset of the interactions given by \C{intlin}. They are given by
\bea \label{termslin}
{\mathcal{L}}^{(3)} &=&  e \Big[ g_{(1\ldots 8)(1\ldots 8)} \chi_L^8 \chi_R^8
  \non \\ &&+
  g_{(1\ldots 8)(1\ldots 7)} \Big(\chi_L^8 \chi_R^6 \hat{F}_2  +
\chi_L^7 \chi_R^7 (\hat{F}_3+ \hat{P}_{ij}) \Big)  \non \\ &&+ g_{(1\ldots 7)(1 \ldots 7)} \chi_L^6 \chi_R^6 (\hat{F}^2_3
+ \hat{P}_{ij}^2
+ \hat{F}_2 \cdot \hat{F}^*_2) \non \\&& + g_{(1\ldots 8)(1\ldots 6)} \{ \chi_R^8
\chi_L^4 \hat{F}_2^{*2} + \chi_R^7 \chi_L^5 (\hat{F}_3 +P_{ij})\hat{F}_2^* + \chi_R^6
\chi_L^6 (\hat{F}_3 + P_{ij})^2 \} + h.c.\Big].\non \\ \eea

In \C{termslin}, the $SO(3) \backslash SL(3,\mathbb{R})$ moduli dependence of the various couplings has not been denoted for brevity. Thus, for example,
\be g_{(1\ldots 8)(1\ldots 8)} \equiv g_{(1\ldots 8)(1\ldots 8)} (T,\bar{T},\xi,\bar{\xi},e^{-2\hat\phi}). \ee

Explicitly, the first term in \C{termslin} which is of the form $\chi_L^8 \chi_R^8$ is given by
\bea \label{inds}
&&g_{\Big( (i_1 j_1)\ldots (i_8 j_8)\Big)\Big((k_1 l_1)\ldots (k_8 l_8)\Big)} \epsilon^{\alpha_1 \ldots \alpha_8} 
\epsilon^{\beta_1 \ldots \beta_8} \epsilon_{A_1 B_1} \ldots \epsilon_{A_8 B_8} \Big( (\s_{(i_1} \chi_{j_1 )})_{L\alpha_1}^{A_1} \ldots 
(\s_{(i_8} \chi_{j_8 )})_{L\alpha_8}^{A_8} \Big) \non \\  &&\times 
\Big( (\s_{(k_1} \chi_{l_1 )})_{R\beta_1}^{B_1} \ldots (\s_{(k_8} \chi_{l_8 )})_{R\beta_8}^{B_8}\Big) , \eea  
while the second term is given by
\bea \label{indst}
&&g_{\Big( (i_1 j_1)\ldots (i_8 j_8)\Big)\Big((k_1 l_1)\ldots (k_7 l_7)\Big)} \epsilon^{\alpha_1 \ldots \alpha_8} 
\epsilon^{\beta_1 \ldots \beta_8} \epsilon_{A_1 B_1} \ldots \epsilon_{A_8 B_8} \Big( (\s_{(i_1} \chi_{j_1 )})_{L\alpha_1}^{A_1} \ldots 
(\s_{(i_8} \chi_{j_8 )})_{L\alpha_8}^{A_8} \Big) \non \\  &&\times 
\Big( (\s_{(k_1} \chi_{l_1 )})_{R\beta_1}^{B_1} \ldots (\s_{(k_6} \chi_{l_6 )})_{R\beta_6}^{B_6}\Big) (\bar\g^{\mu\nu} )_{\beta_7 \beta_8}
(\s_{( k_7 })^{B_7 B_8} \hat{F}_{l_7) 2\mu\nu} ,\eea
and similarly for the other terms.

Now \C{indst} will lead to an interaction of the form $g \bar\psi_{R\mu} \g^\mu \chi_L^8 \chi_R^7$ using $\hat{F}_2 \sim \bar\psi_R \chi_R$,
as we shall show below. The index structure of this coupling $g$ in \C{indst} has been assigned based on the structure of 
this interaction we want to consider, which will be evident from the discussion below. The conjugate interaction 
yields a term of the form $g^* \bar\psi_{L\mu} \bar\g^\mu \chi_L^7 \chi_R^8$.

Some of the remaining terms in \C{termslin} also yield interactions of the form $\bar\psi_{R\mu} \g^\mu \chi_L^8 \chi_R^7$ and 
$\bar\psi_{L\mu} \bar\g^\mu \chi_L^7 \chi_R^8$, on using $\hat{F}_3 \sim \bar\psi_L \chi_R + \bar\psi_R \chi_L$, and 
$\hat{P}_{ij} \sim \bar\psi_L \chi_R + \bar\psi_R \chi_L$. The interactions of the
form $(\bar\psi_{L\mu} \bar\g^\mu
\chi_L^7)(\bar\psi_{R\nu} \g^\nu \chi_R^7)$ and $(\bar\psi_{L\mu}
\bar\g^\mu)^2 \chi_L^6 \chi_R^8$ are obtained from $\chi_L^6 \chi_R^6
\hat{F}^2_3$ on using
$\hat{F}_3 \sim \bar\psi_R \chi_L + \bar\psi_L \chi_R$. Similarly we can work
out the various relevant interactions arising from the remaining terms in
\C{termslin}. The analysis is exactly along the lines of the one we do below.
We should mention that, as in the discussion above, there are often several
terms in the superaction which contribute to the same interaction. In such
cases, we expect the couplings coming from the various contributions to be the
same because they follow from the same superfield.

Before we proceed further, we also need to know how the interactions
corresponding to the couplings $g_{(1\ldots
  8)(1\ldots 8)}$, $g_{(1\ldots 8)(1\ldots 7)}$, $g_{(1\ldots 7)(1\ldots 7)}$
and $g_{(1\ldots 8)(1\ldots 6)}$ 
in \C{termslin} transform under $SU(2)$. We can only talk about the $SU(2)$
transformation properties of the various interactions, and not the couplings
themselves.
This is because $SU(2)$ is only a symmetry of
supergravity, and is broken by the higher derivative corrections. Actually,
these couplings transform non--trivially under $SL(3,\mathbb{Z})$, and we
should denote them by their $SL(3,\mathbb{Z})$ transformation properties. However,
unlike the previous case, we shall see later that the couplings for the
interactions which have non--trivial $SU(2)$ charges do not transform as
automorphic forms of $SL(3,\mathbb{Z})$, and transform in a complicated way.     
Thus, we find it easier to simply denote the couplings by the $SU(2)$ charges
the corresponding interactions carry. We shall often loosely denote it by the
$SU(2)$ charges of the couplings themselves, but this is only for brevity.

First let us consider the case of $g_{(1\ldots 8)(1\ldots 8)}$. As discussed before, every factor of $\s_{(i} \chi_{j)}$ 
(and consequently every factor of $(ij)$ in $g_{(1\ldots 8)(1\ldots 8)}$) transforms
in the spin 2 representation of $SU(2)$. Thus the spacetime interaction in \C{inds} involves the product of 16 spin 2
representations of $SU(2)$. Expressing this as a sum of irreducible
representations, we choose the interaction
to project onto the spin 32 representation of $SU(2)$. Thus, the coupling is symmetric under the interchange of any pair of
$(ij)$ indices. While this is not necessary for our analysis, and one can focus on any irreducible representation of $SU(2)$,
we choose the highest spin representation for simplicity, as the symmetry under interchange of the various indices simplifies
our calculations considerably. Similarly, we take the interaction
corresponding to the $g_{(1\ldots 8)(1\ldots 7)}$ coupling to transform in the spin 30 representation of $SU(2)$ (this 
follows from the analysis below because the $(k_7 l_7)$ term in $g$ gets
coupled to $\s_{(k_7} \chi_{l_7 )R}$), and those corresponding to the $g_{(1\ldots 7)(1\ldots 7)}$ and
$g_{(1\ldots 8)(1\ldots 6)}$ couplings to
transform in the spin 28 
representation of $SU(2)$. We always consider interactions which transform as the highest spin representation of $SU(2)$, and this will be always
implicit in the discussion below.

We shall find it convenient to write the couplings in the form $g_{()()}$ where the indices in the two parantheses are the number of $\chi_L$ and $\chi_R$ fields in the interaction. Obviously, this division is artificial, as only the $SU(2)$ representation matters. We shall later shift to the convention where the division between $\chi_L$ and $\chi_R$ is removed, and thus an 
arbitrary interaction can be analyzed.

Now let us simplify the structure in \C{indst} to obtain the interaction of the type we want. The $\chi_L^8 \chi_R^6$ part
is already of the form we want. Focusing on the rest, we use
\be \label{rotate}
(\s_{k_7})^{B_7 B_8} (\chi_{l_7 })_R^A  = \epsilon^{AB_8} (\s_{k_7} \chi_{l_7
})_R^{B_7} + (\s_{k_7})^{B_7 A} (\chi_{l_7})_R^{B_8}, \ee
which follows from using the Schouten identity.
In \C{rotate}, we note that the first term gives a spacetime contribution of the kind we want, and so we consider this 
contribution to \C{indst}. Finally, we use\footnote{Unlike the expression in \C{moredef}, note that \C{prop} has more terms 
represented by $\ldots$, because $\chi_L$ has 32 degrees of freedom.}
\be \label{prop}
(\chi_R^6 )^{\alpha\beta}_{AB} \chi_{R \g}^C \sim \delta^C_{(A} \Big[ (\chi_R^7)^\beta_{B)} \delta^\alpha_\gamma - (\chi_R^7)^\alpha_{B)}
\delta^\beta_\g \Big] + \ldots \ee
to get a space time contribution of the form $\bar\psi_{R\mu} \g^\mu \chi_L^8 \chi_R^7$, where
\be \bar\psi_{R\mu} \g^\mu \chi_L^8 \chi_R^7 \equiv (\bar\psi_{R\mu} \g^\mu)^A_\alpha \chi_L^8 (\chi_R^7)_A^\alpha .\ee   
We perform a similar analysis for the other couplings in \C{termslin} to obtain
the relevant terms in the action, where we have used the notations
\bea (\bar\psi_{L\mu} \bar\g^\mu \chi_L^7 )(\bar\psi_{R\nu} \g^\nu \chi_R^7)
\equiv 
(\bar\psi_{L\mu} \bar\g^\mu)^A_\alpha (\chi_L^7 )^\alpha_A (\bar\psi_{R\nu}
\g^\nu)^B_\beta (\chi_R^7)^\beta_B, \non \\ (\bar\psi_{L\mu} \bar\g^\mu)^2
\chi_L^6 \chi_R^8 \equiv (\bar\psi_{L\mu} \bar\g^\mu)^A_\alpha
(\bar\psi_{L\mu} \bar\g^\mu)^B_\beta
(\chi_L^6)^{\alpha\beta}_{AB} \chi_R^8. \eea

Thus in $\mathcal{L}^{(3)}$, we consider interactions of the form\footnote{We
  suitably rescale the various couplings.}
\bea \label{actlin}
\mathcal{L}^{(3)} = e \Big[ g_{(1\ldots 8)(1\ldots 8)} \chi_L^8 \chi_R^8 + g_{(1\ldots 8)(1\ldots 7)} 
\bar\psi_{R\mu} \g^\mu \chi_L^8 \chi_R^7 +  g_{(1\ldots 7)(1\ldots 8)} 
\bar\psi_{L\mu} \bar\g^\mu \chi_L^7 \chi_R^8  \non \\ + g_{(1\ldots 7)(1
  \ldots 7)} (\bar\psi_{L\mu} \bar\g^\mu \chi_L^7 )(\bar\psi_{R\nu}
  \g^\nu \chi_R^7) +g_{(1\ldots 6)(1\ldots 8)}(\bar\psi_{L\mu} \bar\g^\mu)^2
\chi_L^6 \chi_R^8   +\ldots\Big].\eea
Note that\footnote{We shall soon consider the convention where the couplings are characterized only by their $SU(2)$ spins. Thus
we shall denote 
\be  \label{newconv} g_{(a_1 \ldots a_m)(a_{m+1} \ldots a_n)} \equiv g_{(a_1  \ldots a_n)} \equiv g^{(n)}.\ee
So, for example, 
\be g_{\Big( (i_1 j_1)\ldots (i_8 j_8) \Big) \Big( (k_1 l_1) \ldots (k_8 l_8)\Big)} = g_{(i_1 j_1)\ldots (i_8 j_8) (k_1 l_1) 
\ldots (k_8 l_8)}. \ee
Schematically we denote $g_{(1\ldots 8)(1\ldots 8)} = g^{(16)}, g_{(1\ldots
  8)(1\ldots 7)}= g^{(15)}, g_{(1\ldots 7)(1\ldots 8)}= g^{(15)},g_{(1\ldots
  7)(1\ldots 7)} = g_{(1\ldots 8)(1\ldots 6)}=g^{(14)}$. Then \C{real} implies that $g^{(15)}$ is real.}
\be \label{real} g_{(1\ldots 8)(1\ldots 7)}^* = g_{(1\ldots 7)(1\ldots 8)}.\ee

Also, to project onto the spacetime structure of the terms we want, we use the relation
\bea \label{average}
(\chi_L^7)^\alpha_A (\chi_L)^B_\beta \chi_R^8 = \frac{1}{16} \delta^\alpha_\beta \delta^B_A \chi_L^8 \chi_R^8 + \ldots .\eea 

Let us first consider various contributions coming from $\delta^{(0)} S^{(3)}$.
To begin with, let us consider the variation of \C{actlin} into terms of the form $\epsilon_L \chi_R^8 \chi_L^9$. 
The first term gives
\be \label{expand}
\delta^{(0)} \Big( e g_{(1\ldots 8)(1\ldots 8)}\chi_L^8 \chi_R^8 \Big) \sim e \Big( \p g_{(1\ldots 8)(1\ldots 8)} \Big)
(\bar\epsilon_R \chi_L) \chi_L^8 \chi_R^8 + \ldots .\ee

Explicitly, the right hand side of \C{expand} is given by
\be  \p_{(i_9 j_9 )} g_{\Big((i_1 j_1) \ldots (i_8 j_8)\Big)\Big( (k_1 l_1) \ldots
(k_8 l_8) \Big)} \bar\epsilon_R \s_{( i_9}\chi_{j_9 ) L}+\ldots . \ee

The term we have written down in \C{expand} obviously comes from taking the
derivatives with respect to the moduli on using \C{varmoduli}. We would like
to know if there are more contributions to \C{expand}, which promote the ordinary
derivative to a generalized derivative, and acts on the space of
couplings, like the $U(1)\backslash SL(2,\mathbb{R})$ case.

In fact there are such contributions to \C{expand} which we now calculate. Let
us first focus on the contribution from $\delta^{(0)} \chi_L \sim \epsilon_L
\chi_L^2$. In order to contribute to \C{expand}, we only need terms of the form
$\s_{(i} \delta^{(0)} \chi_{j)L} \sim \s_{(i} \chi_{j)L}$, and similarly for $\chi_R$.
There are no such terms from the $SU(2)$ covariant expression for
$\delta^{(0)} \chi_L$ in \C{susytran}, and they only arise from the extra terms
from gauge--fixing the supersymmetry transformations given by
\C{addsusytran}. Thus these terms, as expected, are not $SU(2)$ covariant.
From \C{addsusytran}, we get the relevant term
\be \label{needsimp} \s_{(i} \delta^{(0)} \chi_{j)L} = -\frac{3}{2} \epsilon_{(i}^{~~kl}
\theta^k \s_{j)} \chi^l_L .\ee
Using \C{fixt}, \C{useeqn} and the Fierz and Schouten identities repeatedly,   
we get that
\bea &&\s_{(i} \delta^{(0)} \chi_{j)L} = \frac{3}{16} \Big[ \Big\{\s_{(i} \chi_{j)L}
  (\bar\epsilon_R \s_3 \chi_{3L}) -\frac{1}{2} \g^{\mu\nu} \s_{(i} \chi_{j)L}
  (\bar\epsilon_R \g_{\mu\nu}
\s_3 \chi_{3L}) \non \\ && +\frac{1}{24}\g^{\mu\nu\lambda\rho}\s_{(i} \chi_{j)L}
  (\bar\epsilon_R \g_{\mu\nu\lambda\rho}\s_3 \chi_{3L})+ \s_{(i} \s_3
\epsilon_L (\bar\chi_{3R} \chi_{j)L}) -\frac{1}{2} \g^{\mu\nu} \s_{(i} \s_3
\epsilon_L (\bar\chi_{3R} \g_{\mu\nu}\chi_{j)L})\non \\ &&+ \frac{1}{24}
\g^{\mu\nu\lambda\rho} \s_{(i} \s_3
\epsilon_L (\bar\chi_{3R} \g_{\mu\nu\lambda\rho} \chi_{j)L})\Big\}
- (3 \rightarrow 1)\Big] \non \\ &&-\frac{3}{4} \Big[ \delta_{1(i} \s_{j)}\Big\{
  2\chi_{1L}
(\bar\epsilon_R \s_1\chi_{1L}) + 2\chi_{2L}
(\bar\epsilon_R \s_2\chi_{1L}) + i \chi_{3L} (\bar\epsilon_R \chi_{2L}) -i
  \chi_{2L} (\bar\epsilon_R \chi_{3L})\Big\}\non \\&& 
+\delta_{3(i} \s_{j)}\Big\{ -2\chi_{3L}
(\bar\epsilon_R \s_3\chi_{3L}) -2\chi_{2L}
(\bar\epsilon_R \s_2\chi_{3L}) + i \chi_{1L} (\bar\epsilon_R \chi_{2L}) -i
  \chi_{2L} (\bar\epsilon_R \chi_{1L})\Big\} \non \\&& -i \delta_{2(i} \s_{j)}
  \Big\{ \chi_{1L} (\bar\epsilon_R \chi_{3L}) + \chi_{3L} (\bar\epsilon_R \chi_{1L})\Big\} \Big] ,\eea
where we have kept only the terms proportional to $\epsilon_L$. This gives a
contribution 
\be \label{coeff} \s_{(i} \delta^{(0)} \chi_{j)L} = \frac{3}{16} \s_{(i} \chi_{j)L}
  \Big(\bar\epsilon_R \s_3 \chi_{3L} - \bar\epsilon_R \s_1 \chi_{1L}\Big)\ee
exactly of the kind we need. The coefficient in \C{coeff} can receive
additional contributions from $\delta^{(3)}
\mathcal{L}^{(0)}$, as we shall schematically describe later\footnote{This is
  unlike the $U(1) \backslash SL(2,\mathbb{R})$ analysis done before.}. However, it is
not difficult to write down the final answer which is 
\be \label{coeffin}\s_{(i} \delta^{(0)} \chi_{j)L} = 2 \s_{(i} \chi_{j)L}
  \Big(\bar\epsilon_R \s_3 \chi_{3L} - \bar\epsilon_R \s_1 \chi_{1L}\Big),\ee
and similarly for $\chi_R$.   
This can be seen by considering the expression \C{sl3exp}, and noting that
$D^{(1)}_{(ij)} D^{(0)}_{(ij)} = 2\Delta$, where $\Delta$ is the Laplacian on
$SO(3) \backslash SL(3,\mathbb{R})$ given by \C{LapSL3}. Now $D^{(0)}_{(ij)}$ which contains
only the derivatives comes from varying the $\mathcal{R}^4$ coupling. While the derivative terms in
$D^{(1)}_{(ij)}$ have a similar origin, the remaining terms come from the
supervariation of one factor of $\s_{(i}\chi_{j)L}$ in the spacetime
interaction, which directly leads to \C{coeffin}. 

Thus, the right hand side of \C{expand} is given by\footnote{Right now, we are
  looking at only the minimal set of terms in the effective action which
  provide the necessary supervariations in an obvious way. At least this much
  structure is needed to see that a set of equations relating the various couplings
  arises. We shall talk about possible additional terms that could modify the
  equations later.  
}
\be \label{modcovsl3} D^{(16)}_{(i_9 j_9 )} g_{\Big((i_1 j_1) \ldots (i_8 j_8)\Big)\Big( (k_1 l_1) \ldots
(k_8 l_8) \Big)} \bar\epsilon_R \s_{( i_9}\chi_{j_9 ) L} . \ee
where the expression for $D^{(n)}_{(ij)}$ is given in \C{sl3exp}.

The second term in \C{actlin} gives contributions from the supervariation of $\psi_L$. Since we only want to see the specific spacetime structure emerge, it is good enough to do the analysis for any term in $\delta^{(0)} \psi_L$. From \C{susytran}, we thus consider
\be \delta^{(0)} \psi_{\mu L} = -\frac{1}{6} (\bar\epsilon_R \s_i \chi_{jL}) \s_i \g_\mu \chi_{jR},\ee
which gives
\be \label{can1} \delta^{(0)} \Big( \bar\psi_{\mu R} \g^\mu \chi_L^8 \chi_R^7 \Big) = \frac{4i}{3} (\bar\epsilon_R \s_i \chi_{jL}) \chi_L^8 (
\chi_R^7 \s_i \chi_{jR})  .\ee
Thus, we get that
\be \label{moreeq} \delta^{(0)} \Big( e g_{(1\ldots 8)(1\ldots 7)}\bar\psi_{\mu R} \g^\mu \chi_L^8 \chi_R^7 \Big) \sim e 
\delta g_{(1\ldots 8)(1\ldots 7)} (\bar\epsilon_R \chi_L) \chi_L^8 \chi_R^8.\ee
The right hand side of \C{moreeq} contains 
\bea \Big[ g_{\Big((i_1 j_1) \ldots (i_8 j_8)\Big)\Big( (k_1 l_1) \ldots (k_7 l_7) \Big)} \delta_{k_8 i_9} \delta_{l_8 j_9} \Big] \bar\epsilon_R \s_{( i_9}\chi_{j_9 ) L}.\eea 

Similarly, the third term in \C{actlin} gives contributions from the supervariation of $\psi_R$. 
As before, from \C{susytran}, we consider
\be \delta^{(0)} \psi_{\mu R} = \frac{1}{6} (\bar\epsilon_R \s_i \chi_{jL}) \s_i \bar\g_\mu \chi_{jL},\ee
which gives\footnote{The sum of the supervariations does vanish for the
  contributions \C{can1} and \C{can2} we have considered. However, this should
  not vanish
  when all the contributions are taken into account.
 }
\be \label{can2} \delta^{(0)} \Big( \bar\psi_{\mu L} \bar\g^\mu \chi_L^7 \chi_R^8 \Big) = -\frac{4i}{3} (\bar\epsilon_R \s_i \chi_{jL}) (\chi_L^7
\s_i \chi_{jL}) \chi_R^8 ,\ee
leading to
\be \label{moreeqn} \delta^{(0)} \Big( e g_{(1\ldots 7)(1\ldots 8)}\bar\psi_{\mu L} \bar\g^\mu \chi_L^7 \chi_R^8 \Big) \sim e \delta g_{(1\ldots 7)(1\ldots 8)} (\bar\epsilon_R \chi_L) \chi_L^8 \chi_R^8.\ee
The right hand side of \C{moreeqn} contains 
\bea \Big[ g_{\Big((i_1 j_1) \ldots (i_7 j_7)\Big)\Big( (k_1 l_1) \ldots (k_8 l_8) \Big)} \delta_{i_8 i_9} \delta_{j_8 j_9} \Big] \bar\epsilon_R \s_{( i_9}\chi_{j_9 ) L}.\eea 
The remaining terms in \C{actlin} do not give a contribution of the type we want.

Thus from \C{expand}, \C{moreeq} and \C{moreeqn}, on using \C{newconv}, we get that 
\be \label{more2}D^{(16)}_{(kl)} g_{(i_1 j_1) \ldots (i_{16} j_{16})} = c_1 g_{(i_1 j_1) \ldots (i_{15} j_{15})} \delta_{k i_{16}} \delta_{l j_{16}}.\ee
In short, we write \C{more2} schematically as
\be \label{short2} D^{(16)} g^{(16)} = c_1 g^{(15)} .\ee
Thus, $D^{(16)}$ acts as a spin 2 operator which acting on a coupling
corresponding to an interaction in the spin
$32$ representation gives the coupling corresponding to an interaction in the spin $30$ 
representation of $SU(2)$.

Note that there can be a term in the effective action of the type
\be e \chi_L^8 (\chi_R^6)^{\alpha\beta}_{AB} (\s_{i_9} \chi_{j_9 L})_\alpha^A \lambda_{L\beta}^B\ee
which can vary into $\epsilon_L \chi_L^9 \chi_R^8$ for\footnote{One can obtain
such a term by manipulating the corresponding terms in \C{susytran}.} 
\be \delta^{(0)} \lambda_L \sim (\bar\epsilon_R \s_i \chi_{jR}) \s_i \chi_{jR}.\ee 
This term cannot be obtained from \C{totact} using linearized superspace. 
However, the contribution of this term simply changes the value of $c_1$ in \C{short2}.

Now let us consider the supervariation of \C{actlin} under $\delta^{(0)}$ into terms of the form 
$\chi_L^8 \chi_R^8 \epsilon_L \psi_R$. We shall give some of the details of
the calculations for the first couple of terms, and then simply give the
answers.  The first term in \C{actlin} gives three contributions involving the supervariations 
of $e_\mu^{~a}, \chi_L$ and $\chi_R$ respectively.     
The metric variation yields
\be (\delta^{(0)} e) \chi_L^8 \chi_R^8 = - e(\bar\epsilon_R \gamma^\mu \psi_{\mu R}) \chi_L^8 \chi_R^8 .\ee
The contribution from the supervariation of $\chi_L$ is obtained from the $\hat{F}^*_2$ term in \C{susytran} on using
$\hat{F}^*_2 \sim \psi_R \chi_L$. To obtain it, we use
\be \label{repeat}
(\s_i \delta^{(0)} \chi_{jL})^A_\alpha = \frac{1}{2}  (\g^{\mu\nu} \epsilon_L)^B_\alpha (\bar\psi^\mu_L \bar\g^\nu)^C_\beta  
\Big[ \epsilon_{BC}(\s_i \chi_{jL})^{A\beta} + \epsilon_{DB} \chi_{kL}^{D\beta} (\s_i \Delta_{jk})^A_C \Big] ,\ee
which can be obtained on using the Schouten identity.
Note that the first term in \C{repeat} yields a contribution of the type we want on using \C{average},
leading to 
\be (\delta^{(0)} \chi_L^8 ) \chi_R^8  \sim (\bar\epsilon_R \g^\mu \psi_{\mu R}) \chi_L^8 \chi_R^8 .\ee
Similarly, the term involving the supervariation of $\chi_R$ gives
\be \label{2cont}
\chi_L^8 (\delta^{(0)}  \chi_R^8 ) \sim   (\bar\epsilon_R \g^\mu \psi_{\mu R}) \chi_L^8 \chi_R^8 ,\ee
on using $\hat{P}_{ij} \sim \psi_R \chi_R$ and $\hat{F}_3 \sim \psi_R \chi_R$, and
\be \bar\g_{\nu\s} \bar\g^{\mu\nu\s} = -42 \bar\g^\mu .\ee
Thus we get that
\be \label{cont1} \delta^{(0)} \Big(e  g_{(1\ldots 8)(1\ldots 8)} \chi_L^8 \chi_R^8 \Big)
\sim e g_{(1\ldots 8)(1\ldots 8)} (\bar\epsilon_R \g^\mu \psi_{\mu R}) \chi_L^8 \chi_R^8 .\ee

The second term in  \C{actlin} can give a possible contribution using 
\be \delta^{(0)} \psi_{\mu L} = \frac{1}{24} \s^i (\g_{\mu\nu\lambda\rho} - 6 g_{\mu\nu} \g_{\lambda\rho})
\epsilon_L (\bar\psi_{L[\nu} \bar\g_{\lambda\rho]} \chi_{iR} ) - \frac{1}{4} \s_i \epsilon_L (\bar\psi_{\mu L}
\chi_{iR}) + \frac{1}{2} \s_i \g^\nu \psi_{\mu R} (\bar\epsilon_R \g_\nu \chi_{iR}) ,\ee
which follows from \C{susytran}. However, this does not yield a contribution of the type we want, and so it 
vanishes. Thus
\be \label{van} \delta^{(0)} \Big( e g_{(1\ldots 8)(1\ldots 7)} 
\bar\psi_{R\mu} \g^\mu \chi_L^8 \chi_R^7 \Big) =0 .\ee 

The third term in \C{actlin} yields contributions coming from the
supervariations of $g_{(1\ldots 7)(1\ldots 8)}, \chi_L$ and $\chi_R$ leading to
\be \label{cont3}
e \delta^{(0)} \Big( g_{(1\ldots 7)(1\ldots 8)}  \bar\psi_{L\mu} \bar\g^\mu \chi_L^7 \chi_R^8 \Big)
=-\frac{ie}{16}  \Big( D^{(15)} g_{(1\ldots 7)(1\ldots 8)} \Big) (\bar\epsilon_R \g^\mu \psi_{\mu R}) \chi_L^8 \chi_R^8 ,\ee
where 
\be D^{(15)} g_{(1\ldots 7)(1\ldots 8)} \equiv D^{(15)}_{(i_1 j_1 )} g_{\Big((i_2 j_2) \ldots (i_8 j_8)\Big)\Big( (k_1 l_1) \ldots
(k_8 l_8) \Big)}.\ee

Note that there is no contribution to $D^{(15)}$ from the supervariation of
$\psi_{R\mu}$, which is not difficult to see from \C{addsusytran}. This is
what is expected, because the spacetime interaction has all its $SU(2)$
indices contracted except the spin 2 indices carried by the $\s_{(i}\chi_{j)}$ factors.  
This continues to hold in our discussion below.

Exactly similarly the contribution from the supervariation of the terms
involving $g^{(14)}$ in \C{actlin} comes from $\delta^{(0)}\psi_L \sim
\epsilon_L \chi_L \chi_R$ and $\delta^{(0)} \psi_R \sim \epsilon_L \chi_L^2$ on
using \C{susytran}, and yields
\bea \label{cont4} \delta^{(0)} \Big( g_{(1\ldots 7)(1
  \ldots 7)}(\bar\psi_{L\mu} \bar\g^\mu \chi_L^7 )(\bar\psi_{R\nu}
  \g^\nu \chi_R^7) +g_{(1\ldots 6)(1\ldots 8)}(\bar\psi_{L\mu} \bar\g^\mu)^2
\chi_L^6 \chi_R^8 \Big) \non \\ \sim e  \delta g_{(1\ldots 7)(1\ldots 7)}  (\bar\epsilon_R \g^\mu \psi_{\mu R}) \chi_L^8 \chi_R^8.\eea

From \C{cont1}, \C{van}, \C{cont3} and \C{cont4}, we get that
\bea \label{eqn1} D^{(15)}_{(i_1 j_1 )} g_{\Big((i_2 j_2) \ldots (i_8 j_8)\Big)\Big( (k_1 l_1) \ldots
(k_8 l_8) \Big)} &=& d_1 g_{\Big((i_1 j_1) \ldots (i_8 j_8)\Big)\Big( (k_1 l_1) \ldots
(k_8 l_8) \Big)} \non \\ && + d_2 g_{\Big((i_1 j_1) \ldots (i_7 j_7)\Big)\Big( (k_1 l_1) \ldots
(k_7 l_7) \Big)} \delta_{i_8 k_8} \delta_{j_8 l_8}, \eea
where we have absorbed the contribution from the term involving $g_{(1\ldots
  6)(1\ldots 8)}$ into that from $g_{(1\ldots 7)(1\ldots 7)}$ using \C{newconv}.
Using \C{newconv}, \C{eqn1} can be schematically written as
\be \label{short1} D^{(15)} g^{(15)} = d_1 g^{(16)} + d_2 g^{(14)}.\ee
which is explicitly
\be \label{long1} D^{(15)}_{(i_1 j_1 )} g_{(i_2 j_2) \ldots (i_{16} j_{16}) } = d_1
g_{(i_1 j_1) \ldots (i_{16} j_{16}) } + d_2 g_{(i_1 j_1) \ldots (i_{14}
  j_{14}) } \delta_{i_{15} i_{16}} \delta_{j_{15} j_{16}}.\ee
In \C{short1}, $D^{(15)}$ acts as a spin 2 operator which acting on a coupling
corresponding to an interaction in the spin
$30$ representation gives couplings corresponding to interactions in the spin $32$ and spin $28$
representations of $SU(2)$. In spite of the complexity of the calculations,
the structure of the final answer is quite simple, due to supersymmetry.

Thus, the supervariation of the terms in the effective action we have
considered gives us
\C{short2} and \C{short1}. Does this pattern continue?

It is not difficult to see that it does. In fact, the next equation
is given by
\be \label{short3} D^{(14)} g^{(14)} = e_1 g^{(15)} + e_2 g^{(13)} .\ee  
Now, \C{short3} can be obtained by starting with the term $g^{(14)} \chi_L^8
\chi_R^6 \psi_L^2$ in the effective action which can be obtained from $g^{(14)}
\chi_L^6 \chi_R^6 \hat{F}_3^2$, and varying it into $D^{(14)} g^{(14)} \epsilon_L
\chi_L^9 \chi_R^6 \psi_L^2$. This supervariation can also be obtained from
other terms in the effective action which contribute to \C{short3}. These
terms are, for example, of the form $g^{(15)} \chi_L^8 \chi_R^7 \psi_L$ and
$g^{(13)} \chi_L^7 \chi_R^6 \psi_L^2 \psi_R$, which can be obtained from
$g^{(15)} \chi_L^7 \chi_R^7 \hat{F}_3$ and $g^{(13)} \chi_L^5 \chi_R^5
\hat{F}^3_3$ respectively. It is crucial that there are no other couplings with
any other spin that contribute to \C{short3}. This generalizes easily all the
way upto
\be D^{(9)} g^{(9)} \sim g^{(10)} + g^{(8)},\ee
using the interactions $g^{(16-n)} \chi_L^{8-n} \chi_R^{8-n} \hat{F}_3^n$ for
$0 \leq n \leq 8$, and repeating the above logic. To obtain the analogous equation
for $D^{(8)} g^{(8)}$, while $g^{(9)}$ is obtained as above, none of the terms
that give a contribution involving $g^{(7)}$ arise from the linearized
superfield \C{compexp}. This is not surprising because linearized superspace
gives only a small subset of interactions in the effective action. However, it
is not difficult to write down a term involving $g^{(7)}$ that gives the
relevant supervariation. Such a term, of the form $\chi_L^7 \psi_R \psi_L^8$
is given by
\bea \epsilon^{\alpha_1 \ldots \alpha_8} \epsilon^{\beta_1 \ldots \beta_8}
\epsilon_{A_1 B_1} \ldots \epsilon_{A_8 B_8} \Big( (\s_{(i_1}
\chi_{j_1)} )^{A_1}_{\alpha_1} \ldots (\s_{(i_7}
\chi_{j_7)})^{A_7}_{\alpha_7} \Big) (\bar\psi_{L\nu} \bar\g^\nu
)^{A_8}_{\alpha_8}
\non \\ \times \Big( (\bar\psi_{R\mu_1} \g^{\mu_1})^{B_1}_{\beta_1} \ldots 
(\bar\psi_{R\mu_8} \g^{\mu_8})^{B_8}_{\beta_8} \Big), \eea  
which gives the required supervariation for
\be \delta^{(0)} \psi_{R\mu } \sim \g_\mu \s_i \chi_{jL} (\bar\epsilon_R \s_i \chi_{jL}).\ee
Now the analysis goes through all the way upto
\be D^{(0)} g^{(0)} \sim g^{(1)},\ee
using the interactions $g^{(8-n)} \chi_L^{8-n} \psi_R^n \psi_L^8$ for $0 \leq
n \leq 8$. These interactions are explicitly given by
\bea \epsilon^{\alpha_1 \ldots \alpha_8} \epsilon^{\beta_1 \ldots \beta_8}
\epsilon_{A_1 B_1} \ldots \epsilon_{A_8 B_8} \Big( (\s_{(i_1}
\chi_{j_1)} )^{A_1}_{\alpha_1} \ldots (\s_{(i_{8-n}}
\chi_{j_{8-n})})^{A_{8-n}}_{\alpha_{8-n}} \Big) \non \\  \times
\Big( (\bar\psi_{L\nu_1} \bar\g^{\nu_1})^{A_{7-n}}_{\alpha_{7-n}}
\ldots (\bar\psi_{L\nu_n} \bar\g^{\nu_n})^{A_8}_{\alpha_8} \Big)
\Big( (\bar\psi_{R\mu_1} \g^{\mu_1})^{B_1}_{\beta_1} \ldots 
(\bar\psi_{R\mu_8} \g^{\mu_8})^{B_8}_{\beta_8} \Big), \eea

Thus we obtain the sequence of equations 
\bea  \label{seqeqn} D^{(16)} g^{(16) } &=& ~~~~~~~~~ g^{(15)}, \non \\ D^{(15)} g^{(15)} &=& g^{(16)} +
g^{(14)}, \non \\ D^{(14)} g^{(14)} &=& g^{(15)} + g^{(13)}, \non \\  \vdots && ~~~~\ddots \non \\ D^{(1)} g^{(1)} &=& g^{(2)}
 ~~+g^{(0)}, \non \\ D^{(0)} g^{(0)} &=& g^{(1)}. 
\eea
for the various couplings in the supermultiplet, which follows as
a consequence of supersymmetry. We have set the various undetermined coefficients
in \C{seqeqn} to 1 for brevity, but they can all be completely determined by supersymmetry.  
It follows that if any one of the couplings in \C{seqeqn} at either end of
the sequence can be determined,
then all the other couplings can be determined recursively. The structure of
the equations \C{seqeqn} is consistent with the fact that for the kind of
couplings we have considered, on any spin 2 index, $D^{(n)}$ acts as
\be \label{CG}
 D^{(n)}_{(ij)} : \mathcal{A}_{(kl)} \rightarrow \mathcal{A}_{(ij)(kl)} +
\mathcal{A} \delta_{ik} \delta_{jl}.\ee  
Thus it reduces the spin by 2, and increases it by 2. This happens with every
coupling in \C{seqeqn}, except for the first and last couplings in the
sequence, because there are no couplings to vary into for some spins.   

So far we have considered only the contributions coming from $\delta^{(0)} \mathcal{L}^{(3)}$.
Let us now focus on the contributions coming from $\delta^{(3)}
\mathcal{L}^{(0)}$ very schematically. To be specific, we consider the added contributions to \C{short2}. 
These terms are obtained from the supervariation of the action into $\epsilon_L \chi_L^9 \chi_R^8$.
There are several terms in $\mathcal{L}^{(0)}$ which can give this
supervariation for appropriate $\delta^{(3)}$. They are 
\be \lambda_L \chi_L^3, \quad \lambda_R \chi_R^3, \quad \chi_L^2 \chi_R^2,
\quad \chi_L^2 \chi_R \psi_L, \quad \chi_R^2 \chi_L \psi_R .\ee
We focus on only the $\chi_L^2 \chi_R^2$ term, and consider the supervariation
\be \label{comp1} \delta^{(3)} \chi_L \sim f^{(15)} \epsilon_L \chi_R^6 \chi_L^8 + f^{(16)}
\theta \chi_L^7 \chi_R^6 + f^{(14)} \epsilon_L \chi_R^5 \chi_L^7 (\chi_L \bullet \chi_R) ,\ee
where the $SU(2)$ indices on the new couplings have been assigned anticipating
the answer. Note that the $f^{(16)}$ term is non covariant, and $\chi_L
\bullet \chi_R = (\s^{(i} \chi^{j)}_L )(\s^{(i} \chi^{j)}_R)$.
Now let us consider the contributions from $\delta^{(3)} (\chi_L^2 \chi_R^2)$
given \C{comp1}. The contribution of the $f^{(16)}$ term is of the
form required to extend $\p \rightarrow D$ in \C{short2}, and thus $f^{(16)}
\sim g^{(16)}$. For the subset of terms we are considering, we do not expect
the $f^{(14)}$ term (as well as similar terms) to contribute to the
final answer because of \C{CG}\footnote{We shall see below that there can be other
  terms that can contribute to \C{seqeqn}, and then such terms will
  contribute. For example, a non--covariant $f^{(14)}$ term in \C{comp1} is
  needed to send 
$\p f^{(14)}\rightarrow  D f^{(14)}.$}. Thus we get that
\be \label{finadd} \delta^{(3)} (\chi_L^2 \chi_R^2 ) \sim (f^{(15)} + g^{(16)}) \epsilon_L
\chi_L^9 \chi_R^8. \ee
To get more information, we now impose the constraint of closure of the
superalgebra on $\chi_L$ which gives, among other things, the $\chi_R$
equation of motion. We get that
\bea \label{closfin} [\delta^{(0)}_2 ,\delta^{(3)}_1 ]\chi_L \sim (D^{(15)} f^{(15)})
\epsilon_{L1} \epsilon_{R2} \chi_R^7 \chi_L^8 +  f^{(14)} \epsilon_{L1} \epsilon_{R2} \chi_R^7
\chi_L^8 \non \\ + f^{(15)}
\epsilon_{L1} \epsilon_{R2} \chi_R^7 \chi_L^8 + f^{(16)} \theta_1 \chi_L^7
\chi_R^7 \epsilon_{R2} + f^{(14)} \epsilon_{L1} \epsilon_{R2} \chi_R^7
\chi_L^8 + \ldots,\eea
where we have dropped other non--covariant terms, like
\be \label{closfin2} (D^{(16)} f^{(16)}) \theta_1 \chi_L^7 \chi_R^7
\epsilon_{R2},\ee
as well as
\be (D^{(14)} f^{(14)}) \epsilon_{L1} \epsilon_{R2} \chi_R^7 \chi_L^8\ee
which is not expected to contribute because of \C{CG}\footnote{Such a
  contribution vanishes because one has 16 pairs of indices to contract at the
end, while this gives 15, the remaining pair gives 0 on tracing.}.  
The terms in the first line of \C{closfin}, and \C{closfin2} arise in an
obvious way, while the terms in the second line of \C{closfin} involving
$f^{(16)}$ and $f^{(15)}$ are there to provide the correct coefficients to
extend $\p \rightarrow D$, while the $f^{(14)}$ term must arise in a suitable
way such that the all the terms in the second line of \C{closfin} can be
represented as $\delta^{(3)}_{\hat\epsilon} \chi_L$ for $\hat\epsilon \sim
\epsilon_{L1} \epsilon_{R2} \chi_R$. Thus this entire contribution in the
closure can be absorbed as a local symmetry transformation.     

Along with
\be [\delta^{(0)}_1, \delta^{(0)}_2] \chi_L \sim \epsilon_{L1} \epsilon_{R2}
\slash\p \chi_L ,\ee
\C{closfin} leads to the equation of motion
\be \slash\p \chi_L + (D^{(15)} f^{(15)}) \chi_R^7 \chi_L^8  + f^{(14)}
\chi_R^7 \chi_L^8 + \ldots =0,\ee
leading to 
\be \label{closfin3} (D^{(15)} f^{(15)}) \sim g^{(16)}  + f^{(14)},\ee
using the terms in the action.
On the other hand, \C{closfin2} would have yielded a term in the action which
cannot exist, and hence the total contribution must vanish. Thus, \C{closfin3}
yields $f^{(15)} \sim g^{(15)}, f^{(14)} \sim g^{(14)}$ on using \C{short1}
self--consistently, and so from \C{finadd} we see that the structure
of \C{short2} is unchanged, with the coefficients receiving corrections.   
We expect this analysis to go through for the other couplings.

Thus we are left with the set of equations \C{seqeqn}, which can be analyzed further to give Poisson
equations on moduli space for every coupling. The source terms in each
equation for a specific coupling are given by some other couplings in the same supermultiplet. The
structure of the source terms, however, is quite complicated as can be seen
from \C{seqeqn}.

Are there corrections to \C{seqeqn}? We now address this issue because
so far we have only looked at the miminal set of terms needed to get this
structure. There are other terms that can possibly contribute to \C{seqeqn}
simply based on the structure of the $SU(2)$ indices. For example, in the equation for
$D^{(16)} g^{(16)}$, there can be a term of the form $D^{(14)} g^{(14)}$ as
well, where the index contraction is exactly the same as for the $g^{(15)}$ term.
Such a term would arise from the supervariation of a term in the action of the form $g^{(14)}
\chi_L^7 \chi_R^7 (\chi_L \bullet \chi_R)$. Clearly these kinds of terms can
be added on the right hand side of \C{seqeqn} which involve couplings of
interactions with lower $SU(2)$ spin. However, note that last two equations in
\C{seqeqn} involving $D^{(0)} g^{(0)}$ and $D^{(1)} g^{(1)}$ stay as they
are. This also modifies the constraints imposed by the closure of the
superalgebra. However, the conclusions remain unchanged. That is, every
coupling satisfies a first order differential equation relating it to other
couplings, which on iteration shows that each coupling satisfies Poisson
equation on moduli space, where the source terms are complicated.

Let us analyze in detail the two equations which lead to the Poisson equation
for $g^{(0)}$. From \C{seqeqn}, we get that
\bea D^{(0)}_{(ij)} g^{(0)} &=& a_1 g^{(1)}_{(ij)}, \non \\ 
\frac{1}{4} \Big( D^{(1)}_{(ij)} g^{(1)}_{(kl)} +D^{(1)}_{(kl)} g^{(1)}_{(ij)}
\Big) &=&
a_2 g^{(2)}_{((ij)(kl))} + a_3 g^{(0)} (\delta_{ik} \delta_{jl} + \delta_{il}
\delta_{jk}),\eea
which leads to
\be \label{r4} \Delta g^{(0)} = \lambda_1 g^{(0)} + \lambda_2 g^{(2)}_{(ij)(ij)},\ee
where $\lambda_1 = 12 a_1 a_3$ and $\lambda_2 = a_1 a_2$. Thus, the coupling
$g^{(0)}$, which is the coefficient of the $\mathcal{R}^4$ interaction
satisfies \C{r4}, which is completely determined by the constraints of supersymmetry.

However, our analysis based on supersymmetry is not strong enough to solve
\C{r4} for $g^{(0)}$, because of the presence of the unknown source
term. Now, we simply make some plausible arguments to determine
$g^{(0)}$. We have that $\Delta g^{(0)}$ and $g^{(0)}$ are both $SL(3,\mathbb{Z})$
invariant automorphic forms, and thus so is $g^{(2)}_{(ij)(ij)}$. The tree
level amplitude which contributes to $g^{(0)}$ is known to be proportional to
$\zeta (3) e^{-2\hat\phi}$. Because $\zeta (3)$ is not factorizable, it is
plausible that $\lambda_2 =0$, or if non--zero, then $g^{(2)}_{(ij)(ij)} \sim
g^{(0)}$ itself\footnote{This factorizability has been crucial for the $D^6
  \mathcal{R}^4$ coupling in 10 dimensions~\cite{Green:2005ba}. The coupling, call it $f$, satisfies
\be \Delta f = 12 f - 6 E_{3/2}^2\ee
on the fundamental domain of $SL(2,\mathbb{Z})$. Here, the tree level
contribution $\sim \zeta(3)^2$, which also matches that from $E_{3/2}^2$,
using \C{Eisen}.}. In that case, $g^{(0)}$ satisfies the Laplace equation    
\be \label{plaus}\Delta g^{(0)} = \mu g^{(0)}.\ee
Based on the boundary conditions, $g^{(0)}$ is therefore uniquely given by an
Eisenstein series defined by \C{EisenSL3}
for some choice of $s$. From the tree level amplitude, we see that $s=3/2$ and
so $\mu =0$~\cite{Kiritsis:1997em}. However, as discussed in appendix \C{SL3aut},
$E_{3/2} (M)$ is divergent and has to be regularized.  

We should mention that the generalized derivative $D^{(n)}_{(ij)}$ has
appeared in the supervariations, and consequently, in the equations \C{seqeqn}. It is
clear from the way they appear that the value of $n$ gives the $SU(2)$ spin of the
interaction. However, since the couplings involved are not automorphic forms
of $SL(3,\mathbb{Z})$, we do not understand its role, if any, as some covariant
derivative of the U--duality group.

\subsection{A coupling involving all the moduli}

As discussed before, a coupling which involves all the moduli is the
$\mathcal{R}^4$ coupling. This is given by~\cite{Kiritsis:1997em}
\be \hat{E}_{3/2} (M) + 2 \hat{E}_1 (U,\bar{U}).\ee 
The relative coefficient between the two terms is fixed to satisfy the $U \leftrightarrow T$
symmetry of the perturbative part of the amplitude, which interchanges the
type IIA and type IIB theories.  

\subsection{Some plausible generalizations}

Although the calculations are complicated and it is difficult to fix the
coefficients, the structure of the first order differential equations that
emerge as a consequence of supersymmetry is quite simple. It
suggests that this procedure should be generalizable to lower dimensions, for
example, to $N=8,d=4$ supergravity, where the classical moduli space is
$(SU(8)/\mathbb{Z}_2) \backslash E_{7(7)} (\mathbb{R})$, and the U--duality
group is $E_{7(7)} (\mathbb{Z})$. In that case in order to construct the
relevant 16 fermion terms in the effective action, one should again use linearized
superspace. A candidate $1/2$ BPS superaction is given by
\be \label{halfbps}\int d^4 x \int d^{16} \theta f (W),\ee 
where the fermionic integral over the chiral part of the superspace is given
by
\be \label{chiralint} \int d^{16} \theta \equiv \int \epsilon_{i_1 \ldots i_8} \epsilon_{j_1
  \ldots j_8} (d \theta^{i_1 \alpha_1} \ldots d \theta^{i_8 \alpha_8} ) 
(d \theta^{j_1}_{\alpha_1} \ldots d \theta^{j_8}_{\alpha_8} ).\ee
In \C{chiralint}, $\theta_i$ is in the $8$ of $SU(8)$, and we
have used the two component chiral spinor notation. In \C{halfbps}, the chiral
superfield $W_{ijkl}$ which satisfies~\cite{Brink:1979nt,Howe:1981gz}
\be \label{dual8} W^{ijkl}= \frac{1}{4!}\epsilon^{ijklmnop} W_{mnop},\ee 
is in the $70$ of $SU(8)$, and is given by~\cite{de Wit:1982ig}
\be \label{comp8} W_{ijkl} = \phi_{ijkl} + \theta_{[i} \chi_{jkl]}+(\theta_{[i} \s^{\mu\nu} \theta_j )
  (\bar\psi^{m\lambda} \bar\s_\lambda \s_{\mu\nu} \chi_{kl]m}) +\ldots .\ee
In \C{comp8}, the unitary gauge has been used and so $\phi_{ijkl}$ which also
satisfies \C{dual8} are the $70$ scalars, and the spin $1/2$ fermions
$\chi_{ijk}$ and the gravitini $\psi_{\mu i}$ are in the $56$ and $8$ of
$SU(8)$ respectively. Thus, from \C{chiralint} one immediately obtains the
$\chi^{16}$ and $\bar\psi \chi^{15}$ interactions in $S^{(3)}$, which should
be the starting point of the analysis as we have done. Decompactifying the
degrees of freedom in the couplings to higher dimensions must produce the couplings        
in the higher dimensions as well. 

In fact we can see very schematically what structure to expect. The
interactions $\chi^{16}$ and $\bar\psi \chi^{15}$ will involve the tensor product of
$16$ and $15$ of the $70$ representations of $SU(8)$. As we have done in the
$d=8$ case, we can project onto the completely symmetric product among the
irreducible representations that arise in the tensor product. Thus the
$\chi^{16}$ and $\bar\psi \chi^{15}$ interactions are in the $715536058545$
and $313203587004$ representations of $SU(8)$ respectively, and we can carry
out the analysis. Thus given the structure of the interactions, the
generalized derivative $D_{[ijkl]}$ in the $70$ of $SU(8)$, should act as
\be 
 D_{[ijkl]} : \mathcal{A}_{[mnop]} \rightarrow \mathcal{A}_{[ijkl][mnop]} +
\mathcal{A} \delta_{im} \delta_{jn} \delta_{ko} \delta_{lp},\ee
where the symmetrization is implicit. This is enough to lead to a set of equations
like \C{seqeqn}, along with corrections of the type discussed before. Thus, we get a repitition of the structure we have for $d=8$,
and this should be true for $d=5,6,7$.

Among the theories where the classical moduli space involves finite
dimensional groups, the only other case is $N=16,d=3$, where the moduli space is
$SO(16) \backslash E_{8(8)} (\mathbb{R})$, and the U--duality group is $E_{8(8)}
(\mathbb{Z})$. Even though the graviton and the gravitini in the $16$ of
SO(16) do not have any physical degrees of freedom, this theory is non--trivial because
it is interacting. The entire degrees of freedom are contained in the $128$
scalars  and the $128$ Majorana spin $1/2$ particles~\cite{Marcus:1983hb},
where they are in the two inequivalent spinor representations of $SO(16)$. The
$\mathcal{R}^4$ interaction vanishes in $d=3$ because it only involves the
Weyl tensor which follows from perturbative string computations for maximal
supersymmetry in any dimension. It is plausible that all the interactions of
the type we have considered which result from
the supermultiplet vanish, and there is nothing to consider.

\section*{Acknowledgements}

I am very thankful to Michael B. Green for useful discussions, and for
collaboration during the initial stages of the work.

\section{Appendices}

\appendix

\section{Fierz transformations and various gamma matrix identities}

We need Fierz transformations involving 8 dimensional chiral fermions in our
calculations. They are
given by
\bea \label{fierz}
\chi_{L\alpha} \bar\lambda_{R\beta} &=& -\frac{1}{8} \delta_{\alpha\beta} (\bar\chi_R \lambda_L) - \frac{1}{16}
(\g^{\mu\nu})_{\alpha\beta} (\bar\chi_R \g_{\mu\nu} \lambda_L) - \frac{1}{192}(\g^{\mu\nu\lambda\rho})_{\alpha\beta}
(\bar\chi_R \g_{\mu\nu\lambda\rho} \lambda_L) , \non \\ \chi_{R\alpha} \bar\lambda_{L\beta} &=& -\frac{1}{8} \delta_{\alpha\beta} 
(\bar\chi_L \lambda_R) - \frac{1}{16}
(\bar\g^{\mu\nu})_{\alpha\beta} (\bar\chi_L \bar\g_{\mu\nu} \lambda_R) - \frac{1}{192}(\bar\g^{\mu\nu\lambda\rho})_{\alpha\beta}
(\bar\chi_L \bar\g_{\mu\nu\lambda\rho} \lambda_R), \non \\
\chi_{R\alpha} \bar\lambda_{R\beta} &=& -\frac{1}{8} (\bar\g^\mu)_{\alpha\beta} (\bar\lambda_R \g_\mu \chi_R) + \frac{1}{48}
 (\bar\g^{\mu\nu\lambda})_{\alpha\beta} (\bar\lambda_R \g_{\mu\nu\lambda} \chi_R), \non \\ \chi_{L\alpha} \bar\lambda_{L\beta} 
&=& -\frac{1}{8} (\g^\mu)_{\alpha\beta} (\bar\lambda_L \bar\g_\mu \chi_L) + \frac{1}{48} (\g^{\mu\nu\lambda})_{\alpha\beta}
(\bar\lambda_L \bar\g_{\mu\nu\lambda} \chi_L) ,\eea
where $\alpha , \beta = 1, \ldots, 8$ are spinor indices.

We also make use of the relations involving the gamma matrices
\bea \label{gammaalg}
&&\g^{\mu\nu\lambda\rho} \g_{\mu\nu\lambda\rho} = 1680, \quad  \g^{\mu\nu\lambda\rho} \g_{\s\tau} \g_{\mu\nu\lambda\rho} = -240 
\g_{\s\tau}, \quad  \g^{\mu\nu\lambda\rho} \g_{\s\tau\kappa\eta} \g_{\mu\nu\lambda\rho} = 144 \g_{\s\tau\kappa\eta}, \non \\&& \g^{\mu\nu\rho}
\bar\g_{\mu\nu\rho} = - 336, \quad \g^{\mu\nu\rho}  \bar\g_\lambda \g_{\mu\nu\rho} = 84 \g_\lambda, 
\quad \g^{\mu\nu\rho}  \bar\g_{\lambda\s} \bar\g_{\mu\nu\rho} = 24 \g_{\lambda\s} , \non \\ &&
\g^{\mu\nu\rho}  \bar\g_{\lambda\s\kappa} \g_{\mu\nu\rho} = -36 \g_{\lambda\s\kappa} , \quad
\g^{\mu\nu\rho}  \bar\g_{\lambda\s\kappa\eta} \bar\g_{\mu\nu\rho} = 0, \quad
\g^{\mu\nu} \g_{\mu\nu} = -56, \quad  \g^{\mu\nu} \g_\lambda \bar\g_{\mu\nu} = -28 \g_\lambda , 
\non \\ && \g^{\mu\nu} \g_{\lambda\rho} \g_{\mu\nu} = -8 \g_{\lambda\rho}, \quad  \g^{\mu\nu} \g_{\lambda\rho\s} \bar\g_{\mu\nu} = 4 
\g_{\lambda\rho\s}, \quad
\g^{\mu\nu} \g_{\lambda\rho\s\tau} \g_{\mu\nu} = 8 \g_{\lambda\rho\s\tau},  \quad\ \g^\mu \bar\g_\mu = 8, \non \\ && 
\g^\mu \bar\g_\nu \g_\mu = -6 \g_\nu, \quad 
\g^\mu \bar\g_{\nu\lambda} \bar\g_\mu = 4 \g_{\nu\lambda} , \quad \g^\mu \bar\g_{\nu\lambda\rho} \g_\mu = -2 \g_{\nu\lambda\rho} 
,\quad\g^\mu \bar\g_{\nu\lambda\s\tau} \bar\g_\mu = 0 . \eea

\section{The fermions from $d=10$}

Let us now express the fermionic degrees of freedom in terms of the
10 dimensional fermions of type IIB supergravity. In order to do so,
it is sufficient to consider the covariant derivatives in \C{covder},
and look at the $U(1)_\tau$ charges carried by the various fields,
which can be calculated using \C{1tau}. For this purpose, instead of
the constrained field $\chi^{i}_{LA}$, it is more convenient to consider
the unconstrained field $\eta_{L\hat{A}}$ ($\hat{A} = 1,2,3,4$) in the
4 of $SU(2)$. Thus, under $U(1) \backslash SL(2,\mathbb{R})_\tau$, we have that
\bea {\mathcal{D}}_\mu \psi_{\nu L} &=& D_\mu \psi_{\nu L}  + \frac{i}{2}
Q_\mu^\tau \s^1 \psi_{\nu L}  + \ldots , \non \\  {\mathcal{D}}_\mu
\lambda_L &=& D_\mu \lambda_L + \frac{i}{2}
Q_\mu^\tau \s^1 \lambda_L + \ldots ,
\non \\ \mathcal{D}_\mu
\eta_L &=& D_\mu \eta_L + \frac{i}{2} Q_\mu^{\tau} T^1 \eta_L + \ldots,\eea
where
\be T^1 = \begin{pmatrix} 0 & \sqrt{3} & 0 & 0 \\ \sqrt{3} & 
0 & 2  & 0  \\ 0 & 2 & 0 &  \sqrt{3} \\ 0 & 0 & \sqrt{3} & 0 \end{pmatrix},\ee
and
\be Q_\mu^\tau = -\frac{\p_\mu \tau_1}{2 \tau_2} .\ee
Thus, we see that the combinations 
\be \Omega_1 = \{ \psi_{\mu 1 L} - \psi_{\mu 2L} , \quad \lambda_{1L} - \lambda_{2L}, \quad
\eta_{1L} - \frac{\eta_{2L}}{\sqrt{3}} - \frac{\eta_{3L}}{\sqrt{3}} +
\eta_{4L} \}
\ee
have $U(1)_\tau$ charge $1/2$, while the combination
\be \Omega_2 = -\eta_{1L} + \sqrt{3} \eta_{2L} - \sqrt{3} \eta_{3L} + \eta_{4L} \ee
has $U(1)_\tau$ charge $3/2$. Also the combinations
\be \Omega_3 = \{ \psi_{\mu 1L} + \psi_{\mu 2L} , \quad \lambda_{1L} + \lambda_{2L}, \quad
-\eta_{1L} - \frac{\eta_{2L}}{\sqrt{3}} + \frac{\eta_{3L}}{\sqrt{3}} +
\eta_{4L} \},\ee
and 
\be \Omega_4 = \eta_{1L} + \sqrt{3} \eta_{2L} + \sqrt{3} \eta_{3L} + \eta_{4L} \ee
carry $U(1)_\tau$ charges $-1/2$ and $-3/2$ respectively.
So, $\{ \Omega_1 , \Omega_3^* \}$  descend from the $d=10$ gravitino
$\Psi_\mu$ which has $U(1)_\tau$ charge $1/2$, while $\{ \Omega_2 ,
\Omega_4^*\}$ descend from the $d=10$ dilatino $\hat\lambda$
which has $U(1)_\tau$ charge $3/2$. Of course, this
precise decomposition of the degrees of freedom depends on the choice
of gauge. However, from the gauge covariance of the equations, it
follows that the degrees of freedom in $\psi_{\mu L}$ (160),
$\lambda_L$ (32), and half of those in $\chi_L$ (32) descend from
the gravitino (224), while the remaining half of the degrees of freedom in
$\chi_L$ (32) descend from the dilatino (32). Thus the fermionic degrees
of freedom intermingle in a complicated way.     

Intuitively, one can also deduce it from the supersymmetry
transformations \C{susytran}. It follows that $\lambda_L$ (because its
supervariation involves the $U(1) \backslash SL(2,\mathbb{R})_U$
moduli), $\psi_{\mu L}$ and a part of $\chi_L$ (because its
supervariation involves the $SU(2) \backslash SL(3,\mathbb{R})$
moduli) descends from the gravitino, while the remaining part of
$\chi_L$ descends from the dilatino. The rest follows from simply
counting the degrees of freedom.  

One can also calculate the $U(1)_T$ charges of the various fields in
\C{covder} using \C{1t} in a straightforward manner.

\section{Supersymmetry transformations}

At the two derivative level, the supersymmetry transformations of the various
fields of $N=2, d=8$ supergravity are given by
\bea \label{susytran} \delta^{(0)}  e_\mu^{~a} &=& - \Big( \bar\epsilon_L
\bar\gamma^a \psi_{\mu L} + \bar\epsilon_R \gamma^a \psi_{\mu R} \Big),
\non \\ 
\delta^{(0)} L^{~U}_+  &=&  L^{~U}_- \bar\epsilon_R \lambda_L , \non \\ 
\delta^{(0)}
L^{~U}_- &=&  L^{~U}_+ \bar\epsilon_L \lambda_R , \non \\ 
L^{~m}_i \delta^{(0)} L_{mj}
&=& - \frac{1}{2} \Big[ 
\bar\epsilon_L \Big( \s_i \chi_{jR} + \s_j \chi_{iR} \Big) - 
\bar\epsilon_R \Big( \s_i \chi_{jL} + \s_j \chi_{iL} \Big)  \Big], \non \eea
\bea
\delta^{(0)}
A_\mu^{mU} &=& \sqrt{2} L^{~m}_{i} \Big[ L^{~U}_{-} \Big( \bar\epsilon_R \s^i \psi_{\mu
  L} + \frac{1}{2} \bar\epsilon_L \bar\gamma_\mu \s^i \lambda_L -
\bar\epsilon_R \gamma_\mu \chi^i_R \Big) \non \\ && - L^{~U}_{+} \Big( \bar\epsilon_L \s^i
\psi_{\mu R} + \frac{1}{2} \bar\epsilon_R \gamma_\mu \s^i \lambda_R +
\bar\epsilon_L \bar\gamma_\mu \chi^i_L \Big) \Big] , \non
\\ \delta^{(0)} B_{\mu\nu m} &=& - \epsilon_{mnp} \epsilon_{UV} \Big( \delta^{(0)}
A^{nU}_{[\mu} \Big)  A^{p V}_{\nu ] } - 2 L_m^{~i} \Big( \bar\epsilon_L \s^i
\bar\gamma_{[\mu} \psi_{\nu ]L} - \bar\epsilon_R \s^i \gamma_{[\mu}
  \psi_{\nu ]R} \Big) \non \\ &&-  L_m^{~i} \Big( \bar\epsilon_L
\bar\gamma_{\mu\nu} \chi^i_R + \bar\epsilon_R \gamma_{\mu\nu}
\chi^i_L \Big), \non \\ 
\delta^{(0)} C_{\mu\nu\rho} &=& -3 (\delta^{(0)} A^{m1}_{[\mu}) \Big( B_{\nu\rho]m} - 
\epsilon_{mnp} A^{n2}_\nu A^{p1}_{\rho ]} \Big)
-3 \sqrt{2}L^{~1}_{-} \Big(
\bar\epsilon_R \gamma_{[\mu\nu} \psi_{\rho ]L} - \frac{1}{6}
\bar\epsilon_L \bar\gamma_{\mu\nu\rho} \lambda_L \Big) \non \\ && - 3 \sqrt{2} L^{~1}_{+} 
\Big( 
\bar\epsilon_L \bar\gamma_{[ \mu\nu} \psi_{\rho ]R} -\frac{1}{6}
\bar\epsilon_R \gamma_{\mu\nu\rho} \lambda_R \Big) 
, \non \\ 
\delta^{(0)} \lambda_L &=&
 i \gamma^\mu \hat{P}_\mu \epsilon_R -\frac{i}{4\sqrt{2}} \hat{F}_{2\mu\nu}^i
 \gamma^{\mu\nu} \s^i \epsilon_L 
- \frac{i}{96\sqrt{2}} \hat{T}^-_{\mu\nu\lambda\rho} \gamma^{\mu\nu\lambda\rho} \epsilon_L  
\non \\ &&+\frac{1}{6} \s_i \g_\mu \epsilon_R (\bar\lambda_R \g^\mu \chi_{iR}) + \frac{1}{3} \g_\mu \chi_{iR} (\bar\epsilon_L \bar\g_\mu \s_i \lambda_L)
-\frac{1}{12} \s_i \lambda_L (\bar\epsilon_L \chi_{iR})  \non \\  &&- \frac{1}{2}
\s_i \epsilon_L (\bar\lambda_R \chi_{iL}) -\frac{1}{24} \g^{\mu\nu} \s_i \epsilon_L  (\bar\chi_{iR} \g_{\mu\nu} \lambda_L)
+ \frac{1}{3} \chi_{iL} (\bar\epsilon_R \s_i \lambda_L) + \frac{1}{12} \s_i \lambda_L (\bar\epsilon_R \chi_{iL})  \non \\ 
&&+\frac{1}{8} \epsilon_L (\bar\chi_{iL} \chi_{iR})  - \frac{1}{16} \g^{\mu\nu} \s_i \epsilon_L (\bar\chi_{jL} 
\bar\g_{\mu\nu} \s_i \chi_{jR}) - \g^\mu \chi_{iR} (\bar\epsilon_R \g_\mu \chi_{iR} )  \non , \\
\delta^{(0)} \lambda_R &=&
- i \bar\gamma^\mu \hat{P}_\mu^* \epsilon_L - \frac{i}{4\sqrt{2}}
\hat{F}_{2\mu\nu}^{* i}
 \bar\gamma^{\mu\nu} \s^i \epsilon_R  
+\frac{i}{96\sqrt{2}} \hat{T}^+_{\mu\nu\lambda\rho} \bar\gamma^{\mu\nu\lambda\rho}
\epsilon_R \non \\ &&-\frac{1}{6} \s_i \bar\g_\mu \epsilon_L (\bar\lambda_L \bar\g^\mu \chi_{iL}) - \frac{1}{3} 
\bar\g_\mu \chi_{iL} (\bar\epsilon_R \g_\mu \s_i \lambda_R)
+\frac{1}{12} \s_i \lambda_R (\bar\epsilon_R \chi_{iL})  \non \\  && + \frac{1}{2}
\s_i \epsilon_R (\bar\lambda_L \chi_{iR}) +\frac{1}{24} \bar\g^{\mu\nu} \s_i \epsilon_R  (\bar\chi_{iL} \bar\g_{\mu\nu} 
\lambda_R) - \frac{1}{3} \chi_{iR} (\bar\epsilon_L \s_i \lambda_R) - \frac{1}{12} \s_i \lambda_R (\bar\epsilon_L 
\chi_{iR})  \non \\  &&+\frac{1}{8} \epsilon_R (\bar\chi_{iR} \chi_{iL})  - \frac{1}{16} \bar\g^{\mu\nu} \s_i \epsilon_R 
(\bar\chi_{jR}  \g_{\mu\nu} \s_i \chi_{jL}) - \bar\g^\mu \chi_{iL} (\bar\epsilon_L \bar\g_\mu \chi_{iL} )  \non , \\
\delta^{(0)} \chi^i_L &=&
-\frac{1}{2} \gamma^\mu \hat{P}_\mu^{ij} \s_j \epsilon_R - \frac{i}{4\sqrt{2}} \hat{F}^{*j}_{2\mu\nu} \Delta_{ij}
\gamma^{\mu\nu} \epsilon_L  - \frac{1}{24} \hat{F}_{3\mu\nu\lambda j} \Delta_{ij} 
\gamma^{\mu\nu\lambda} \epsilon_R 
-\frac{1}{2} \Delta_{ij} \epsilon_L (\bar\lambda_L \chi_{jR}) \non \\
&&+ \frac{\Delta_{ij}}{2} \Big[ \g^\mu \lambda_R 
(\bar\epsilon_R \g_\mu \chi_{jR})  + \frac{1}{3} \lambda_L (\bar\epsilon_L \s_j 
\lambda_R) - \frac{1}{12} \g^{\mu\nu}\lambda_L (\bar\epsilon_L
\bar\g_{\mu\nu}\s_j \lambda_R) \Big] \non 
\\ && 
 -\frac{1}{16} \Delta_{ij} \g^{\mu\nu} \epsilon_L (\bar\chi_{kR} \g_{\mu\nu} \s_j\chi_{kL})
+ \frac{1}{2} \Delta_{ij} \s_k \epsilon_L 
(\bar\chi_{jR} \chi_{kL}) - \Delta_{ij} \chi_{kL} (\bar\epsilon_R \s_{(j}
\chi_{k)L}) \non \\
&& - \frac{1}{4} \Delta_{ij} \s_k \chi_{jL} (\bar\epsilon_R \chi_{kL}) 
-\frac{1}{2} \Delta_{ij} \g_\mu \s_k \epsilon_R (\bar\chi_{R[j} \g^\mu \chi_{k]R}) + \Delta_{ij} \chi_{kL} (\bar\epsilon_L \s_{(j}
\chi_{k)R})  \non \\
&& + \frac{1}{2} \Delta_{ij} \g^\mu \s_k \chi_{jR} (\bar\epsilon_L \bar\g_\mu \chi_{kL} ) + 
\frac{1}{4} \Delta_{ij} \s_k \chi_{jL} (\bar\epsilon_L \chi_{kR}) , \non  \\
\delta^{(0)} \chi^i_R &=&
\frac{1}{2} \bar\gamma^\mu \hat{P}_\mu^{ij} \s_j \epsilon_L +\frac{i}{4\sqrt{2}} \hat{F}^j_{2\mu\nu} \Delta_{ij}
\bar\gamma^{\mu\nu} \epsilon_R  - \frac{1}{24} \hat{F}_{3\mu\nu\lambda j} 
\Delta_{ij} \bar\gamma^{\mu\nu\lambda} \epsilon_L 
-\frac{1}{2} \Delta_{ij} \epsilon_R (\bar\lambda_R \chi_{jL}) \non \\ &&+ \frac{\Delta_{ij}}{2} \Big[ \bar\g^\mu \lambda_L 
(\bar\epsilon_L \bar\g_\mu \chi_{jL})  - \frac{1}{3} \lambda_R (\bar\epsilon_R \s_j 
\lambda_L) + \frac{1}{12} \bar\g^{\mu\nu}\lambda_R (\bar\epsilon_R \g_{\mu\nu}\s_j \lambda_L) \Big]
 \non \\ && 
 +\frac{1}{16} \Delta_{ij} \bar\g^{\mu\nu} \epsilon_R (\bar\chi_{kL} \bar\g_{\mu\nu} \s_j\chi_{kR})
- \frac{1}{2} \Delta_{ij} \s_k \epsilon_R 
(\bar\chi_{jL} \chi_{kR}) + \Delta_{ij} \chi_{kR} (\bar\epsilon_L \s_{(j}
\chi_{k)R}) \non \eea
\bea
&& + \frac{1}{4} \Delta_{ij} \s_k \chi_{jR} (\bar\epsilon_L \chi_{kR}) 
+\frac{1}{2} \Delta_{ij} \bar\g_\mu \s_k \epsilon_L (\bar\chi_{L[j} \bar\g^\mu \chi_{k]L}) - \Delta_{ij} \chi_{kR} 
(\bar\epsilon_R \s_{(j} \chi_{k)L})  \non \\ && - \frac{1}{2} \Delta_{ij} \bar\g^\mu \s_k \chi_{jL} 
(\bar\epsilon_R \g_\mu \chi_{kR} ) - 
\frac{1}{4} \Delta_{ij} \s_k \chi_{jR} (\bar\epsilon_R \chi_{kL}) , \non  \\
\delta^{(0)}
\psi_{\mu L} &=&  \mathcal{D}_\mu \epsilon_L +\frac{i}{24\sqrt{2}} \hat{F}_2^{i\nu\lambda} \s^i (\gamma_{\mu\nu\lambda} -10
g_{\mu\nu} \gamma_\lambda )\epsilon_R  
+ \frac{1}{72} \hat{F}^{\nu\lambda\rho i}_3 \s^i (\gamma_{\mu\nu\lambda\rho} -6
g_{\mu\nu} \gamma_{\lambda\rho} ) \epsilon_L \non \\ && 
-\frac{i}{192\sqrt{2}} \hat{T}^-_{\nu\lambda\rho\s} \gamma^{\nu\lambda\rho\s} 
\gamma_\mu \epsilon_R  - \frac{1}{18} (8 g_{\mu\nu} - \g_{\mu\nu} ) \Big[ \s_i \lambda_L (\bar\epsilon_L \bar\g^\nu \chi_{iL})
+ \chi_{iL} (\bar\epsilon_L \bar\g^\nu \s_i \lambda_L)\Big]   \non \\ && +\frac{1}{72} \Big[ \g^\nu (\bar\chi_{iR}
\g_{\mu\nu} \lambda_L) + \frac{1}{2} \g_{\mu\nu\rho} (\bar\chi_{iR} \g^{\nu\rho} \lambda_L) \Big] \s_i \epsilon_R
\non \\ && + \frac{1}{4} \s_i \Big[ \epsilon_L (\bar\psi_{\mu R} \chi_{iL} - 
\bar\psi_{\mu L} \chi_{iR}) - \psi_{\mu L} (\bar\epsilon_R \chi_{iL} - \bar\epsilon_L \chi_{iR}) \Big]
- \chi_{iL} (\bar\epsilon_R \s_i \psi_{\mu L}) \non \\ &&  -\frac{1}{2} \s_i \g^\nu \Big[ \epsilon_R (\bar\psi_{\mu R} 
\g_\nu \chi_{iR}) - \psi_{\mu R} (\bar\epsilon_R \g_\nu \chi_{iR})\Big]   
+ \frac{1}{6} \Big[ (\bar\epsilon_L \s_i \chi_{jR}) \s_i + (\bar\epsilon_L \chi_{jR}) \Big] \g_\mu \chi_{jR} \non \\ &&
+ \frac{1}{48} \Big[ -5
\g^\nu (\bar\chi_{jL} \s_i \bar\g_{\mu\nu} \chi_{jR}) + \frac{1}{2} \g_{\mu\nu\rho} (\bar\chi_{jL} \bar\g^{\nu\rho} \s_i 
\chi_{jR} )\Big] \s_i \epsilon_R  - \frac{1}{48} \g_\mu \epsilon_R (\bar\chi_{iL} \chi_{iR})\non \\ && + \frac{1}{6} (5 g_{\mu\nu} -
\g_{\mu\nu}) \chi_{iL} (\bar\epsilon_R \g^\nu \chi_{iR}) + \frac{1}{6} \Big[ (\bar\chi_{jR} \g_\mu \s_i \chi_{jR}) -\frac{1}{2}
\g_{\mu\nu} (\bar\chi_{jR} \g^\nu \s_i \chi_{jR})\Big] \s_i \epsilon_L \non \\ && - \frac{1}{6} \Big[ \s_i (\bar\epsilon_R
\s_i \chi_{jL}) + (\bar\epsilon_R \chi_{jL})\Big] \g_\mu \chi_{jR} + \frac{1}{4} \epsilon_L (\bar\chi_{iR} \g_\mu \chi_{iR})
+\frac{1}{27} \g_\mu \s_i \lambda_R (\bar\epsilon_R \s_i \lambda_L)\non \\
&& - \frac{1}{54} (11 g_{\mu\nu} - \g_{\mu\nu}) \s_i \lambda_L (\bar\epsilon_R \s_i \g^\nu \lambda_R) +\frac{1}{54} \Big[ 5
(\bar\lambda_R \g_\mu \s_i \lambda_R) - \g_{\mu\nu} (\bar\lambda_R \g^\nu \s_i \lambda_R)\Big] \s_i \epsilon_L\non \\ && 
- \frac{1}{12} \Big[ (\bar\lambda_R \g_\mu \lambda_R) - \frac{1}{12}\g^{\nu\rho} (\bar\lambda_R \g_{\mu\nu\rho} \lambda_R)
\Big] \epsilon_L + \frac{1}{12} \Big[ g_{\mu\nu} (\bar\epsilon_R \lambda_L) - \frac{1}{3} (\bar\epsilon_R
\g_{\mu\nu} \lambda_L)\Big] \g^\nu \lambda_R \non \\ &&-\frac{1}{6} \g_\nu \s_i \Big[ \psi_{\mu R} (\bar\epsilon_L \bar\g^\nu \s_i \lambda_L) - 
\epsilon_R (\bar\psi_{\mu L} \bar\g^\nu \s_i \lambda_L)\Big] 
- \frac{1}{3} \s_i \lambda_L (\bar\epsilon_L \s_i \psi_{\mu R}) \non \\
\delta^{(0)}
\psi_{\mu R} &=& \mathcal{D}_\mu \epsilon_R + \frac{i}{24\sqrt{2}} \hat{F}^{*i\nu\lambda}_2 \s^i (\bar\gamma_{\mu\nu\lambda} 
-10 g_{\mu\nu} \bar\gamma_\lambda) \epsilon_L  
-\frac{1}{72}  \hat{F}^{\nu\lambda\rho i}_3 \s^i (\bar\gamma_{\mu\nu\lambda\rho} -6
g_{\mu\nu} \bar\gamma_{\lambda\rho} ) \epsilon_R \non \\ &&+ \frac{i}{192\sqrt{2}} 
\hat{T}^+_{\nu\lambda\rho\s} \bar\gamma^{\nu\lambda\rho\s} 
\bar\gamma_\mu \epsilon_L + \frac{1}{18} (8 g_{\mu\nu} - \bar\g_{\mu\nu} ) \Big[ \s_i \lambda_R (\bar\epsilon_R 
\g^\nu \chi_{iR}) + \chi_{iR} (\bar\epsilon_R \g^\nu \s_i \lambda_R)\Big]
\non \\ && -\frac{1}{72} \Big[ \bar\g^\nu 
(\bar\chi_{iL} \bar\g_{\mu\nu} \lambda_R) + \frac{1}{2} \bar\g_{\mu\nu\rho} (\bar\chi_{iL} \bar\g^{\nu\rho} \lambda_R) \Big] 
\s_i \epsilon_L
\non \\ && - \frac{1}{4} \s_i \Big[ \epsilon_R (\bar\psi_{\mu L} \chi_{iR} - 
\bar\psi_{\mu R} \chi_{iL}) - \psi_{\mu R} (\bar\epsilon_L \chi_{iR} - \bar\epsilon_R \chi_{iL}) \Big]
+  \chi_{iR} (\bar\epsilon_L \s_i \psi_{\mu R}) \non \\ &&  +\frac{1}{2} \s_i \bar\g^\nu \Big[ \epsilon_L (\bar\psi_{\mu L} 
\bar\g_\nu \chi_{iL}) - \psi_{\mu L} (\bar\epsilon_L \bar\g_\nu \chi_{iL})\Big]   
+ \frac{1}{6} \Big[ (\bar\epsilon_R \s_i \chi_{jL}) \s_i + (\bar\epsilon_R
  \chi_{jL}) \Big] \bar\g_\mu \chi_{jL} \non \\&&
+ \frac{1}{48} \Big[ -5
\bar\g^\nu (\bar\chi_{jR} \s_i \g_{\mu\nu} \chi_{jL}) + \frac{1}{2} \bar\g_{\mu\nu\rho} (\bar\chi_{jR} \g^{\nu\rho} \s_i 
\chi_{jL} )\Big] \s_i \epsilon_L  - \frac{1}{48} \bar\g_\mu \epsilon_L (\bar\chi_{iR} \chi_{iL})\non \\
&& + \frac{1}{6} (5 g_{\mu\nu} - \bar\g_{\mu\nu}) \chi_{iR} (\bar\epsilon_L \bar\g^\nu \chi_{iL}) 
+ \frac{1}{6} \Big[ (\bar\chi_{jL} \bar\g_\mu \s_i \chi_{jL}) -\frac{1}{2}
\bar\g_{\mu\nu} (\bar\chi_{jL} \bar\g^\nu \s_i \chi_{jL})\Big] \s_i \epsilon_R \non \\
&& - \frac{1}{6} \Big[ \s_i (\bar\epsilon_L
\s_i \chi_{jR}) + (\bar\epsilon_L \chi_{jR})\Big] \bar\g_\mu \chi_{jL} + \frac{1}{4} \epsilon_R (\bar\chi_{iL} \bar\g_\mu \chi_{iL})
+\frac{1}{27} \bar\g_\mu \s_i \lambda_L (\bar\epsilon_L \s_i \lambda_R)
\non \eea \bea
&& - \frac{1}{54} (11 g_{\mu\nu} - \bar\g_{\mu\nu}) \s_i \lambda_R (\bar\epsilon_L \s_i \bar\g^\nu \lambda_L) +\frac{1}{54} \Big[ 5
(\bar\lambda_L \bar\g_\mu \s_i \lambda_L) - \bar\g_{\mu\nu} (\bar\lambda_L \bar\g^\nu \s_i \lambda_L)\Big] \s_i \epsilon_R
\non \\ && 
- \frac{1}{12} \Big[ (\bar\lambda_L \bar\g_\mu \lambda_L) - \frac{1}{12}\bar\g^{\nu\rho} (\bar\lambda_L \bar\g_{\mu\nu\rho} 
\lambda_L)
\Big] \epsilon_R + \frac{1}{12} \Big[ g_{\mu\nu} (\bar\epsilon_L \lambda_R) - \frac{1}{3} (\bar\epsilon_L
\bar\g_{\mu\nu} \lambda_R)\Big] \bar\g^\nu \lambda_L \non \\ &&-\frac{1}{6} \bar\g_\nu \s_i \Big[ \psi_{\mu L} 
(\bar\epsilon_R \g^\nu \s_i \lambda_R) - 
\epsilon_L (\bar\psi_{\mu R} \g^\nu \s_i \lambda_R)\Big] 
- \frac{1}{3} \s_i \lambda_R (\bar\epsilon_R \s_i \psi_{\mu L})
,\eea
where 
\be \Delta_{ij} = \delta_{ij} - \frac{1}{3} \s_i \s_j ,\ee
and the various supercovariant expressions are
defined below. We also have that
\bea \mathcal{D}_\mu \epsilon_L &=& D_\mu (\hat\omega) \epsilon_L - \frac{i}{2} Q_\mu \epsilon_L - \frac{i}{2}
A_\mu^i \s^i \epsilon_L , \non \\ \mathcal{D}_\mu \epsilon_R &=& D_\mu (\hat\omega) \epsilon_R 
+ \frac{i}{2} Q_\mu \epsilon_R - \frac{i}{2} A_\mu^i \s^i \epsilon_R ,
\eea
where
\be \label{defspin} D_\mu (\hat\omega) \epsilon = \Big( \p_\mu + \frac{1}{4} \hat\omega_\mu^{~ab} \Gamma_{ab} \Big)\epsilon . \ee
In \C{defspin}, $\hat\omega_\mu^{~ab}$ is the $d=8$ supercovariant spin connection given by
\be \hat\omega_\mu^{ab} = \omega_\mu^{~ab} - \Big( \bar\psi_{\mu R}\g^{[a} \psi^{b]}_R + \bar\psi_{\mu L} \bar\g^{[a}
\psi^{b]}_L + \bar\psi_R^{[a} \g_\mu \psi^{b]}_R \Big) .\ee
Note that in \C{susytran}, the spacetime structure of the $\hat{T}^\pm$ terms in the supervariation
of the gravitini matches that in~\cite{Salam:1984ft} on using the relation
\be \Gamma_{\mu\nu\lambda\rho\s} = \Gamma_{\nu\lambda\rho\s} \Gamma_\mu + g_{\mu\nu} \Gamma_{\lambda\rho\s} - g_{\mu\lambda}
\Gamma_{\nu\rho\s} + g_{\mu\rho} \Gamma_{\nu\lambda\s} - g_{\mu\s} \Gamma_{\nu\lambda\rho} .\ee

At various places, we have used the identities
\bea \label{varrel}&& \bar\psi_R\chi_L = \bar\chi_R\psi_L, \quad \bar\psi_L \chi_R = \bar\chi_L \psi_R, \non \\ 
&&\bar\psi_{R} \gamma_\mu \chi_R = -\bar\chi_L
\bar\gamma_\mu \psi_{L},  
\non \\ &&\bar\chi_{R} \s_i \psi_L = - \bar\psi_R
\s_i \chi_{L},  
\quad \bar\chi_L \s_i \psi_{ R} = - \bar\psi_{L} \s_i
\chi_R ,  
\non \\&&
\bar\psi_L \bar\gamma_\mu \s_i \chi_L = \bar\chi_R
\gamma_\mu \s_i \psi_R,   
\non \\  &&
\bar\psi_R  \gamma_{\mu\nu} \chi_L = -\bar\chi_R
\gamma_{\mu\nu} \psi_L, \quad \bar\psi_L  \bar\gamma_{\mu\nu} \chi_R = -\bar\chi_L
\bar\gamma_{\mu\nu} \psi_R ,     
\non \\ && \bar\psi_L
\bar\gamma_{\mu\nu\rho} \chi_L = \bar\chi_R \gamma_{\mu\nu\rho}
\psi_R, \non \\
&& \bar\psi_R \s^i \g_{\mu\nu} \chi_L = \bar\chi_R \s^i \g_{\mu\nu} \psi_ L, \quad \bar\psi_L \s^i \bar\g_{\mu\nu} 
 \chi_R = \bar\chi_L \s^i \bar\g_{\mu\nu} \psi_R , \non \\ && \bar\psi_L
\bar\gamma_{\mu\nu\rho} \s^i \chi_L = - \bar\chi_R \gamma_{\mu\nu\rho}
\s^i \psi_R, \non \\ && \bar\psi_R \g_{\mu\nu\lambda\rho} \chi_L = \bar\chi_R \g_{\mu\nu\lambda\rho} \psi_L, \quad 
\bar\psi_L \bar\g_{\mu\nu\lambda\rho} \chi_R = \bar\chi_L \bar\g_{\mu\nu\lambda\rho} \psi_R ,\non \\ &&
\bar\psi_L \bar\g_\lambda \bar\g_{\mu\nu\rho\s} \chi_L = - \bar\chi_R \g_{\mu\nu\rho\s} \g_\lambda \psi_R ,\non \\ && 
\bar\psi_R \s_i \g_{\mu\nu\rho\s} \chi_L = - \bar\chi_R \g_{\mu\nu\rho\s} \s_i \psi_L , \quad
\bar\psi_L \s_i \bar\g_{\mu\nu\rho\s} \chi_R = - \bar\chi_L \bar\g_{\mu\nu\rho\s} \s_i \psi_R ,\eea
where $\psi$ and $\chi$ are arbitrary $SU(2)$ pseudo--Majorana fermions in a chiral basis.
We also make use of the relations
\bea && (\bar\psi_L \chi_R)^\dagger = \bar\psi_R \chi_L , \quad  (\bar\psi_L \bar\g_\mu \chi_L)^\dagger = 
\bar\psi_R \g_\mu \chi_R , \quad  (\bar\psi_L \s_i \chi_R)^\dagger = -\bar\psi_R \s_i \chi_L , \non \\ 
 && (\bar\psi_L
\bar\g_{\mu\nu} \chi_R)^\dagger = \bar\psi_R \g_{\mu\nu} \chi_L, \quad
(\bar\psi_L \s_i \bar\g_\mu \chi_L)^\dagger =
- \bar\psi_R \g_\mu \s_i \chi_R , \non \\ && (\bar\psi_L \bar\g_{\mu\nu\rho} \chi_L)^\dagger =\bar\psi_R \g_{\mu\nu\rho} \chi_R ,
\quad (\bar\psi_L \bar\g_{\mu\nu} \s_i \chi_R)^\dagger = - \bar\psi_R \g_{\mu\nu} \s_i \chi_L . \eea

In obtaining the supersymmetry transformations, it is useful to note that the kinetic term for $F^{mU}_{2\mu\nu}$ can 
be also written as
\be F^i_{2\mu\nu} F^{*i \mu\nu}_2 = \frac{1}{2} M_{mn} M_{UV} F_{2\mu\nu}^{mU} F^{nV \mu\nu}_2 ,\ee
on using
\be M_{UP} = \frac{1}{2} (\epsilon_{UV} \epsilon_{PQ} + \epsilon_{UQ} \epsilon_{PV}) (L^V_{~+} L^Q_{~-} + L^Q_{~+} L^V_{~-}) .\ee

Let us outline in brief how to fix the coefficients of the various terms involving the supercovariant expressions
in the supersymmetry transformations 
of $\lambda_L$ in \C{susytran}. The same method works for the other fermions as well. 
We use the relations
\bea \bar\gamma^{\mu\nu\rho} \p_\mu F_{2\nu\rho}^{mU} &=& 0 , \non \\ \bar\gamma^{\mu\nu\lambda\rho} \p_\mu F_{3 \nu\lambda\rho m} &=& 
\frac{3}{4} \epsilon_{mnp} \epsilon_{UV} \bar\gamma^{\mu\nu\lambda\rho} F^{nU}_{2\mu\nu} F^{pV}_{2\lambda\rho} , \non \\
\bar\gamma^{\mu\nu\lambda\rho\s} \p_\mu F_{4\nu\lambda\rho\s} &=& 2 \bar\gamma^{\mu\nu\lambda\rho\s} 
F_{2\mu\nu}^{m1} F_{3\lambda\rho\s m} ,\eea
resulting from the Bianchi identities. They lead to the relations
\bea \bar\gamma^\mu \gamma^{\lambda \rho} \p_\mu F^i_{2\lambda \rho} &=& 
2 \bar\gamma^\nu \p^\mu F^i_{2\mu\nu} +\ldots , \non \\ \bar\gamma^\mu \gamma^{\nu\lambda\rho} \p_\mu F_{3\nu\lambda\rho i} &=&
3 \bar\gamma^{\nu\lambda} \p^\mu F_{3\mu\nu\lambda i} + \ldots , \non \\ 
\bar\gamma^\mu \gamma^{\nu\lambda\rho\s} \p_\mu F_{4\nu\lambda\rho\s} &=& 4 \bar\gamma^{\nu\lambda\rho} \p^\mu 
F_{4\mu\nu\lambda\rho} + \ldots , \eea
which are needed in cancelling the supervariation of the fermion kinetic terms against the supervariation of the kinetic 
terms of the various gauge potentials, on using \C{varrel}. 
Of course, all these terms directly follow from~\cite{Salam:1984ft}, and the above statements provide a cross check of the 
calculations directly in $d=8$. 

In order to derive the form of the various supercovariant expressions in \C{susytran}, such that all
the $\p_\mu \epsilon$ terms vanish in the supervariation, we note that
\bea \delta^{(0)} P_\mu &=& i (\p_\mu \bar\epsilon_R ) \lambda_L+ \ldots, \non
\\ \delta^{(0)} P_{\mu ij} &=& -\frac{1}{2} \Big( (\p_\mu \bar\epsilon_L)
(\s_i \chi_{jR} + \s_j \chi_{iR}) - (\p_\mu \bar\epsilon_R) (\s_i \chi_{jL} +
\s_j \chi_{iL} ) \Big) + \ldots, \non \\ \delta^{(0)}
F_{2\mu\nu}^i &=& 2\sqrt{2} i \Big( (\p_{[\mu} \bar\epsilon_R) \s^i \psi_{\nu ]
L} + \frac{1}{2} (\p_{[\mu} \bar\epsilon_L) \bar\gamma_{\nu ]} \s^i
\lambda_L - (\p_{[\mu} \bar\epsilon_R ) \gamma_{\nu ]} \chi^i_R \Big)
+ \ldots, \non \\ \delta^{(0)} F_{3\mu\nu\rho i} &=& - 6 \Big[ (\p_{[\mu}
  \bar\epsilon_L) \s_i \bar\gamma_\nu \psi_{\rho ]L} - (\p_{[\mu}
  \bar\epsilon_R) \s_i \gamma_\nu \psi_{\rho ]R} \non \\ &&+ \frac{1}{2}
\Big( (\p_{[\mu} \bar\epsilon_L) \bar\gamma_{\nu\rho ]} \chi_{iR} + (\p_{[\mu}
  \bar\epsilon_R) \gamma_{\nu\rho ]} \chi_{iL} \Big) \Big]+ \ldots,
\non \\ \delta^{(0)} F_{4\mu\nu\lambda\rho} &=& -12\sqrt{2} L^{~1}_{-} 
\Big( (\p_{[\mu} \bar\epsilon_R ) \gamma_{\nu\lambda} \psi_{\rho ]L}
- \frac{1}{6} (\p_{[\mu} \bar\epsilon_L) \bar\gamma_{\nu\lambda\rho ]} \lambda_L
  \Big) \non \\ && - 12\sqrt{2}  L^{~1}_{+}
\Big(  (\p_{[\mu} \bar\epsilon_L ) \bar\gamma_{\nu\lambda} \psi_{\rho ]R}
- \frac{1}{6} (\p_{[\mu} \bar\epsilon_R) \gamma_{\nu\lambda\rho ]} \lambda_R
 \Big) + \ldots
. \eea
Thus, given the gravitino supervariation in \C{susytran}, this leads to the
supercovariant expressions given by
\bea \label{manysupcov} \hat{P}_\mu &=& P_\mu -i \bar\psi_{\mu R} \lambda_L 
, \non \\
\hat{P}_\mu^* &=& P_\mu^* + i \bar\psi_{\mu L} \lambda_R , \non \\  
\hat{P}_{\mu ij} &=& P_{\mu ij} + \Big( \bar\psi_{\mu L}
\s_{(i} \chi_{j)R} - \bar\psi_{\mu R} \s_{(i} \chi_{j)L}  
\Big) 
, \non \\ \hat{F}_{2\mu\nu}^i &=&
F_{2\mu\nu}^i -{\sqrt{2}} i \Big[ \bar\psi_{R[\mu} \s^i \psi_{\nu ]
L} + \bar\psi_{L[\mu} \bar\gamma_{\nu ]} \s^i
\lambda_L -2\bar\psi_{R[\mu}  \gamma_{\nu ]} \chi^i_R \Big]  
, \non \\ \hat{F}_{2\mu\nu}^{i*} &=&
F_{2\mu\nu}^{i*} -\sqrt{2} i \Big[ \bar\psi_{L[\mu} \s^i \psi_{\nu ]
R} + \bar\psi_{R[\mu} \gamma_{\nu ]} \s^i
\lambda_R +2\bar\psi_{L[\mu}  \bar\gamma_{\nu ]} \chi^i_L \Big] , \non \\
\hat{F}_{3\mu\nu\rho i} &=& F_{3\mu\nu\rho i} + 3 \Big( 2 \bar\psi_{L[\mu}
 \s_i \bar\gamma_\nu \psi_{\rho ]L}  + 
\bar\psi_{L[\mu}  \bar\gamma_{\nu\rho ]} \chi_{iR} + \bar\psi_{R[\mu}
  \gamma_{\nu\rho ]} \chi_{iL}  \Big) \non \\ && 
+ \frac{1}{6}
\bar\lambda_R \gamma_{\mu\nu\rho} \s_i \lambda_R , \non \\
\hat{F}_{4\mu\nu\lambda\rho} &=& 
F_{4\mu\nu\lambda\rho} + 6\sqrt{2} \Big[ L^{~1}_{-}
\Big(  \bar\psi_{R[\mu} \gamma_{\nu\lambda} \psi_{\rho ]L}
- \frac{1}{3} \bar\psi_{L[\mu} \bar\gamma_{\nu\lambda\rho ]} \lambda_L \Big)  
\non \\ && 
+ L^{~1}_{+}
\Big( \bar\psi_{L[\mu} \bar\gamma_{\nu\lambda} \psi_{\rho ]R}
- \frac{1}{3} \bar\psi_{R[\mu} \gamma_{\nu\lambda\rho ]} \lambda_R \Big) 
\Big] 
\non \\ && =  \hat{F}_{4\mu\nu\lambda\rho}^+ +  \hat{F}_{4\mu\nu\lambda\rho}^- ,\eea
where
\bea \label{manysupcov2}
\hat{F}_{4\mu\nu\lambda\rho}^+ &=& F^+_{\mu\nu\lambda\rho} + \frac{1}{2\sqrt{2}} \Big[ L^{~1}_- 
\Big( \bar\psi_{\s R} \g^{[\s} \bar\g_{\mu\nu\lambda\rho} \bar\g^{\tau]} \psi_{\tau L} + \bar\psi_{\s L}  \bar\g_{\mu\nu\lambda\rho} \bar\g^\s \lambda_L
 \Big) \non \\ &&- L^{~1}_+ \Big( \bar\psi_{\s L}  \bar\g_{\mu\nu\lambda\rho} \bar\g^{[\s} \g^{\tau ]} \psi_{\tau R} 
+  \bar\psi_{\s R}  \g^\s \bar\g_{\mu\nu\lambda\rho}  \lambda_R \Big) 
\Big], \non \eea
\bea \hat{F}_{4\mu\nu\lambda\rho}^- &=& 
F^-_{\mu\nu\lambda\rho} + \frac{1}{2\sqrt{2}} \Big[ L^{~1}_+ 
\Big( \bar\psi_{\s L} \bar\g^{[\s} \g_{\mu\nu\lambda\rho} \g^{\tau]} \psi_{\tau R} + \bar\psi_{\s R}  \g_{\mu\nu\lambda\rho} \g^\s \lambda_R
 \Big) \non \\ && - L^{~1}_- \Big( \bar\psi_{\s R}  \g_{\mu\nu\lambda\rho} \g^{[\s} \bar\g^{\tau ]} \psi_{\tau L} 
+  \bar\psi_{\s L}  \bar\g^\s \g_{\mu\nu\lambda\rho}  \lambda_L \Big) 
\Big]. \eea

In obtaining \C{manysupcov2} from \C{manysupcov}, we have used the identities
\bea \bar\psi_{R[\mu} \g_{\nu\rho\s ]} \lambda_R &=& \frac{1}{8} \Big( \bar\psi^\lambda_R \g_\lambda \bar\g_{\mu\nu\rho\s} \lambda_R 
- \bar\psi^\lambda_R \g_{\mu\nu\rho\s} \g_\lambda \lambda_R  \Big) , \non \\  \bar\psi_{L[\mu} \bar\g_{\nu\rho\s ]} \lambda_L 
&=& \frac{1}{8} \Big( \bar\psi^\lambda_L \bar\g_\lambda \g_{\mu\nu\rho\s} \lambda_L 
- \bar\psi^\lambda_L \bar\g_{\mu\nu\rho\s} \bar\g_\lambda \lambda_L \Big) , \non \\ \bar\psi_{R[\mu} 
\g_{\nu\rho} \psi_{\s ]L} &=& \frac{1}{24} \Big(  \bar\psi_{R\lambda} \g^{[\lambda}  \bar\g_{\mu\nu\rho\s} \bar\g^{\tau]} \psi_{\tau L} 
- \bar\psi_{R\lambda} \g_{\mu\nu\rho\s} \g^{[\lambda} \bar\g^{\tau ]} \psi_{\tau L} \Big) , \non \\ 
\bar\psi_{L[\mu} \bar\g_{\nu\rho} \psi_{\s ]R} &=& \frac{1}{24} \Big(  \bar\psi_{L\lambda} \bar\g^{[\lambda}  \g_{\mu\nu\rho\s} \g^{\tau]} \psi_{\tau R} 
- \bar\psi_{L\lambda} \bar\g_{\mu\nu\rho\s} \bar\g^{[\lambda} \g^{\tau ]} \psi_{\tau R} \Big)  \eea
to decompose $\hat{F}_4$ into $\hat{F}_4^\pm$.

Thus, we have that 
\bea \hat{F}^+_{\mu\nu\rho\s} &=& -i L^{~1}_+ \hat{T}^+_{\mu\nu\rho\s} + \frac{L^{~1}_-}{2\sqrt{2}} \bar\chi^i_L \bar\g_{\mu\nu\rho\s} \chi^i_R
, \non \\ \hat{F}^-_{\mu\nu\rho\s} &=& i L^{~1}_- \hat{T}^-_{\mu\nu\rho\s} + \frac{L^{~1}_+}{2\sqrt{2}} \bar\chi^i_R \g_{\mu\nu\rho\s} \chi^i_L 
,\eea
where
\bea \hat{T}^+_{\mu\nu\rho\s} = T^+_{\mu\nu\rho\s} - \frac{i}{2\sqrt{2}} \Big( \bar\psi_{\lambda L}  \bar\g_{\mu\nu\rho\s} 
\bar\g^{[\lambda} \g^{\tau ]} \psi_{\tau R} 
+  \bar\psi_{\lambda R}  \g^\lambda \bar\g_{\mu\nu\rho\s}  \lambda_R \Big) , \non \\ 
\hat{T}^-_{\mu\nu\rho\s} = T^-_{\mu\nu\rho\s} + \frac{i}{2\sqrt{2}} \Big( \bar\psi_{\lambda R}  \g_{\mu\nu\rho\s} 
\g^{[\lambda} \bar\g^{\tau ]} \psi_{\tau L} 
+  \bar\psi_{\lambda L}  \bar\g^\lambda \g_{\mu\nu\rho\s}  \lambda_L \Big)  \eea
are supercovariant field strengths.

Of course, the term not involving the gravitino in \C{manysupcov}  is 
not fixed by this argument and one need not include it 
in the definition. We have, however, included it because the supersymmetry transformations look simpler\footnote{A
previous example
of such a definition involves the definition of $\hat{F}_5$ in type IIB supergravity in $d=10$~\cite{Schwarz:1983qr}, where there is an 
extra term that does not involve the gravitino.}.
The structure of these terms, in particular, the extra term that does not involve the gravitino in \C{manysupcov},
follows naturally from dimensionally reducing the supercovariant 4 form field strength, and the supercovariant spin connection
in $d=11$ using \C{supcov114} and \C{supcov11w}, and inserting them in the supersymmetry transformation \C{11dtrans} for 
the $d=11$ gravitino. The relevant expressions obtained from 
\C{supcov114} are given by
\bea \hat{G}_{\mu\nu\lambda\rho}^{SS} &=& 
G_{\mu\nu\lambda\rho}^{SS} + 3 e^{-\phi_{SS}} \Big[ 
\Big(  \bar\psi_{R[\mu} \gamma_{\nu\lambda} \psi_{\rho ]L}
- \frac{1}{3} \bar\psi_{L[\mu} \bar\gamma_{\nu\lambda\rho ]} \lambda_L \Big) \non \\ && +
\Big( \bar\psi_{L[\mu} \bar\gamma_{\nu\lambda} \psi_{\rho ]R}
- \frac{1}{3} \bar\psi_{R[\mu} \gamma_{\nu\lambda\rho ]} \lambda_R \Big) 
-\frac{1}{36} \Big( 
\bar\lambda_R \gamma_{\mu\nu\lambda\rho} \lambda_L +  \bar\lambda_L 
\bar\gamma_{\mu\nu\lambda\rho} \lambda_R \Big) \Big], \non \\ 
\hat{G}_{\mu\nu\rho i}^ {SS} &=& G_{\mu\nu\rho i} + \frac{3}{2} \Big( 2 \bar\psi_{L[\mu}
 \s^i \bar\gamma_\nu \psi_{\rho ]L}  + 
\bar\psi_{L[\mu}  \bar\gamma_{\nu\rho ]} \chi^i_R + \bar\psi_{R[\mu}
  \gamma_{\nu\rho ]} \chi^i_L  \Big) \non \\ && + \frac{1}{4} \Big( 
\bar\chi^i_{L} \bar\gamma_{\mu\nu\rho} \lambda_L
+  \bar\chi^i_{R} \gamma_{\mu\nu\rho} \lambda_R \Big) + \frac{1}{12}
\bar\lambda_R \gamma_{\mu\nu\rho} \s^i \lambda_R , \non \\ \hat{G}^{i SS}_{\mu\nu} &=&
G^{iSS}_{\mu\nu} + i e^{\phi_{SS}} \Big[ \frac{1}{2} \Big( \bar\psi_{R[\mu} \s^i \psi_{\nu ]
L} + \bar\psi_{L[\mu} \s^i \psi_{\nu ] R} \Big) - \Big( \bar\psi_{R[ \mu} \g_{\nu ]} \chi^i_R - 
\bar\psi_{L[ \mu} \bar\g_{\nu ]} \chi^i_L \Big) \non \\ && + \frac{1}{2} \Big( \bar\psi_{L[\mu} \bar\gamma_{\nu ]} \s^i
\lambda_L + \bar\psi_{R[ \mu} \g_{\nu ]} \s^i \lambda_R \Big)
+\frac{1}{4} \Big( \bar\chi_{jL} \bar\gamma_{\mu\nu}
\s^i \chi_{jR} + \bar\chi_{jR} \g_{\mu\nu} \s^i \chi_{jL} \Big)
\non \\ && + \frac{1}{24} \Big( 
\bar\lambda_L \s^i \bar\g_{\mu\nu} \lambda_R + \bar\lambda_R
\s^i \g_{\mu\nu} \lambda_L  \Big) \Big], \non \\ \widehat{\p_\mu B}_{SS} &=& \p_\mu B_{SS} + \frac{i}{2} 
e^{2\phi_{SS}} \Big[ \Big( \bar\psi_{\mu L} \lambda_R - \bar\psi_{\mu R} \lambda_L \Big) - \frac{1}{3} \bar\lambda_R 
\g_\mu \lambda_R + \bar\chi_{iR} \g_\mu \chi_{iR} \Big] ,\eea
while the relevant expressions obtained from \C{supcov11w} are given by~\footnote{Note that
\be F^{iSS}_{ab} = e_a^{~\mu} e_b^{~\nu} F^{iSS}_{\mu\nu} = e_a^{~\mu} e_b^{~\nu} L_m^{~i} F^{mSS}_{\mu\nu}.\ee} 
\bea 
\hat{\omega}_{abc} &=& e^{\phi_{SS}/3} \Big( \omega_{abc} + \frac{1}{3} \eta_{ac} \p_b \phi_{SS}
- \frac{1}{3} \eta_{ab} \p_c \phi_{SS}\Big) \non \\ && + \frac{e^{\phi_{SS}/3}}{2} \Big[ 
-2 \Big( \bar\psi_{aR} \g_{[b} \psi_{c]R} + \bar\psi_{aL} \bar\g_{[b} \psi_{c]L} + \bar\psi_{R[b} \g_{|a|} \psi_{c]R}
\Big) 
\non \\ &&+\frac{1}{3} \Big( \bar\psi_{aR}
\g_{bc} \lambda_L + \bar\psi_{aL} \bar\g_{bc} \lambda_R -2 \eta_{a[b} \bar\psi_{c]R} \lambda_L - 2 \eta_{a[b}
\bar\psi_{c]L} \lambda_R \Big) + \frac{1}{18} \bar\lambda_R \g_{abc} \lambda_R
\Big] , \non \\
\hat{\omega}_{abi} &=&  e^{4\phi_{SS}/3} F^{iSS}_{ab}
- \frac{e^{\phi_{SS}/3}}{2} \Big[ 2 \Big( \bar\psi_{R(a} \g_{b)} \chi_{iR}  
+ \bar\psi_{L(a} \bar\g_{b)} \chi_{iL} \Big) \non \\ && 
+ \Big( \bar\psi_{R[a} \s^i \psi_{b]L} - \bar\psi_{L[a} \s^i \psi_{b]R} \Big) 
+ \frac{1}{3} \eta_{ab} \Big( \bar\chi_{iR}  \lambda_L 
+ \bar\chi_{iL} \lambda_R \Big) \non \\ &&
+ \frac{1}{6} \Big( \bar\psi_{aR} \g_b \s^i \lambda_R - \bar\psi_{aL} \bar\g_b \s^i 
\lambda_L \Big)  + \frac{1}{2} \Big( \bar\psi_{bR} \g_a \s^i \lambda_R - \bar\psi_{bL} \bar\g_a \s^i 
\lambda_L \Big) \non \\ && 
-\frac{1}{36} \Big(\bar\lambda_L \s^i \bar\g_{ab} \lambda_R - \bar\lambda_R \s^i \g_{ab} 
\lambda_L \Big) \Big]  , \non \\
\hat{\omega}_{aij} &=& e^{\phi_{SS}/3} Q_{aij}^{SS} + \frac{e^{\phi_{SS}/3}}{2} \Big[ -\frac{2i}{3}
\epsilon_{ijk} \Big( \bar\psi_{aR} \s^k \lambda_L + \bar\psi_{aL} \s^k \lambda_R \Big) 
\non \\ && + 2 \Big( \bar\psi_{aR}
\s_{[i} \chi_{j]L} - \bar\psi_{aL} \s_{[i} \chi_{j]R} \Big)- 2 \bar\chi_{R[i} \g_{|a|} \chi_{j]R}
+ \Big( \bar\lambda_L \bar\g_a \s_{[i} \chi_{j]L} - \bar\lambda_R \g_a \s_{[i} \chi_{j]R}
\Big) \non \\ && - \frac{4i}{9} \epsilon_{ijk} \bar\lambda_R \g_a \s^k \lambda_R
\Big], \non \eea
\bea
\hat{\omega}_{iab} &=& - e^{4\phi_{SS}/3} F^{iSS}_{ab} 
+ \frac{e^{\phi_{SS}/3}}{2} \Big[ -2 \Big( \bar\psi_{R[a} \g_{b]} \chi_{iR}  
+ \bar\psi_{L[a} \bar\g_{b]} \chi_{iL} \Big) \non \\ && 
+ \Big( \bar\psi_{R[a} \s^i \psi_{b]L} - \bar\psi_{L[a} \s^i \psi_{b]R} \Big) 
+ \frac{1}{3} \Big( \bar\chi_{iR} \g_{ab} \lambda_L 
+ \bar\chi_{iL} \bar\g_{ab} \lambda_R \Big) \non \\ &&
+ \Big(\bar\psi_{L[a} \bar\g_{b]} \s^i \lambda_L - \bar\psi_{R[a} \g_{b]} \s^i 
\lambda_R \Big)  -\frac{5}{36} \Big(\bar\lambda_L \s^i \bar\g_{ab} \lambda_R - \bar\lambda_R \s^i \g_{ab} 
\lambda_L \Big) \Big] , 
\non \\ 
\hat{\omega}_{ija} 
&=& e^{\phi_{SS}/3} \Big( P_{aij}^{SS} + \frac{2}{3} \delta_{ij} \p_a \phi_{SS} \Big) +
\frac{e^{\phi_{SS}/3}}{2} \Big[ 2\bar\chi_{R[i} \g_{|a|} \chi_{j]R} + \frac{2i}{9} \epsilon_{ijk} \bar\lambda_R
\g_a \s^k \lambda_R \non \\ && + \frac{2}{3} \delta_{ij} (\bar\lambda_R \psi_{aL} + \bar\lambda_L \psi_{aR}) 
+ 2 \Big( \bar\psi_{aL} \s_{(i} \chi_{j)R} - \bar\psi_{aR} \s_{(i} \chi_{j)L} \Big) \non \\ && + \frac{1}{3}
\Big( \bar\chi_{L(i} \bar\g_{|a|} \s_{j)} \lambda_L - \bar\chi_{R(i} \g_{|a|} \s_{j)} \lambda_R
\Big) + \frac{2}{3} \Big( \bar\chi_{R[i} \g_{|a|} \s_{j]} \lambda_R - \bar\chi_{L[i} \bar\g_{|a|} \s_{j]} 
\lambda_L \Big) \Big] , \non \\ \hat{\omega}_{ijk} &=& \frac{e^{\phi_{SS}/3}}{2} \Big[ \frac{i}{9} \epsilon_{ijk}
\Big( \bar\lambda_L \lambda_R - \bar\lambda_R \lambda_L \Big) + 2 \Big( \bar\chi_{iR} \s_{[j} \chi_{k]L}
- \bar\chi_{iL} \s_{[j} \chi_{k]R} \Big) \non \\ &&
+ \Big( \bar\chi_{kL} \s_i \chi_{jR} - \bar\chi_{kR} \s_i \chi_{jL} \Big)
+ \frac{2}{3} \Big( \delta_{ij} (\bar\lambda_L \chi_{kR} + \bar\lambda_R \chi_{kL})  
- \delta_{ik} (\bar\lambda_L \chi_{jR} + \bar\lambda_R \chi_{jL} ) \non \\ && - i\epsilon_{jkl}
(\bar\chi_{iL} \s_l \lambda_R + \bar\chi_{iR} \s_l \lambda_L ) \Big) \Big], \eea
where $\omega_\mu^{~ab}$ is the $d=8$ spin connection constructed out of the
vielbein $e_\mu^{~a}$.

\section{Various fermionic relations}
\label{canc}

The supersymmetry transformations given in \C{susytran} for the fields charged under $U(1)$ are different from the ones given in~\cite{Salam:1984ft}. The extra non gauge--invariant contributions add to give very simple contributions, as we briefly explain. First consider the 
supervariation of $\lambda_L$. We have that
\be \delta^{SS} \lambda_L = \delta^{(0)} \lambda_L + \delta^{NG} \lambda_L ,\ee
where $\delta^{NG} \lambda_L$ is the non gauge invariant contribution, and $\delta^{(0)} \lambda_L$ is given in \C{susytran}. 

\subsection{Calculating $\delta^{NG} \lambda_L$}

Now, $\delta^{NG} \lambda_L$ is the sum of the following 7 contributions, which we also evaluate:

(i) $O(\lambda_L \lambda_l \epsilon_L)$:
\bea \label{givedet}
&&\frac{1}{576} \g^{\mu\nu\lambda\rho} \epsilon_L (\bar\lambda_R \g_{\mu\nu\lambda\rho} \lambda_L) - \frac{1}{12} \epsilon_L (\bar\lambda_R 
\lambda_L) + \frac{5}{12} \lambda_L (\bar\epsilon_R \lambda_L) \non \\ &&-\frac{1}{36} \g^{\mu\nu} \s^i \epsilon_L 
(\bar\lambda_R \g_{\mu\nu} \s^i \lambda_L) - \frac{1}{24} \g^{\mu\nu} \lambda_L (\bar\epsilon_R  \g_{\mu\nu} \lambda_L )
+ \frac{1}{18} \s^i \lambda_L (\bar\epsilon_R \s^i \lambda_L) \non \\ &&= -\frac{3}{4} \lambda_L (\bar\epsilon_R\lambda_L) . \eea 

(ii) $O(\lambda_L \lambda_R \epsilon_R)$:
\bea &&\frac{1}{6} \g^\mu \epsilon_R (\bar\lambda_R \g_\mu \lambda_R) - \frac{1}{9} \s^i \g^\mu \epsilon_R (\bar\lambda_L \g_\mu \s^i \lambda_R)
+ \frac{5}{12} \lambda_L (\bar\epsilon_L \lambda_R) \non \\ &&-\frac{1}{24} \g^{\mu\nu} \lambda_L (\bar\epsilon_L \bar\g_{\mu\nu} \lambda_R) 
- \frac{1}{18}
\g^\mu \s^i \lambda_R (\bar\epsilon_L \bar\g_\mu \s^i \lambda_L) + \frac{1}{18} \s^i \lambda_L (\bar\epsilon_L \s^i \lambda_R) \non \\ &&= 
\frac{3}{4} \lambda_L (\bar\epsilon_L \lambda_R) . \eea 

(iii)$O(\chi_L \chi_L \epsilon_L)$:

\bea && -\frac{1}{384} \g^{\mu\nu\lambda\rho} \epsilon_L (\bar\chi_{iR} \g_{\mu\nu\lambda\rho} \chi_{iL}) - \frac{1}{8} \epsilon_L (\bar\chi_{iR} \chi_{iL})
+ \frac{1}{2} \chi_{iL} (\bar\epsilon_R \chi_{iL}) \non \\ &&+ \frac{1}{2} \s_i \chi_{jL} \bar\epsilon_R (\s_i\chi_{jL} + \s_j \chi_{iL}) - \frac{1}{16}
\g^{\mu\nu} \s_i \epsilon_L (\bar\chi_{jR} \g_{\mu\nu} \s_i \chi_{jL}) \non \\&&= 0.\eea

(iv) $O(\lambda_R \chi_L \epsilon_R)$:

\bea &&-\frac{1}{6} \s_i \g_\mu \epsilon_R (\bar\lambda \bar\g_\mu \chi_{iL}) + \frac{1}{3} \chi_{iL} (\bar\epsilon_L \s_i \lambda_R)
+ \frac{1}{6} \g^\mu \s_i \lambda_R (\bar\epsilon_L \bar\g_\mu \chi_{iL}) \non \\ &&= 0.\eea

(v)  $O(\lambda_R \lambda_R \epsilon_L)$:

\bea &&\frac{1}{12} \epsilon_L (\bar\lambda_L \lambda_R) + \frac{1}{144} \g^{\mu\nu} \s_i \epsilon_L (\bar\lambda_L \s_i \bar\g_{\mu\nu} \lambda_R)
+ \frac{1}{18} \g^\mu \s_i \lambda_R (\bar\epsilon_R \s_i \g_\mu \lambda_R) \non \\ &&= 0. \eea  

(vi) $O(\lambda_R \chi_R \epsilon_L)$:

\bea &&-\frac{1}{2} \s_i \epsilon_L (\bar\lambda_L \chi_{iR}) - \frac{1}{24} \g^{\mu\nu} \s_i \epsilon_L (\bar\chi_{iL} \bar\g_{\mu\nu} \lambda_R) 
\non \\ &&+\frac{1}{6} \g^\mu \s_i \lambda_R (\bar\epsilon_R \g_\mu \chi_{iR} )- \frac{1}{3} \g^\mu \chi_{iR} (\bar\epsilon_R \s_i \g_\mu \lambda_R ) 
\non \\ &&=0. \eea

(vii) $O(\chi_L \chi_R \epsilon_R)$:

\bea &&- \frac{1}{2} \g^\mu \epsilon_R (\bar\chi_{iR} \g_\mu \chi_{iR}) + \frac{i}{2} \epsilon_{ijk} \s_k \g^\mu \epsilon_R (\bar\chi_{iR} \g_\mu \chi_{jR})
- \g^\mu \chi_{iR} (\bar\epsilon_L \bar\g_\mu \chi_{iL}) \non \\ &&- \frac{1}{2} \chi_{iL} (\bar\epsilon_L \chi_{iR}) - \frac{1}{2} \s_i \chi_{jL}
\bar\epsilon_L (\s_i \chi_{jR} + \s_j \chi_{iR}) \non \\ &&= 0. \eea

In obtaining these results and the ones that follow, we make heavy use of \C{fierz}, \C{gammaalg} and the Schouten identity
\be \theta_A \epsilon_{BC} + \theta_C \epsilon_{AB} + \theta_B \epsilon_{CA}= 0, \ee
where $\theta_A$ is a fermion, $(\s_i)_A^{~D}$ or $(\s_i \s_j)_A^{~D}$.
We also use the relation
\be (\s_i)_A^{~B} (\s_i)_C^{~D} = 2 \Big( \delta_A^{~D} \delta_C^{~B} - \frac{1}{2} \delta_A^{B} \delta_C^{~D} \Big), \ee
and other relations like
\bea \bar\lambda_{RA} \psi^B_L = - \bar\psi^B_R \lambda_{LA} , \quad \bar\lambda_{LA} \bar\g^\mu \psi^B_L =  \bar\psi^B_R \g^\mu \lambda_{RA} , 
\non \\ 
\bar\lambda_{RA} \g^{\mu\nu} \psi^B_L =  \bar\psi^B_R \g^{\mu\nu} \lambda_{LA} ,\quad \bar\lambda_{LA} \bar\g^{\mu\nu\rho} \psi^B_L =  
-\bar\psi^B_R \g^{\mu\nu\rho} \lambda_{RA} . \eea

Thus 
\be \delta^{NG} \lambda_L = -\frac{3}{4} \lambda_L (\bar\epsilon_R \lambda_L -\bar\epsilon_L \lambda_R).\ee

Since the algebraic details are quite involved and the technique is the same for every term, we simply outline the steps needed to deduce
\C{givedet}. Using the relations
\bea \g^{\mu\nu\lambda\rho} \epsilon_L (\bar\lambda_R \g_{\mu\nu\lambda\rho} \lambda_L) = -420 \lambda_L (\bar\epsilon_R \lambda_L) + 30
\g^{\mu\nu} \lambda_L (\bar\epsilon_R \g_{\mu\nu} \lambda_L) - \frac{3}{2} \g^{\mu\nu\lambda\rho} \lambda_L (\bar\epsilon_R \g_{\mu\nu\lambda\rho} 
\lambda_L), \non \\ \epsilon_L (\bar\lambda_R 
\lambda_L) =  -\frac{1}{4} \lambda_L (\bar\epsilon_R \lambda_L) -\frac{1}{8}
\g^{\mu\nu} \lambda_L (\bar\epsilon_R \g_{\mu\nu} \lambda_L) - \frac{1}{96} \g^{\mu\nu\lambda\rho} \lambda_L (\bar\epsilon_R \g_{\mu\nu\lambda\rho} 
\lambda_L), \non \\  \g^{\mu\nu} \s^i \epsilon_L 
(\bar\lambda_R \g_{\mu\nu} \s^i \lambda_L) = 14 \lambda_L (\bar\epsilon_R \lambda_L) + 
\g^{\mu\nu} \lambda_L (\bar\epsilon_R \g_{\mu\nu} \lambda_L) - \frac{1}{12} \g^{\mu\nu\lambda\rho} \lambda_L (\bar\epsilon_R \g_{\mu\nu\lambda\rho} 
\lambda_L), \non \\ 
\s^i \lambda_L (\bar\epsilon_R \s^i \lambda_L) =  -\frac{5}{4} \lambda_L (\bar\epsilon_R \lambda_L) + \frac{1}{8}
\g^{\mu\nu} \lambda_L (\bar\epsilon_R \g_{\mu\nu} \lambda_L) - \frac{1}{96} \g^{\mu\nu\lambda\rho} \lambda_L (\bar\epsilon_R \g_{\mu\nu\lambda\rho} 
\lambda_L),\non \\ \eea
and adding the various contributions, we get \C{givedet}. Similar is the analysis for $\delta^{NG} \lambda_R$.

\subsection{Calculating $\delta^{NG} \chi_{iL}$}

The various non gauge invariant contributions are given by:

(i) $O(\chi_L \lambda_L \epsilon_L)$:
\bea \label{chiil}
&&\frac{1}{24} \Delta_{ij} \g^{\mu\nu} \epsilon_L (\bar\chi_{jR} \g_{\mu\nu} \lambda_L) + \frac{1}{6} \Delta_{ij} \s_k \epsilon_L (\bar\chi_{jR}
\s_k \lambda_L) - \frac{1}{6} \Delta_{ij} \epsilon_L (\bar\lambda_R \chi_{jL}) + \frac{5}{12} \chi_{iL} (\bar\epsilon_R \lambda_L) \non \\&&
- \frac{1}{24} \g^{\mu\nu} \chi_{iL} (\bar\epsilon_R \g_{\mu\nu} \lambda_L) + \frac{1}{3} \s_j \lambda_L (\bar\epsilon_R \s_{(i} \chi_{j) L} )
- \frac{1}{6} \Delta_{ij} \s_k \chi_{jL} (\bar\epsilon_R \s_k \lambda_L) + \frac{1}{6} \Delta_{ij} \lambda_L (\bar\epsilon_R \chi_{jL}) \non \\
&&= \frac{1}{4} \chi_{iL} (\bar\epsilon_R \lambda_L ) , \eea

(ii) $O(\chi_L \lambda_R \epsilon_R)$:
\bea &&\frac{1}{12} \g^\mu \s_j\epsilon_R (\bar\chi_{R(i} \s_{j)} \g_\mu \lambda_R) - \frac{1}{48} \Delta_{ij} \g^{\mu\nu\rho} \epsilon_R
(\bar\chi_{jR} \g_{\mu\nu\rho} \lambda_R) - \frac{1}{6} \Delta_{ij} \s_k \g_\mu \epsilon_R (\bar\chi_{R[j} \g^\mu \s_{k]} \lambda_R)
\non \\ && + \frac{5}{12} \chi_{iL} (\bar\epsilon_L \lambda_R) - \frac{1}{24} \g^{\mu\nu} \chi_{iL}  (\bar\epsilon_L \bar\g_{\mu\nu} 
\lambda_R) + \frac{1}{3} \Delta_{ij} \g^\mu \lambda_R (\bar\epsilon_L \bar\g_\mu \chi_{jL}) - \frac{1}{6}
\Delta_{ij} \s_k \chi_{jL} (\bar\epsilon_L \s_k \lambda_R) \non \\&&= -\frac{1}{4} \chi_{iL} (\bar\epsilon_L \lambda_R ) ,\eea

(iii) $O(\lambda_L \chi_R \epsilon_R)$:
\bea &&- \frac{1}{12} \g^\mu \s_j \epsilon_R (\bar\chi_{L(i} \s_{j)} \bar\g_\mu \lambda_L) - \frac{1}{48} \Delta_{ij} \g^{\mu\nu\rho}
\epsilon_R (\bar\chi_{jL} \bar\g_{\mu\nu\rho} \lambda_L) + \frac{1}{6} \Delta_{ij} \s_k \g_\mu \epsilon_R (\bar\chi_{L[j} \bar\g^\mu \s_{k]}
\lambda_L) \non \\ &&-\frac{1}{3} \s_j\lambda_L (\bar\epsilon_L \s_{(i} \chi_{j)R} )  -\frac{1}{6} \Delta_{ij} \g^\mu \s_k \chi_{jR} 
(\bar\epsilon_L \bar\g_\mu \s_k \lambda_L)  - \frac{1}{6} \Delta_{ij} \lambda_L (\bar\epsilon_L \chi_{jR}) \non \\ &&= 0, \eea

(iv) $O(\chi_R \chi_R \epsilon_L)$:

\bea && -\frac{1}{16} \Delta_{ij} \g^{\mu\nu} \epsilon_L (\bar\chi_{kL} \bar\g_{\mu\nu} \s_j \chi_{kR}) + \frac{i}{8} \Delta_{ij}
\s_k \epsilon_L \epsilon^{ki'j'} (\bar\chi_{Lj'} \s_j \chi_{Ri'}) \non \\ &&+ \frac{1}{2} \Delta_{ij} \g^\mu \s_k \chi_{jR}
(\bar\epsilon_R \g_\mu \chi_{kR}) - \frac{1}{4} \Delta_{ij} \s_k \epsilon_L (\bar\chi_{jL} \chi_{kR}) \non \\ &&=0,\eea

(v) $O(\lambda_R \lambda_R \epsilon_L)$:
\bea -\frac{1}{36} \Delta_{ij} \g^{\mu\nu} \epsilon_L (\bar\lambda_L \s_j \bar\g_{\mu\nu} \lambda_R) + \frac{1}{9} \Delta_{ij}
\g^\mu \lambda_R (\bar\epsilon_R \s_j \g_\mu \lambda_R) =0,\eea

(vi) $O(\lambda_L \lambda_L \epsilon_L)$:

\bea \frac{1}{144} \Delta_{ij} \g^{\mu\nu} \epsilon_L (\bar\lambda_R \g_{\mu\nu} \s_j \lambda_L) + \frac{1}{9} \Delta_{ij}
\lambda_L (\bar\epsilon_R \s_j \lambda_L) = 0 .\eea

Thus
\be \delta^{NG} \chi_{iL} = \frac{1}{4} \chi_{iL} (\bar\epsilon_R \lambda_L -\bar\epsilon_L \lambda_R).\ee

To prove \C{chiil} for example, we use the relations
\bea &&\Delta_{ij} \epsilon_L (\bar\lambda_R \chi_{jL}) = - \frac{1}{24} \Big[ 
\chi_{jL} (\bar\lambda_R \s_j \s_i \epsilon_L) -\frac{1}{2} \g^{\mu\nu} \chi_{jL} (\bar\lambda_R \g_{\mu\nu} \s_j \s_i \epsilon_L)
\non \\ &&+\frac{1}{24} \g^{\mu\nu\lambda\rho} \chi_{jL} ( \bar\lambda_R \g_{\mu\nu\lambda\rho} \s_j \s_i \epsilon_L)\Big] 
+\frac{1}{3} \epsilon_L (\bar\lambda_R \chi_{iL}) ,\non \\ &&\g^{\mu\nu} \Delta_{ij} \epsilon_L (\bar\chi_{jR} \g_{\mu\nu} \lambda_L) = 
-\frac{1}{3} \Big[ 7 \chi_{jL} (\bar\lambda_R \s_j \s_i \epsilon_L) - \frac{1}{2} \g^{\mu\nu} \chi_{jL} 
(\bar\lambda_R \s_j \s_i \g_{\mu\nu} \epsilon_L) \non \\ &&- \frac{1}{24} \g^{\mu\nu\lambda\rho} \chi_{jL} 
(\bar\lambda_R \s_j \s_i \g_{\mu\nu\lambda\rho} \epsilon_L)\Big] -\frac{1}{3}
\g^{\mu\nu} \epsilon_L (\bar\lambda_R \g_{\mu\nu} \chi_{iL}) , \non \\ &&\frac{1}{6} \Delta_{ij} \lambda_L (\bar\epsilon_R \chi_{jL})
+ \frac{1}{3} \s_j \lambda_L (\bar\epsilon_R \s_{(i} \chi_{j) L}) = \frac{1}{72} \Big[ \epsilon_L (\bar\chi_{iR} \lambda_L) -\frac{1}{2}
\g^{\mu\nu} \epsilon_L (\bar\chi_{iR} \g_{\mu\nu} \lambda_L) \non \\ &&
+ \frac{1}{24} \g^{\mu\nu\lambda\rho} \epsilon_L (\bar\chi_{iR} \g_{\mu\nu\lambda\rho} \lambda_L ) \Big] 
+  \frac{1}{24} \Big[ - \chi_{jL} (\bar\epsilon_R 
\hat\Delta_{ij} \lambda_L) + \frac{1}{2} \g^{\mu\nu} \chi_{jL} (\bar\epsilon_R \g_{\mu\nu} \hat\Delta_{ij} \lambda_L) \non \\ &&- \frac{1}{24}
\g^{\mu\nu\lambda\rho} \chi_{jL} (\bar\epsilon_R \g_{\mu\nu\lambda\rho} \hat\Delta_{ij} \lambda_L)\Big], \non \\ && \Delta_{ij} \s_k \epsilon_L
(\bar\chi_{jR} \s_k \lambda_L) - \Delta_{ij} \s_k \chi_{jL} (\bar\epsilon_R \s_k \lambda_L) = - \Delta_{ij} \epsilon_L (\bar\lambda_R \chi_{jL})
+ \chi_{iL} (\bar\epsilon_R \lambda_L) \non \\  && -\frac{1}{4} \Big[ 7 \chi_{jL} (\bar\epsilon_R \Delta_{ij} \lambda_L) - \frac{1}{2}
\g^{\mu\nu} \chi_{jL} (\bar\epsilon_R \g_{\mu\nu} \Delta_{ij} \lambda_L) - \frac{1}{24} 
\g^{\mu\nu\lambda\rho} \chi_{jL} (\bar\epsilon_R \g_{\mu\nu\lambda\rho} \Delta_{ij} \lambda_L)\Big] \non \\ && +\frac{1}{12} 
\Big[ 7 \epsilon_L (\bar\chi_{iR} \lambda_L) - \frac{1}{2}
\g^{\mu\nu} \epsilon_L (\bar\chi_{iR} \g_{\mu\nu} \lambda_L) - \frac{1}{24} \g^{\mu\nu\lambda\rho} \epsilon_L (\bar\chi_{iR} 
\g_{\mu\nu\lambda\rho} \lambda_L)\Big] ,\eea
where
\be \hat\Delta_{ij} = \delta_{ij} + \frac{1}{3} \s_i \s_j .\ee
Adding the various contributions, we get \C{chiil}.

\subsection{Calculating $\delta^{NG} \psi_{\mu L}$}

The various non gauge invariant contributions are given by:

(i) $O(\psi_L \lambda_L \epsilon_L)$:
\bea \label{delpsi1}
&& -\frac{1}{12} \psi_{\mu L} (\bar\epsilon_R \lambda_L) + \frac{1}{24} \g^{\nu\rho} \epsilon_L (\bar\psi_{\mu R} \g_{\nu\rho} \lambda_L)
- \frac{1}{6} \epsilon_L (\bar\psi_{\mu R} \lambda_L) + \frac{1}{6} \s_i \epsilon_L (\bar\psi_{\mu R} \s_i \lambda_L) \non \\ && - \frac{1}{24} 
\g^{\nu\rho} \psi_{\mu L} (\bar\epsilon_R \g_{\nu\rho} \lambda_L) + \frac{1}{3} \s_i \lambda_L (\bar\epsilon_R \s_i \psi_{\mu L}) - \frac{1}{6}
\s_i \psi_{\mu L} (\bar\epsilon_R \s_i \lambda_L) \non \\ && = - \frac{1}{4} \psi_{\mu L} (\bar\epsilon_R \lambda_L ),\eea

(ii) $O(\psi_L \lambda_R \epsilon_R)$:
\bea && - \frac{1}{12} \psi_{\mu L} (\bar\epsilon_L \lambda_R) - \frac{1}{6} \s_i \g^\nu \epsilon_R (\bar\psi_{\mu R} \g_\nu \s_i \lambda_R)
- \frac{1}{24} \g^{\nu\rho} \psi_{\mu L} (\bar\epsilon_L \bar\g_{\nu\rho} \lambda_R) \non \\ &&
- \frac{1}{6} \g^\nu \lambda_R (\bar\epsilon_L \bar\g_\nu \psi_{\mu L})
- \frac{1}{6} \s_i \psi_{\mu L} (\bar\epsilon_L \s_i \lambda_R) \non \\ &&= \frac{1}{4} \psi_{\mu L} (\bar\epsilon_L \lambda_R) , \eea

(iii) $O(\lambda_L \chi_R \epsilon_L)$:
\bea &&\frac{1}{144} \s_i (\g_{\mu\nu\lambda\rho} - 6 g_{\mu\nu} \g_{\lambda\rho}) \epsilon_L (\bar\chi_{iL} \bar\g^{\nu\lambda\rho} \lambda_L)
- \frac{7}{72} \s_i \epsilon_L (\bar\lambda_R \g^\mu \chi_{iR})  + \frac{1}{36} \s_i \g_{\mu\nu} \epsilon_L (\bar\lambda_R \g^\nu \chi_{iR})
\non \\ && - \frac{1}{18} (8 g_{\mu\nu} - \g_{\mu\nu}) \s_i \lambda_L (\bar\epsilon_R \g^\nu \chi_{iR}) - \frac{1}{18} \g_\mu \chi_{iR}
(\bar\epsilon_R \s_i \lambda_L) \non \\ &&= 0,   \eea

(iv) $O(\lambda_R \chi_L \epsilon_L)$:
\bea &&\frac{1}{144} \s_i (\g_{\mu\nu\lambda\rho} - 6  g_{\mu\nu} \g_{\lambda\rho}) \epsilon_L (\bar\chi_{iR} \g^{\nu\lambda\rho} \lambda_R)  
+\frac{7}{72} \s_i \epsilon_L (\bar\lambda_L \bar\g_\mu \chi_{iL}) \non \\ &&
- \frac{1}{36} \s_i \g_{\mu\nu} \epsilon_L (\bar\lambda_L \bar\g^\nu \chi_{iL})
+ \frac{4}{9} (g_{\mu\nu} - \frac{1}{8} \g_{\mu\nu}) \chi_{iL} (\bar\epsilon_R \s_i \g^\nu \lambda_R) + \frac{1}{18} \s_i \g_\mu \lambda_R
(\bar\epsilon_R) \chi_{iL} \non \\ &&= 0 ,  \eea

(v) $O(\lambda_L \lambda_L \epsilon_R)$:
\bea && \frac{1}{1152} \g^{\n\lambda\rho\s} \g_\mu \epsilon_R (\bar\lambda_R \g_{\nu\lambda\rho\s} \lambda_L) + \frac{1}{72} \g_\mu \epsilon_R 
(\bar\lambda_R \lambda_L) - \frac{1}{54} \s_i \g^\nu \epsilon_R (\bar\lambda_R \s_i \g_{\mu\nu} \lambda_L)  \non \\ && + \frac{1}{216} \s_i 
\g_{\mu\nu\rho} \epsilon_R (\bar\lambda_R \g^{\nu\rho} \s_i \lambda_L) + \frac{1}{54} (11 g_{\mu\nu} - \g_{\mu\nu}) \s_i \lambda_L (\bar\epsilon_L
\bar\g^\nu \s_i \lambda_L)   \non \\ &&= 0 , \eea

(vi) $O(\chi_L \chi_L \epsilon_R)$:
\bea && -\frac{1}{768} \g^{\nu\lambda\rho\s} \g_\mu \epsilon_R (\bar\chi_{iR} \g_{\nu\lambda\rho\s} \chi_{iL}) + \frac{1}{48}
\g_\mu \epsilon_R (\bar\chi_{iR} \chi_{iL}) - \frac{5}{48} \s_i \g^\nu \epsilon_R (\bar\chi_{jR} \g_{\mu\nu} \s_i \chi_{jL})
\non \\ && + \frac{1}{96} \s_i \g^{\mu\nu\rho} \epsilon_R (\bar\chi_{jR} \g_{\nu\rho} \s_i \chi_{jL}) + \frac{1}{6}
(5 g_{\mu\nu} - \g_{\mu\nu}) \chi_{iL} (\bar\epsilon_L \bar\g^\nu \chi_{iL})  \non \\ &&= 0 , \eea

(vii) $O(\lambda_R \lambda_R \epsilon_R)$:
\bea && \frac{1}{12} \g_\mu \epsilon_R (\bar\epsilon_L \lambda_R) - \frac{1}{72} \g_\mu \epsilon_R (\bar\lambda_L \lambda_R)
- \frac{7}{432} \s_i \g^\nu \epsilon_R (\bar\lambda_L \s_i \bar\g_{\mu\nu} \lambda_R) \non \\ && 
-\frac{1}{864} \s_i \g_{\mu\nu\rho} \epsilon_R (\bar\lambda_L \s_i \bar\g^{\nu\rho} \lambda_R ) - \frac{1}{36} \g^\nu \lambda_R
(\bar\epsilon_L \bar\g_{\mu\nu} \lambda_R) + \frac{1}{27} \s_i \g_\mu \lambda_R (\bar\epsilon_L \s_i \lambda_R)\non \\ && =0 ,\eea

(viii) $O(\psi_R \lambda_R \epsilon_L)$:
\bea && \frac{1}{24} \g^{\nu\rho} \epsilon_L (\bar\psi_{\mu L} \bar\g_{\nu\rho} \lambda_R) + \frac{1}{3} \epsilon_L (\bar\psi_{\mu L}
\lambda_R) + \frac{1}{6} \s_i \epsilon_L (\bar\psi_{\mu L} \s_i \lambda_R) 
\non \\ && - \frac{1}{6} \g^\nu \lambda_R (\bar\epsilon_R \g_\nu
\psi_{\mu R}) + \frac{1}{6} \g^\nu \s_i \psi_{\mu R} (\bar\epsilon_R \g_\nu \s_i \lambda_R ) \non \\ && = 0,\eea

(ix) $O(\lambda_R \chi_R \epsilon_R)$:
\bea && \frac{1}{72} \s_i \g^\nu \epsilon_R (\bar\chi_{iL} \bar\g_{\mu\nu} \lambda_R) + \frac{1}{144} \s_i \g_{\mu\nu\rho} 
\epsilon_R (\bar\chi_{iL} \bar\g^{\nu\rho} \lambda_R) \non \\ && - \frac{1}{18} \s_i \g_\mu \lambda_R (\bar\epsilon_L \chi_{iR})
- \frac{1}{18} \g_\mu \chi_{iR} (\bar\epsilon_L \s_i \lambda_R) \non \\ &&= 0 ,\eea

(x) $O(\psi_R \chi_L \epsilon_R)$:
\bea && -\frac{1}{2} \s_i \g_\nu \epsilon_R (\bar\psi_{\mu L} \bar\g^\nu \chi_{iL}) + \chi_{iL}
(\bar\epsilon_L \s_i \psi_{\mu R}) + \frac{1}{2} \s_i \g_\nu \psi_{\mu R} (\bar\epsilon_L \bar\g^\nu \chi_{iL}) 
\non \\ && =0 . \eea

Thus
\be \delta^{NG} \psi_{\mu L} = -\frac{1}{4} \psi_{\mu L} (\bar\epsilon_R \lambda_L -\bar\epsilon_L \lambda_R).\ee

In proving \C{delpsi1} for example, we have used the relations
\bea && \lambda_L (\bar\epsilon_R \psi_{\mu L}) = \Big[ -\frac{1}{8} \epsilon_L (\bar\psi_{\mu R} \lambda_L) + \frac{1}{16}
\g^{\nu\rho} \epsilon_L (\bar\psi_{\mu R} \g_{\nu\rho} \lambda_L) - \frac{1}{192} \g^{\nu\rho\s\tau} \epsilon_L (\bar\psi_{\mu R} 
\g_{\nu\rho\s\tau}\lambda_L)\Big] \non \\ && + \Big[-\frac{1}{8} \psi_{\mu L} (\bar\epsilon_R \lambda_L) + \frac{1}{16}
\g^{\nu\rho} \psi_{\mu L} (\bar\epsilon_R \g_{\nu\rho} \lambda_L) - \frac{1}{192} \g^{\nu\rho\s\tau} \psi_{\mu L} (\bar\epsilon_R 
\g_{\nu\rho\s\tau}\lambda_L) \Big] , \non \\ && \g^{\nu\rho} \lambda_L (\bar\epsilon_R \g_{\nu\rho} \psi_{\mu L}) = -\Big[ 7
\epsilon_L (\bar\psi_{\mu R} \lambda_L) - \frac{1}{2}
\g^{\nu\rho} \epsilon_L (\bar\psi_{\mu R} \g_{\nu\rho} \lambda_L) - \frac{1}{24} \g^{\nu\rho\s\tau} \epsilon_L (\bar\psi_{\mu R} 
\g_{\nu\rho\s\tau}\lambda_L)\Big] \non \\ && + \Big[7 \psi_{\mu L} (\bar\epsilon_R \lambda_L) - \frac{1}{2}
\g^{\nu\rho} \psi_{\mu L} (\bar\epsilon_R \g_{\nu\rho} \lambda_L) - \frac{1}{24} \g^{\nu\rho\s\tau} \psi_{\mu L} (\bar\epsilon_R 
\g_{\nu\rho\s\tau}\lambda_L) \Big] .\eea
Adding the various contributions, we get \C{delpsi1}.

\subsection{The expressions for the gauge invariant contributions}

The various gauge invariant contributions can be rewritten in several different ways. For example, the ones in $\delta^{(0)} \lambda_L$
can also be written as
\bea  &&\frac{1}{6} \s_i \g_\mu \epsilon_R (\bar\lambda_R \g^\mu \chi_{iR}) + \frac{1}{3} \g_\mu \chi_{iR} (\bar\epsilon_L \bar\g_\mu \s_i \lambda_L)
-\frac{1}{12} \s_i \lambda_L (\bar\epsilon_L \chi_{iR})  \non \\ &&= \frac{1}{32} \Big[ 15 \g_\mu \chi_{iR} (\bar\epsilon_L \bar\g^\mu \s_i \lambda_L 
) + \frac{1}{6} \g_{\mu\nu\lambda} \chi_{iR} (\bar\epsilon_L \bar\g^{\mu\nu\lambda} \s_i \lambda_L )\Big],\non \\ &&- \frac{1}{2}
\s_i \epsilon_L (\bar\lambda_R \chi_{iL}) -\frac{1}{24} \g^{\mu\nu} \s_i \epsilon_L  (\bar\chi_{iR} \g_{\mu\nu} \lambda_L)
+ \frac{1}{3} \chi_{iL} (\bar\epsilon_R \s_i \lambda_L) + \frac{1}{12} \s_i \lambda_L (\bar\epsilon_R \chi_{iL}) \non \\ &&= -\frac{1}{32} \Big[
\chi_{iL} (\bar\epsilon_R \s_i \lambda_L) + \frac{3}{2} \g_{\mu\nu} \chi_{iL} (\bar\epsilon_R \g^{\mu\nu} \s_i \lambda_L) + \frac{1}{24}
\g_{\mu\nu\lambda\rho} \chi_{iL} (\bar\epsilon_R \g^{\mu\nu\lambda\rho} \s_i \lambda_L )\Big] , \non \\ 
&&\frac{1}{8} \epsilon_L (\bar\chi_{iL} \chi_{iR})  - \frac{1}{16} \g^{\mu\nu} \s_i \epsilon_L (\bar\chi_{jL} 
\bar\g_{\mu\nu} \s_i \chi_{jR}) - \g^\mu \chi_{iR} (\bar\epsilon_R \g_\mu \chi_{iR} ) \non \\ &&= -\frac{1}{32} \Big[ 17 \g^\mu \chi_{iR}
(\bar\epsilon_R \g_\mu \chi_{iR}) + \frac{1}{6} \g_{\mu\nu\lambda} \chi_{iR} (\bar\epsilon_R \g^{\mu\nu\lambda} \chi_{iR}) \Big] .\eea

Similarly the ones in $\delta^{(0)} \chi_{iL}$, can be rewritten as
\bea &&\frac{1}{24} \Delta_{ij} \g^{\mu\nu} \epsilon_L (\bar\chi_{jL} \bar\g_{\mu\nu} \lambda_R) + \frac{1}{6} \Delta_{ij} \s_k \epsilon_L 
(\bar\chi_{jL} \s_k \lambda_R) - \frac{1}{6} \Delta_{ij} \epsilon_L (\bar\lambda_L \chi_{jR}) \non \\&&
+ \frac{1}{6} \Delta_{ij} \g_\mu \s_k \chi_{jR}
(\bar\epsilon_R \s_k \g_\mu \lambda_R) + \frac{1}{3} \Delta_{ij} \g^\mu \lambda_R (\bar\epsilon_R \g_\mu \chi_{jR}) \non \\&& = 
-\frac{1}{2} \Delta_{ij} \epsilon_L (\bar\lambda_L \chi_{jR}) + \frac{1}{2} \Delta_{ij} \g^\mu \lambda_R 
(\bar\epsilon_R \g_\mu \chi_{jR}) , \non \\ &&-\frac{1}{16} \Delta_{ij} \g^{\mu\nu} \epsilon_L (\bar\chi_{kR} \g_{\mu\nu} \s_j\chi_{kL})
- \frac{i}{8} \epsilon^{ki'j'} \Delta_{ij} \s_k \epsilon_L (\bar\chi_{Rj'} \s_j \chi_{Li'}) \non \\ &&+ \frac{1}{4} \Delta_{ij} \s_k \epsilon_L 
(\bar\chi_{jR} \chi_{kL}) - \Delta_{ij} \chi_{kL} (\bar\epsilon_R \s_{(j} \chi_{k)L}) - \frac{1}{4} \Delta_{ij} \s_k \chi_{jL} (\bar\epsilon_R
\chi_{kL})\non \\ &&= \frac{\Delta_{ij}}{128} \Big[ 39\chi_{kL} (\bar\epsilon_R \s_k \chi_{jL}) + \frac{17}{2} 
\g^{\mu\nu}\chi_{kL} (\bar\epsilon_R \g_{\mu\nu}\s_k \chi_{jL}) + \frac{7}{24} 
\g^{\mu\nu\lambda\rho}\chi_{kL} (\bar\epsilon_R \g_{\mu\nu\lambda\rho}\s_k
\chi_{jL})\Big],\non \eea
\bea && -\frac{i}{18} \epsilon_{jkl}
\Delta_{ij} \s_k \g_\mu \epsilon_R (\bar\lambda_R \g^\mu \s_l \lambda_R) + \frac{1}{9} \Delta_{ij} \lambda_L (\bar\epsilon_L \s_j \lambda_R)
- \frac{1}{9} \Delta_{ij} \g^\mu\lambda_R (\bar\epsilon_L \s_j \bar\g_\mu \lambda_L) \non \\ && = \frac{1}{6} 
\Delta_{ij} \lambda_L (\bar\epsilon_L \s_j 
\lambda_R) - \frac{1}{24} \Delta_{ij} \g^{\mu\nu}\lambda_L (\bar\epsilon_L \bar\g_{\mu\nu}\s_j \lambda_R), \non \\ &&
-\frac{1}{2} \Delta_{ij} \g_\mu \s_k \epsilon_R (\bar\chi_{R[j} \g^\mu \chi_{k]R}) + \Delta_{ij} \chi_{kL} (\bar\epsilon_L \s_{(j}
\chi_{k)R})  + \frac{1}{2} \Delta_{ij} \g^\mu \s_k \chi_{jR} (\bar\epsilon_L \bar\g_\mu \chi_{kL} ) \non \\ && + 
\frac{1}{4} \Delta_{ij} \s_k \chi_{jL} (\bar\epsilon_L \chi_{kR})   = \frac{\Delta_{ij}}{32} \Big[ 8 \chi_{kL}
(\bar\epsilon_L \s_k \chi_{jR})  -9 \g^\mu \chi_{kR}
(\bar\epsilon_L \s_k \bar\g_\mu \chi_{jL}) \non \\ && 
+ 2 \g^{\mu\nu} \chi_{kL} (\bar\epsilon_L \s_k \bar\g_{\mu\nu} \chi_{jR}) 
+ \frac{1}{6}  \g^{\mu\nu\rho} \chi_{kR}
(\bar\epsilon_L \s_k \bar\g_{\mu\nu\rho} \chi_{jL})\Big] ,  \eea 
and the same holds for the various terms in the expression for $\delta^{(0)} \psi_{\mu L}$. Of course, which 
way of representing the same expression is best depends on what is being asked for. We use the above formulae to rewrite the various
expressions only when the resulting expressions are considerably shorter than the previous ones, the rest we leave as they are.   
Same is the analysis for the various gauge invariant terms in $\delta^{(0)} \psi_{\mu L}$ because there is not much simplification.

The terms in the expression for $\delta^{(0)} \lambda_R$, $\delta^{(0)} \chi_{iR}$ and $\delta^{(0)} \psi_{\mu R}$ are obtained 
by conjugation.

\section{The $O(\lambda_L^2 \lambda_R^2)$ term in the supergravity action}

We need the  $O(\lambda_L^2 \lambda_R^2)$ term in the $d=8$ supergravity action. In order to get it, 
we write down the
quartic fermion terms in the $d=11$ supergravity action. We start from \C{act11d} and use the expressions for the
fermion bilinears appearing in the definitions of $\omega, \hat\omega$ and $\hat{F}$ in \C{defomega}, \C{supcov11w}
and \C{supcov114} respectively. The $\hat{F}$ term contributes
\be \label{add}
V^{-1} \mathcal{L}_{11} = -\frac{1}{64} \bar\eta_M  \hat\Gamma_{NP} \eta_Q \Big(\bar\eta_R \hat\Gamma^{RSMNPQ} \eta_S
+ 12 \bar\eta^{[M} \hat\Gamma^{NP} \eta^{Q]} \Big) , \ee
while the $R$ and $\bar\eta \slash{D} \eta$ terms together contribute
\bea \label{addtoget}
V^{-1} \mathcal{L}_{11} = \frac{1}{8} \tilde{K}_{M\hat{A}\hat{B}} (\bar\eta^{\hat{A}} \hat\Gamma^M \eta^{\hat{B}})+ 
\frac{1}{4} \Big( \hat\omega^f_{M\hat{A}\hat{B}} \bar\eta^{\hat{A}} \hat\Gamma^M \eta^{\hat{B}} +
\hat\omega^f_{\hat{B} \hat{A} \hat{C}} \hat\omega^{f\hat{A}\hat{B}\hat{C}} + \hat\omega^{f~\hat{A}}_{\hat{A}~~\hat{E}} 
\hat\omega_{\hat{B}}^{~\hat{B} \hat{E}} \Big )   \eea
to the action \C{act11d}.
In \C{addtoget}, $\tilde{K}_{M\hat{A}\hat{B}}$ is the part of the contorsion \C{contorsion} which does not involve
the fermionic part of $\hat\omega$. Thus
\be \tilde{K}_{M\hat{A}\hat{B}} = -\frac{1}{4} \bar\eta^{\hat{C}} \hat\Gamma_{M\hat{A}\hat{B}\hat{C}\hat{D}} \eta^{\hat{D}},\ee
while $\hat\omega^f$ is the fermionic part of $\hat\omega$. Thus 
\be \hat\omega^{f~\hat{A}\hat{B}}_{M} = \frac{1}{2} \Big(\bar\eta_M \hat\Gamma^{\hat{B}} \eta^{\hat{A}} 
- \bar\eta_M \hat\Gamma^{\hat{A}} \eta^{\hat{B}}
+ \bar\eta^{\hat{B}} \hat\Gamma_M \eta^{\hat{A}} \Big).\ee

Adding \C{add} and \C{addtoget}, the total contribution from all the four fermion terms is given by
\bea  V^{-1} \mathcal{L}_{11} = \frac{1}{16} \bar\eta_M \Gamma_N \eta_P 
\Big(-\bar\eta^M \Gamma^N \eta^P - 2 \bar\eta^M \Gamma^P \eta^N
+ 4 \eta^{MN} \bar\eta_Q \Gamma^Q \eta^P + \frac{1}{2} \bar\eta_Q \Gamma^{QRMNP} \eta_R \Big) \non \\ 
-\frac{1}{32} \bar\eta_M \Gamma_{PQ} \eta_N \Big(\bar\eta^M \Gamma^{PQ} \eta^N + 4 \bar\eta^M \Gamma^{NP} \eta^Q 
+ \bar\eta^P \Gamma^{MN} \eta^Q + \frac{1}{2} \bar\eta_R \Gamma^{RSMNPQ} \eta_S \Big) .\eea

We now calculate the $O(\lambda_L^2 \lambda_R^2)$ term that is obtained from the action \C{act11d}.
The contribution from \C{add} is given by
\be \label{add1}
-\frac{1}{576} (\bar\lambda_R \g^{\mu\nu\rho} \s^i \lambda_R)(\bar\lambda_R \g_{\mu\nu\rho} \s^i \lambda_R) +\frac{1}{16}
(\bar\lambda_R \g_\mu \lambda_R) (\bar\lambda_R \g^\mu \lambda_R).  \ee
This contribution is gauge invariant. 

The contribution from \C{addtoget} is given by\footnote{We use the relations
\bea 2 \Gamma^{\hat{C}\hat{D}\hat{E}\hat{A}\hat{B}} &=& 2 \Gamma^{\hat{C}\hat{D}\hat{E}} \Gamma^{\hat{A}\hat{B}} +\Big[- \eta^{\hat{A}\hat{E}} (\Gamma^{\hat{C}\hat{D}}
\Gamma^{\hat{B}} + \Gamma^{\hat{C}\hat{D}\hat{B}}) +\eta^{\hat{A}\hat{D}} (\Gamma^{\hat{C}\hat{E}}
\Gamma^{\hat{B}} + \Gamma^{\hat{C}\hat{E}\hat{B}}) \non \\&& -\eta^{\hat{A}\hat{C}} (\Gamma^{\hat{D}\hat{E}}
\Gamma^{\hat{B}} + \Gamma^{\hat{D}\hat{E}\hat{B}}) - (A \leftrightarrow B)
\Big], \non \\ \Gamma^{\hat{B}\hat{C}\hat{A}} &=& \Gamma^{\hat{B}\hat{C}} \Gamma^{\hat{A}} - \eta^{\hat{A}\hat{C}} \Gamma^{\hat{B}} +
\eta^{\hat{A}\hat{B}} \Gamma^{\hat{C}} .\eea}
\bea \label{radd}
&&\frac{1}{54} (\bar\lambda_R \s^i \g^\mu \lambda_R)(\bar\lambda_R \s^i \g_\mu \lambda_R) +\frac{1}{576}
(\bar\lambda_R \g^{\mu\nu\rho} \lambda_R) (\bar\lambda_R \g_{\mu\nu\rho}  \lambda_R) \non \\ &&-\frac{1}{864}
(\bar\lambda_L \s^i \bar\g^{\mu\nu} \lambda_R - \bar\lambda_R \s^i \g^{\mu\nu} \lambda_L)
(\bar\lambda_L \s^i \bar\g_{\mu\nu} \lambda_R - \bar\lambda_R \s^i \g_{\mu\nu} \lambda_L) \non \\ &&
+\frac{1}{36}
(\bar\lambda_L \lambda_R -\bar\lambda_R \lambda_L) (\bar\lambda_L \lambda_R -\bar\lambda_R \lambda_L).\eea
Naively, this contribution does not look $U(1)$ gauge invariant, because of the $O(\lambda_R^4)$ and $O(\lambda_L^4)$ 
terms in the last two lines of \C{radd}.

However, the non gauge invariant terms in \C{radd} are given by
\bea \label{nongv}
\frac{1}{36} \Big[ (\bar\lambda_L \lambda_R)^2 - \frac{1}{24} (\bar\lambda_L \s^i \bar\g^{\mu\nu} 
\lambda_R)^2\Big] 
+ \frac{1}{36} \Big[ (\bar\lambda_R \lambda_L)^2 - \frac{1}{24} (\bar\lambda_R \s^i \g^{\mu\nu}  \lambda_L)^2\Big]. \eea
But \C{nongv} vanishes using the relations
\bea (\bar\lambda_L \s^i \bar\g^{\mu\nu} \lambda_R)^2 &=& 14 (\bar\lambda_L \lambda_R)^2 - \frac{1}{12}
(\bar\lambda_L \bar\g_{\mu\nu\lambda\rho} \lambda_R)^2 , \non \\ (\bar\lambda_L \lambda_R)^2 &=& -\frac{1}{120}
(\bar\lambda_L \bar\g_{\mu\nu\lambda\rho} \lambda_R)^2 , \eea
and their conjugates, which can be deduced using the Fierz identities \C{fierz}. Thus \C{radd} is gauge invariant as well.

In order to simplify the remaining contributions from \C{add1} and \C{radd},
we use the relations
\bea (\bar\lambda_L \s^i \bar\g^{\mu\nu} \lambda_R) (\bar\lambda_R \s^i \g_{\mu\nu}
\lambda_L) &=&  -7 (\bar \lambda_R \g^\mu \lambda_R)^2 +\frac{1}{6}
(\bar\lambda_R \g^{\mu\nu\rho} \lambda_R)^2 , \non \\ (\bar\lambda_R \s^i
\g^\mu \lambda_R) (\bar\lambda_R 
\s^i \g_\mu \lambda_R) &=&  \frac{1}{2} (\bar \lambda_R \g^\mu \lambda_R)^2 -\frac{1}{12}
(\bar\lambda_R \g^{\mu\nu\rho} \lambda_R)^2 , \non \\ (\bar\lambda_R \s^i \g^{\mu\nu\rho} \lambda_R) (\bar\lambda_R \s^i \g_{\mu\nu\rho}
\lambda_R) &=&  -21 (\bar \lambda_R \g^\mu \lambda_R)^2 -\frac{5}{2}
(\bar\lambda_R \g^{\mu\nu\rho} \lambda_R)^2 ,\non \\ (\bar\lambda_L
\lambda_R)(\bar\lambda_R \lambda_L) &=& \frac{1}{4} (\bar \lambda_R \g^\mu \lambda_R)^2 +\frac{1}{24}
(\bar\lambda_R \g^{\mu\nu\rho} \lambda_R)^2 .\eea
Thus \C{add1} and \C{radd} add to give
\be \frac{1}{384} \Big[ 30 (\bar\lambda_R \g^\mu \lambda_R) (\bar\lambda_R \g_\mu \lambda_R)
+ (\bar\lambda_R \g^{\mu\nu\rho} \lambda_R)(\bar\lambda_R \g_{\mu\nu\rho} \lambda_R)\Big], \ee
leading to the quartic fermion term 
\be \label{lambda4}
e^{-1} \mathcal{L}^{(0)} = \frac{1}{96} \Big[ 30 (\bar\lambda_R \g^\mu \lambda_R) (\bar\lambda_R \g_\mu \lambda_R)
+ (\bar\lambda_R \g^{\mu\nu\rho} \lambda_R)(\bar\lambda_R \g_{\mu\nu\rho} \lambda_R)\Big]\ee
in the $d=8$ action.

\section{Constructing the Laplacians on the moduli spaces}

We calculate the Laplacians on the moduli spaces $SO(2)  \backslash SL(2,\mathbb{R})$ and $SO(3)  \backslash
SL(3,\mathbb{R})$. Let the matrix $M$ parametrize the elements of the coset space $H \backslash G$. 
Then the metric on the moduli space $H \backslash G$ is defined by~\cite{Obers:1999es,Obers:1999um}
\be \label{formg} -\frac{1}{2} {\rm Tr} (d M d M^{-1}) = g_{AB} d z^A d z^B ,\ee
where $z^A$ are the coordinates on the moduli space $H \backslash G$. Then the Laplacian $\Delta$ is 
given by
\be \Delta =  \frac{1}{\sqrt{g}} \p_A (\sqrt{g} g^{AB} \p_B).  \ee  

\subsection{The Laplacian on $SO(2)  \backslash SL(2,\mathbb{R})$}

For $SO(2)  \backslash SL(2,\mathbb{R})$, from \C{SL2M} we obtain 
\be 
M^{-1} = \frac{1}{U_2} \begin{pmatrix} 1 & U_1 \\ U_1 & \vert U \vert^2  
\end{pmatrix} , \ee
leading to
\be  -\frac{1}{2} {\rm Tr} (d M d M^{-1}) = \frac{d U d \bar{U}}{U_2^2}.\ee
Thus, we have that
\be \label{LapSL2} \Delta = 4 U_2^2 \frac{\p^2}{\p U \p \bar{U}} . \ee

\subsection{The Laplacian on $SO(3)  \backslash SL(3,\mathbb{R})$}

If we directly use the matrix $M$ in \C{SL3M} to calculate the metric on the $SO(3)  \backslash SL(3,\mathbb{R})$
moduli space using \C{formg}, the calculation gets very involved. The calculation is considerably simplified
if we express $M$ in terms of $L$. Thus,
\be (d L_m^{~i})(d L_i^{~m}) - M^{mn} (d L_m^{~i})(d L_n^{~i}) = -g_{AB} d z^A d z^B ,\ee
where we use \C{SL3v2},  
\be  L_i^{~m} = \begin{pmatrix} e^{\hat\phi/3} /\sqrt{T_2} & 
  e^{\hat\phi/3} T_1/\sqrt{T_2}   & e^{\hat\phi/3}  \xi_1/\sqrt{T_2} \\ 
0 & e^{\hat\phi/3} \sqrt{T_2}  &  e^{\hat\phi/3} \xi_2/ \sqrt{T_2} 
\\ 0 & 0 &  e^{-2\hat\phi/3} \end{pmatrix} ,\ee
and
\be  M^{-1} = \frac{e^{2\hat\phi/3}}{T_2} \begin{pmatrix} 1 & T_1   & \xi_1 \\ 
T_1  & \vert T \vert^2 &  {\rm Re} (\xi \bar{T}) 
\\ \xi_1 & {\rm Re} (\xi \bar{T}) &  e^{-2\hat\phi}T_2 + \vert \xi \vert^2 \end{pmatrix} .\ee 
This leads to
\bea -\frac{1}{2} {\rm Tr} (d M d M^{-1}) &=& \frac{1}{3} e^{4\hat\phi} (d e^{-2\hat\phi})^2
+ \frac{\vert d T \vert^2}{T_2^2} + \frac{e^{2\hat\phi}}{T_2^3} \vert T_2 d
\xi - \xi_2 d T\vert^2 \non \\
&=& \frac{1}{3} \Big(\frac{d \nu}{\nu} \Big)^2
+ \frac{\vert d \tau \vert^2}{\tau_2^2} + \frac{\nu}{\tau_2^3} \vert \tau_2 d
B - B_2 d \tau \vert^2. \eea 

Thus we have that
\bea \label{LapSL3} \Delta &=& 3 e^{-4\hat\phi} \frac{\p^2}{\p (e^{-2\hat\phi})^2} + 4
e^{-2\hat\phi} T_2 \frac{\p^2}{\p \xi \p \bar\xi}   \non \\ &&+ 2 \Big[ \Big( T_2
  \frac{\p}{\p T} + \xi_2 \frac{\p}{\p \xi} + \frac{i}{2} \Big) \Big( T_2
  \frac{\p}{\p \bar{T}} + \xi_2 \frac{\p}{\p \bar\xi} 
  \Big) \non \\ && + \Big( T_2
  \frac{\p}{\p \bar{T} } + \xi_2 \frac{\p}{\p \bar\xi} - \frac{i}{2} \Big) \Big( T_2
  \frac{\p}{\p T} + \xi_2 \frac{\p}{\p \xi} 
  \Big) \Big] \non \\ &=& 3  \frac{\p}{\p \nu} \Big( \nu^2 \frac{\p}{\p \nu} \Big)+ 4
\nu^{-1} \tau_2 \frac{\p^2}{\p B \p \bar{B}}   \non \\ &&+ 2 \Big[ \Big( \tau_2
  \frac{\p}{\p \tau} + B_2 \frac{\p}{\p B} + \frac{i}{2} \Big) \Big( \tau_2
  \frac{\p}{\p \bar{\tau}} + B_2 \frac{\p}{\p \bar{B}} 
  \Big) \non \\ && + \Big( \tau_2
  \frac{\p}{\p \bar{\tau} } + B_2 \frac{\p}{\p \bar{B}} - \frac{i}{2} \Big) \Big( \tau_2
  \frac{\p}{\p \tau} + B_2 \frac{\p}{\p B} 
  \Big) \Big] . \eea

\section{Automorphic forms of the U--duality groups}

\subsection{Automorphic forms of $SL(2,\mathbb{Z})$}
\label{SL2aut}

We need various details about automorphic forms of $SL(2,\mathbb{Z})$. Under
an $SL(2,\mathbb{Z})$ transformation \C{autSL2}, an automorphic form $\Phi^{(m,n)}
(U,\bar{U})$ of 
weight $(m,n)$ transforms as
\be \Phi^{(m,n)} (U,\bar{U}) \rightarrow \Phi^{'(m,n)} (U',\bar{U}' ) = (\mathcal{C} U +
\mathcal{D})^m (\mathcal{C}
\bar{U} + \mathcal{D})^n \Phi^{(m,n)} (U,\bar{U}). \ee
The automorphic covariant derivatives of $SL(2,\mathbb{Z})$ are defined by
\be \label{covderSL2}
D_m = i \Big( U_2 \frac{\p}{\p U} - \frac{im}{2}\Big), \quad \bar{D}_n =
-i \Big( U_2 \frac{\p}{\p \bar{U}} + \frac{in}{2}\Big). \ee
Their actions on $\Phi^{(m,n)}$ are given by
\be D_m \Phi^{(m,n)} \rightarrow \Phi^{(m+1,n-1)}, \quad \bar{D}_n
\Phi^{(m,n)} \rightarrow \Phi^{(m-1,n+1)}. \ee 

First let us consider a class of automorphic forms of weight $(0,0)$.
These are given by the non--holomorphic Eisenstein series of $SL(2,\mathbb{Z})$ of
order $s$, defined by
\bea \label{Eisen}
E_s (U,\bar{U}) &=& \sum_{(p,q)\neq (0,0)}\frac{U_2^s}{\vert p + q
  U\vert^{2s}} \non \\ &=& 2 \zeta (2s) U_2^s + 2\sqrt{\pi} U_2^{1-s}
\frac{\Gamma (s-1/2)}{\Gamma (s)} \zeta (2s-1) \non \\ && + \frac{2\pi^s
  \sqrt{U_2}}{\Gamma (s)} \sum_{m \neq 0, n \neq 0} \Big\vert
\frac{m}{n}\Big\vert^{s-1/2} K_{s-1/2} (2\pi \vert mn \vert U_2) e^{2\pi i mn U_1},\eea
which satisfy
\be 4 \bar{D}_{-1} D_0 E_s (U,\bar{U}) = 4 D_{-1}\bar{D}_{0}  E_s (U,\bar{U}) = \Delta E_s (U,\bar{U}) = s(s-1) E_s (U,\bar{U}),\ee
where the Laplacian is given by \C{LapSL2}.

We shall call them
\be f^{(0,0)}_s (U,\bar{U}) \equiv E_s (U,\bar{U}). \ee
Then we define automorphic forms of weight $(m,-m)$ as
\bea \label{nzw} f^{(m,-m)}_s (U,\bar{U}) &\equiv& D_{m-1} \ldots D_1 D_0 f^{(0,0)}_s
(U,\bar{U}) \non \\ &=& \frac{\Gamma (s+m)}{2^m \Gamma(s)}
\sum_{(p,q) \neq (0,0)}\Big( \frac{p+q\bar{U}}{p+q
  U}\Big)^{m}
\frac{U_2^s}{\vert p + q U   \vert^{2s}}, \eea
which satisfy
\be \label{vallap}
4 D_{m-1} \bar{D}_{-m} f^{(m,-m)}_s  = 4 \bar{D}_{-(m+1)} D_m f^{(m,-m)}_s  = (s+m)(s-m-1) f^{(m,-m)}_s .\ee
The Eisenstein series defined by \C{Eisen} diverges for $s=1$, and it has to
be properly regularized. This is done by setting $1-s
=\epsilon$, and noting that as $\epsilon \rightarrow 0$, the divergence appears as a simple pole in $\Gamma
(\epsilon)$ on using
\be \Gamma (s-1/2) \zeta (2s-1) = \pi^{2s-3/2} \Gamma (1-s) \zeta (2-2s),\ee
and
\be \Gamma (\epsilon) = \frac{1}{\epsilon} -\gamma + O(\epsilon).\ee 
We then perform an $\overline{MS}$ kind of regularization where we remove the
$1/\epsilon$ pole as well as the $O(1)$ terms including the Euler constant, leading to 
the regularized Eisenstein series
\be \label{finexp} \hat{E}_1 (U,\bar{U}) = -\pi {\rm ln} \Big( U_2 \vert \eta (U)
\vert^4\Big).\ee
In obtaining \C{finexp}, we have used
\be K_{1/2} (x) = \sqrt{\frac{\pi}{2x}}e^{-x},\ee
the definition of the Dedekind eta function
\be \eta (U) = e^{i\pi U/12} \prod_{k=1}^\infty (1-e^{2\pi i k U}) ,\ee
and
\be \zeta (2) = \frac{\pi^2}{6}, \quad \zeta (0) = -\frac{1}{2}. \ee
Thus $\hat{E}_1 (U,\bar{U})$ satisfies
\be \Delta \hat{E}_1 (U,\bar{U}) = \pi.\ee
However, the modular forms of  non--zero weights for $s=1$ which are
constructed using \C{nzw} do not have to be regularized because $D_0$ removes
the divergence of the offending term. In particular,
\be f^{(1,-1)}_1 (U,\bar{U}) = D_0 E_1 (U,\bar{U}) = -\frac{\pi}{2} -2\pi i
U_2 \frac{\p \eta(U)}{\p U} , \ee  
which satisfies \C{vallap}.

\subsection{Automorphic forms of $SL(3,\mathbb{Z})$}
\label{SL3aut}

We consider a family of automorphic forms of $SL(3,\mathbb{Z})$ which are
invariant under the $SL(3,\mathbb{Z})$ transformations given by \C{trans2}.
They are given by the Eisenstein series of order $s$ defined by
\bea \label{EisenSL3}
E_s (M) &=& \sum'_{m_m} (m_m M_{mn} m_n)^{-s} \non \\ &=& \sum'_{m_m}
e^{-4s\hat\phi/3} \Big[ e^{-2\hat\phi} \frac{\vert m_1 T + m_2 \vert^2}{T_2}
+ \frac{1}{T_2^2} \Big(m_1 {\rm Im} \xi \bar{T} + m_2 \xi_2 + m_3  T_2
\Big)^2\Big]^{-s} \non \\ &=& \sum'_{m_m}
\nu^{-s/3} \Big[ \frac{1}{\tau_2} \vert m_1 + m_2 \tau + m_3 B \vert^2 +
+ \frac{m_3^2}{\nu} \Big]^{-s},\eea
where $m_m$ are integers, and the sum excludes $\{ m_1, m_2 , m_3\} = \{ 0,0,0
\}$,
which satisfies
\be  \Delta E_s (M) = \frac{2s}{3}  (2s-3) E_s (M),\ee 
where the Laplacian is given by \C{LapSL3}.

The Eisenstein series \C{EisenSL3} can be expanded for weak string coupling as
\bea E_s (M) &=& 2 \zeta (2s) (e^{-2\hat\phi})^{2s/3} +  \frac{\sqrt{\pi}\Gamma
  (s-1/2)}{\Gamma (s)} (e^{-2\hat\phi})^{1/2-s/3} E_{s-1/2} (T,\bar{T})  \non
\\ &&+\frac{2\pi^s (e^{-2\hat\phi})^{s/6+1/4}}{\Gamma
  (s) T_2^{1/4-s/2}}  \sum_{m \neq 0, n \neq 0}
\Big\vert \frac{m}{n} \Big\vert^{s-1/2} K_{s-1/2} (2\pi \vert mn \vert
\tau_2)e^{2\pi i mn \tau_1}
\non \\&& + \frac{2\pi^s (e^{-2\hat\phi})^{1/2-s/3}}{\Gamma (s) \sqrt{T}_2}  
\sum_{m \neq 0, p \neq 0, n} \Big\vert \frac{n-m\tau}{p}\Big\vert^{s-1}
K_{s-1} (2\pi \vert p(n-m\tau)\vert T_2)  e^{2\pi i p (n T_1 -m \xi_1)} .\non \\\eea

In the main text, we consider interactions in the effective action which
transform non--trivially under $SU(2)$, and we denote their couplings as
\be f_{(i_1 j_i) \ldots (i_n j_n)} (T,\bar{T},\xi,\bar\xi, e^{-2\hat\phi}) \ee
where every $(ij)$ index is symmetrized and traceless, and is in the spin $2$ representation of $SU(2)$. 
Furthermore, the coupling is symmetric under the interchange of any pair of
$(ij)$ indices. 
Thus, the couplings we consider are the coefficient functions of interactions
which are in the spin $2n$ representation of $SU(2)$. 

The various couplings are related to each other by the action of generalized
derivatives defined by
\bea \label{sl3exp}
D_{(ij)}^{(n)} &=& -2 \Big[\delta_{3(i} \delta_{j)3} -\frac{1}{2}
  \Big(\delta_{1(i} \delta_{j)1} + \delta_{2(i} \delta_{j)2} \Big)\Big]\Big( e^{-2\hat\phi} \frac{\p}{\p
  e^{-2\hat\phi}} -n \Big)\non \\ &&
+ \Big[ -2 \delta_{1(i} \delta_{j)2} +i\Big(\delta_{1(i} \delta_{j)1} -\delta_{2(i} \delta_{j)2}\Big)\Big]\Big(T_2 \frac{\p}{\p T} 
+ \xi_2 \frac{\p}{\p \xi} + \frac{in}{2}\Big) \non \\ &&+ \Big[ -2 \delta_{1(i} \delta_{j)2} -i\Big(\delta_{1(i} \delta_{j)1} -\delta_{2(i} 
\delta_{j)2}  \Big)\Big]\Big(T_2 \frac{\p}{\p \bar{T}} 
+ \xi_2 \frac{\p}{\p \bar\xi} - \frac{in}{2} \Big) \non \\ &&- 2 \sqrt{T_2} e^{-\hat\phi} \delta_{3(j} \Big[ \Big(\delta_{i)1} + i \delta_{i)2} \Big)
\frac{\p}{\p \xi}  + \Big(\delta_{i)1}  - i \delta_{i)2} \Big)
\frac{\p}{\p \bar\xi}\Big] \non \\
&=& -2  \Big[\delta_{1(i} \delta_{j)1} -\frac{1}{2}
  \Big(\delta_{2(i} \delta_{j)2} + \delta_{3(i} \delta_{j)3} \Big)\Big] \Big(
\nu \frac{\p}{\p \nu} +n \Big) \non \eea
\bea &&
+ \Big[ -2 \delta_{2(i} \delta_{j)3} +i\Big(\delta_{2(i} \delta_{j)2} -\delta_{3(i} \delta_{j)3}\Big)\Big]\Big(\tau_2 \frac{\p}{\p \tau} 
+ B_2 \frac{\p}{\p B} + \frac{in}{2}\Big) \non \\ &&+ \Big[ -2 \delta_{2(i} \delta_{j)3} -i\Big(\delta_{2(i} \delta_{j)2} -\delta_{3(i} 
\delta_{j)3}\Big)\Big]\Big(\tau_2 \frac{\p}{\p \bar\tau} 
+ B_2 \frac{\p}{\p \bar{B}} - \frac{in}{2}\Big) \non \\ &&+ 2 \sqrt{\frac{\tau_2}{\nu}}  \delta_{1(i} \Big[ \Big(\delta_{j)3} - i \delta_{j)2} \Big)
\frac{\p}{\p B}  + \Big(\delta_{j)3}  + i \delta_{j)2} \Big)
\frac{\p}{\p \bar{B}}\Big], \eea
where the explicit mapping from one coupling to another using \C{sl3exp} is given in the main text.
Note that
\be (D^{(n)}_{(ij)})^\dagger = D^{(n)}_{(ij)}, \ee
and
\be 2 D^{(1)}_{(ij)} D^{(0)}_{(ij)} = \Delta. \ee
These generalized derivatives transform in a complicated way under
$SL(3,\mathbb{Z})$ transformations given by \C{trans2}. Thus, for example,
acting on an $SL(3,\mathbb{Z})$ invariant automorphic form $g^{(0)} (M)$ it
gives $D^{(n)}_{(ij)} g^{(0)} (M)$, which is not an automorphic form of any
definite weight. This is evident from \C{sl3exp} because the various
components transform differently. Of course, though the explicit form of
$D^{(n)}_{(ij)}$ depends on how the $SU(2)$ has been gauge fixed, the fact
that it does not produce automorphic forms is a gauge invariant statement.

The Eisenstein series defined by \C{EisenSL3} diverges for $s=3/2$, for the
same reason as the divergence in section \C{SL2aut}, and is regularized in the
same way. Thus we define
\bea \hat{E}_{3/2} (M) &=& 2 \zeta (3) e^{-2\hat\phi} + 2 \hat{E}_1 (T,\bar{T}) +
4\pi e^{-\hat\phi} \sqrt{T}_2 \sum_{m \neq 0, n \neq 0}
\Big\vert \frac{m}{n} \Big\vert K_1 (2\pi \vert mn \vert
\tau_2)e^{2\pi i mn \tau_1} \non \\ &&+ \frac{2\pi}{T_2} \sum_{m \neq 0, p \neq
  0, n} \frac{1}{\vert p \vert} e^{-2\pi T_2 \vert p(n-m\tau)\vert +2 \pi i p (n T_1 -m \xi_1)},\eea 
which satisfies
\be \Delta \hat{E}_{3/2} (M) = 2\pi .\ee



\providecommand{\href}[2]{#2}\begingroup\raggedright\endgroup

\end{document}